\documentclass[a4paper,11pt]{article}
\pdfoutput=1 % if you are submitting a pdflatex (i.e. if you have
\usepackage{jcappub} % for details on the use of the package, please
\usepackage[T1]{fontenc}

\usepackage{aas_macros} % for the bib shorthand commands from ADS

\usepackage{array}
\usepackage{graphicx}
\usepackage{bm}
\usepackage[dvipsnames]{xcolor}     
\usepackage{makecell}
\usepackage{hyperref}
\usepackage[normalem]{ulem}
\usepackage{multicol,multirow}

\usepackage{lineno}
\usepackage{orcidlink} % ORCID links for author list
\usepackage{doi} % supplementary material link

\newcommand{\frachalf}{\frac{1}{2}}

\newcommand{\hMpc}{h^{-1}\text{Mpc}}

\newcommand{\hGpc}{\ h^{-1}\text{Gpc}}

\newcommand{\be}{\begin{equation}}
\newcommand{\ee}{\end{equation}}

\newcommand{\sL}{\mathcal{L}}

\newcommand{\vone}{\mathbf{1}}

\newcommand{\vn}{\mathbf{n}}
\newcommand{\vbn}{\mathbf{\bn}}

\newcommand{\vdelta}{\bm{\delta}}

\newcommand{\vS}{\mathbf{S}}

\newcommand{\vN}{\mathbf{N}}

\newcommand{\vM}{\mathbf{M}}

\newcommand{\vB}{\mathbf{B}}

\newcommand{\bn}{\bar{n}}

\usepackage[nameinlink,noabbrev]{cleveref}
\crefname{equation}{Eq.}{Eqs.}
\crefname{section}{Section}{Sections}
\crefname{figure}{Fig.}{Figs.}
\crefname{table}{Table}{Tables}
\crefname{appendix}{Appendix}{Appendices}
\Crefname{figure}{Figure}{Figures}
\Crefname{equation}{Equation}{Equations}
\Crefname{section}{Section}{Sections}
\Crefname{table}{Table}{Tables}

\newcommand{\aperp}{\alpha_\perp}
\newcommand{\apar}{\alpha_\parallel}
\newcommand{\aiso}{\alpha_{\rm iso}}

\newcommand{\alap}{\alpha_{\rm AP}}

\newcommand{\wfkpcl}{w_{\rm FKPeff,L}}
\newcommand{\wfkpce}{w_{\rm FKPeff,E}}

\newcommand{\ihMpc}{\,h{\rm Mpc}^{-1}}
\newcommand{\ihGpc}{\,h^{-1}{\rm Gpc}}

\newcommand{\rascalc}{{\sc RascalC}}
\newcommand{\abacus}{{\tt Abacus}}
\newcommand{\abacussecond}{{\tt Abacus-2}}
\newcommand{\ezmock}{{\tt EZmock}}
\newcommand{\ezmocks}{{\tt EZmocks}}
\newcommand{\cross}{{$\times$}}

\newcommand{\lrgxelg}{{\tt LRG$\times$ELG}}
\newcommand{\lrgelg}{{\tt LRG$+$ELG}}

\newcommand{\desidrone}{{DESI DR1}}

\newcommand{\comntile}{\langle C_{\rm assign} \rangle(n_{\rm tile})}
\newcommand{\comntilei}{\langle C_{\rm assign}^i \rangle(n_{\rm tile})}

\newcommand{\elgo}{{\tt ELG1}}
\newcommand{\elgt}{{\tt ELG2}}

\newcommand{\elg}{{\tt ELG}}
\newcommand{\lrgo}{{\tt LRG1}}
\newcommand{\lrgt}{{\tt LRG2}}
\newcommand{\lrgth}{{\tt LRG3}}
\newcommand{\lrg}{{\tt LRG}}

\newcommand{\supplementarymaterial}{\doi{10.5281/zenodo.19561326}}

\title{Combined tracer analysis for DESI 2024 BAO}

\affiliation{Affiliations are in \cref{sec:affiliations}.}

\author[1,2]{{D.~Valcin}\orcidlink{0000-0003-0129-0620},}
\author[3,4,5]{{M.~Rashkovetskyi}\orcidlink{0000-0001-7144-2349},}
\author[1]{{H.~Seo}\orcidlink{0000-0002-6588-3508},}
\author[6]{{F.~Beutler}\orcidlink{0000-0003-0467-5438},}
\author[7]{{P.~McDonald}\orcidlink{0000-0001-8346-8394},}
\author[8]{{A.~de~Mattia}\orcidlink{0000-0003-0920-2947},}
\author[1]{{A.~J.~Rosado-Mar\'{i}n}\orcidlink{0000-0001-7545-3504},}
\author[4,9,10]{{A.~J.~Ross}\orcidlink{0000-0002-7522-9083},}
\author[11]{{N.~Padmanabhan},}
\author[7]{{J.~Aguilar},}
\author[12]{{S.~Ahlen}\orcidlink{0000-0001-6098-7247},}
\author[13,14]{{U.~Andrade}\orcidlink{0000-0002-4118-8236},}
\author[15,16]{{D.~Bianchi}\orcidlink{0000-0001-9712-0006},}
\author[17]{{D.~Brooks},}
\author[7]{{E.~Chaussidon}\orcidlink{0000-0001-8996-4874},}
\author[18]{{S.~Chen}\orcidlink{0000-0002-5762-6405},}
\author[11]{{X.~Chen}\orcidlink{0000-0003-3456-0957},}
\author[7]{{T.~Claybaugh},}
\author[7]{{A.~Cuceu}\orcidlink{0000-0002-2169-0595},}
\author[19]{{K.~S.~Dawson}\orcidlink{0000-0002-0553-3805},}
\author[20]{{A.~de la Macorra}\orcidlink{0000-0002-1769-1640},}
\author[21,22]{{Biprateep~Dey}\orcidlink{0000-0002-5665-7912},}
\author[23]{{Z.~Ding}\orcidlink{0000-0002-3369-3718},}
\author[17]{{P.~Doel},}
\author[7,2]{{S.~Ferraro}\orcidlink{0000-0003-4992-7854},}
\author[24]{{A.~Font-Ribera}\orcidlink{0000-0002-3033-7312},}
\author[25,26]{{J.~E.~Forero-Romero}\orcidlink{0000-0002-2890-3725},}
\author[27,28,29]{{E.~Gaztañaga}\orcidlink{0000-0001-9632-0815},}
\author[7,30]{{S.~Gontcho A Gontcho}\orcidlink{0000-0003-3142-233X},}
\author[31]{{G.~Gutierrez},}
\author[32]{{C.~Hahn}\orcidlink{0000-0003-1197-0902},}
\author[4,5,10]{{K.~Honscheid}\orcidlink{0000-0002-6550-2023},}
\author[33]{{C.~Howlett}\orcidlink{0000-0002-1081-9410},}
\author[34]{{M.~Ishak}\orcidlink{0000-0002-6024-466X},}
\author[35]{{R.~Kehoe},}
\author[36]{{D.~Kirkby}\orcidlink{0000-0002-8828-5463},}
\author[7]{{T.~Kisner}\orcidlink{0000-0003-3510-7134},}
\author[7]{{A.~Kremin}\orcidlink{0000-0001-6356-7424},}
\author[17]{{O.~Lahav},}
\author[7]{{A.~Lambert},}
\author[7]{{M.~Landriau}\orcidlink{0000-0003-1838-8528},}
\author[7]{{M.~E.~Levi}\orcidlink{0000-0003-1887-1018},}
\author[37,24]{{M.~Manera}\orcidlink{0000-0003-4962-8934},}
\author[38]{{A.~Meisner}\orcidlink{0000-0002-1125-7384},}
\author[39]{{J.~Mena-Fern\'andez}\orcidlink{0000-0001-9497-7266},}
\author[40,24]{{R.~Miquel},}
\author[41]{{J.~Moustakas}\orcidlink{0000-0002-2733-4559},}
\author[28]{{S.~Nadathur}\orcidlink{0000-0001-9070-3102},}
\author[42,32]{{E.~Paillas}\orcidlink{0000-0002-4637-2868},}
\author[8,7]{{N.~Palanque-Delabrouille}\orcidlink{0000-0003-3188-784X},}
\author[43,44,45]{{W.~J.~Percival}\orcidlink{0000-0002-0644-5727},}
\author[46]{{F.~Prada}\orcidlink{0000-0001-7145-8674},}
\author[47]{{I.~P\'erez-R\`afols}\orcidlink{0000-0001-6979-0125},}
\author[48]{{G.~Rossi},}
\author[49]{{R.~Ruggeri}\orcidlink{0000-0002-0394-0896},}
\author[50,51,52]{{L.~Samushia}\orcidlink{0000-0002-1609-5687},}
\author[53]{{E.~Sanchez}\orcidlink{0000-0002-9646-8198},}
\author[54]{{C.~Saulder}\orcidlink{0000-0002-0408-5633},}
\author[7]{{D.~Schlegel},}
\author[55,14]{{M.~Schubnell},}
\author[7]{{J.~Silber}\orcidlink{0000-0002-3461-0320},}
\author[38]{{D.~Sprayberry},}
\author[14]{{G.~Tarl\'{e}}\orcidlink{0000-0003-1704-0781},}
\author[38]{{B.~A.~Weaver},}
\author[56]{{J.~Yu}\orcidlink{0009-0001-7217-8006},}
\author[7]{{R.~Zhou}\orcidlink{0000-0001-5381-4372},}
\author[57]{{H.~Zou}\orcidlink{0000-0002-6684-3997}}

\emailAdd{dvalcin@berkeley.edu}

\abstract{
This paper demonstrates how the Dark Energy Spectroscopic Instrument (DESI) Data Release 1 (DR1) and future baryon acoustic oscillations (BAO) analyses can optimally combine overlapping tracers (galaxies of distinct types) in the same redshift range.
We make a unified catalog of Luminous Red Galaxies (LRGs) and Emission Line Galaxies (ELGs) in the redshift range $0.8<z<1.1$ and investigate the impact on the BAO constraints.
DESI DR1 contains $\sim 30\%$ of the final DESI LRG sample and less than $25\%$ of the final ELG sample, and the combination of LRGs and ELGs increases the number density and reduces the shot noise. 
We developed a pipeline to merge the overlapping tracers using galaxy bias as an approximately optimal weight and tested the pipeline on a suite of $\abacus$ simulations, calibrated on the final version of the DESI Early Data Release.
When applying our pipeline to the DESI DR1 catalog, we find an improvement in the BAO constraints of $11\%$ for $\aiso$ and $\sim 7.0\%$ for $\alap$ consistent with our findings in mock catalogs.
Our analysis was integrated into the DESI DR1 BAO analysis to produce the \lrgelg\ result in the $0.8<z<1.1$ redshift bin, which provided the most precise BAO measurement from DESI DR1 with a 0.86\% constraint on the BAO distance scale and a 9.1$\sigma$ detection of the isotropic BAO feature.
}


\begin{document}

% \label{firstpage}
% \pagerange{\pageref{firstpage}--\pageref{lastpage}}

\maketitle
\flushbottom

% \tableofcontents

\section{Introduction}
\label{sec:intro}

Baryon Acoustic Oscillations (BAO) are sound waves that propagated in the photon-baryon plasma in the early Universe due to the competing effects of gravity and radiation pressure~\citep{1970ApJ...162..815P,1970Ap&SS...7...20S,1987MNRAS.226..655B}.
The opposing effects of these two forces created acoustic waves, which in turn moved the material into concentric shells with a characteristic radius based on the time these waves were able to travel until near the epoch of recombination.
This radius represents a special scale in the distribution of matter (or galaxies), which can be used as a standard ruler to map out the expansion history of the Universe~\citep{2003ApJ...594..665B,2003ApJ...598..720S}.
Since the first detection of the BAO feature about 20 years ago~\citep{2001MNRAS.327.1297P,2005MNRAS.362..505C,2005ApJ...633..560E}, it has established itself as one of the most reliable low-redshift observables at redshifts $z < 3$ in cosmology~\citep{2017MNRAS.470.2617A,2021PhRvD.103h3533A}.

BAO measurements have been reported in multiple galaxy redshift surveys from redshift $z = 0.1$~\citep{2011MNRAS.416.3017B,2018MNRAS.481.2371C} to redshift $z = 1.5$~\citep{2020MNRAS.499..210N} using galaxies and quasars, and at even higher redshift ($z=2.3$) using the Lyman-alpha forest~\citep{2020ApJ...901..153D}.
Because the BAO is a very large-scale feature ($\approx 150$~Mpc today), there are very few known processes that can easily remove or alter this signal.
The large-scale bulk flow has been shown to dampen the BAO feature and shift its peak position~\citep{2008PhRvD..77b3533C,2007ApJ...664..660E,2008PhRvD..77d3525S,2007MNRAS.375.1329G}, with the effect being largest at low redshift.
However, since the bulk flow (and the related displacement) is sourced by the density field, one can use the measured galaxy distribution to estimate the displacements and reverse most of these effects, a procedure known as density field reconstruction~\citep{2007ApJ...664..675E,2012MNRAS.427.2132P}.
Even though density-field reconstruction only works on linear and quasi-linear scales, it has been shown to improve the signal-to-noise ratio substantially, especially at low redshift, and effectively remove the shift in the BAO scale well below percent level~\citep{2011ApJ...734...94M,Ding2017:1708.01297v2}.

This paper is part of a series on the BAO analysis of the first data release (DR1) of the Dark Energy Spectroscopic Instrument (DESI). 
The DESI survey \citep{Snowmass2013.Levi,DESI2016a.Science} covers a very large redshift range from $z \sim 0.2$ to $z \sim 3.5$ using different tracers at different redshift intervals.
This naturally leads to multiple tracers being present at the same redshift range.
Here we investigate how best to combine such tracers in light of a BAO analysis.
We focus on the redshift range $0.8 < z <1.1$ where we have both Luminous Red Galaxies \citep[LRGs,][]{LRG.TS.Zhou.2023} and Emission Line Galaxies \citep[ELGs,][]{ELG.TS.Raichoor.2023} available as tracers.
An optimal analysis relies on combining the LRGs and ELGs for both the clustering estimator and the density field reconstruction.
Studying the clustering signal of two different tracer types in the same volume also allows for tests of potential galaxy sample-specific systematic effects, including redshift space distortions, non-linear clustering and non-linear galaxy bias.
Whereas most of these tracer-dependent effects have been studied carefully in N-body simulations (e.g.~\citep{2024arXiv240214070C, KP4s10-Mena-Fernandez,KP4s11-Garcia-Quintero, Ross2017}), systematic studies directly on the data have the advantage of relying on fewer assumptions, e.g., about the fiducial cosmology or the models for the galaxy-halo connection.

One interesting potential source of systematic bias for BAO is the relative velocity effect~\citep{2010PhRvD..82h3520T}.
The fact that dark matter and baryons have a different velocity profile directly after decoupling can affect early galaxy formation processes.
For example, wherever the relative velocity is large, baryons can escape the gravitational potential, reducing the chance of galaxy formation.
Hence, the selection effects due to the relative velocity can introduce additional terms in the large-scale power spectrum~\citep{2013PhRvD..88j3520Y}, the extent of which depends on the galaxy populations through halo mass and stellar mass~\citep{barreira_baryon-cdm_2020}.
Given that the relative velocity effect is sourced by pre-recombination physics, just like the BAO itself, this effect can bias BAO measurements if not accounted for.
Investigations using the two-point and three-point clustering of the Baryon Oscillation Spectroscopic Survey (BOSS) and the WiggleZ Dark Energy Survey have not been able to detect this effect~\citep{2016MNRAS.455.3230B,2017MNRAS.470.2723B,2018MNRAS.474.2109S}.
Testing such effects in mock datasets is challenging, as it requires understanding the impact of the relative velocity effect on galaxy formation processes.
The tests on hydrodynamic simulations have detected the relative baryon-to-dark matter overdensity effect in galaxies \citep{barreira_baryon-cdm_2020}, while tests on the Lyman alpha forest have shown that the effect there from the relative velocity effect is small but potentially important at DESI precision~\citep{2018MNRAS.474.2173H}.
In \cite{KP4s2-Chen}, we include the estimated shift on the BAO scale for DESI galaxies and quasars in our systematic budget based on the former measurements, finding that the nonlinear shift due to the overdensity effect is dominant unless the relative velocity effect is at the very highest range explored in the literature. 
With the overlapping tracers, we can test the level of this systematics directly using the data by comparing BAO scales from different combinations of tracers that are likely subject to different levels of the relative velocity effect. In this paper, we will test if there is a significant difference in the DESI DR1 BAO scales when using different combinations of LRGs and ELGs clustering in the same redshift bin.

\begin{figure*}
    \center
    \includegraphics[width= 0.45\textwidth]{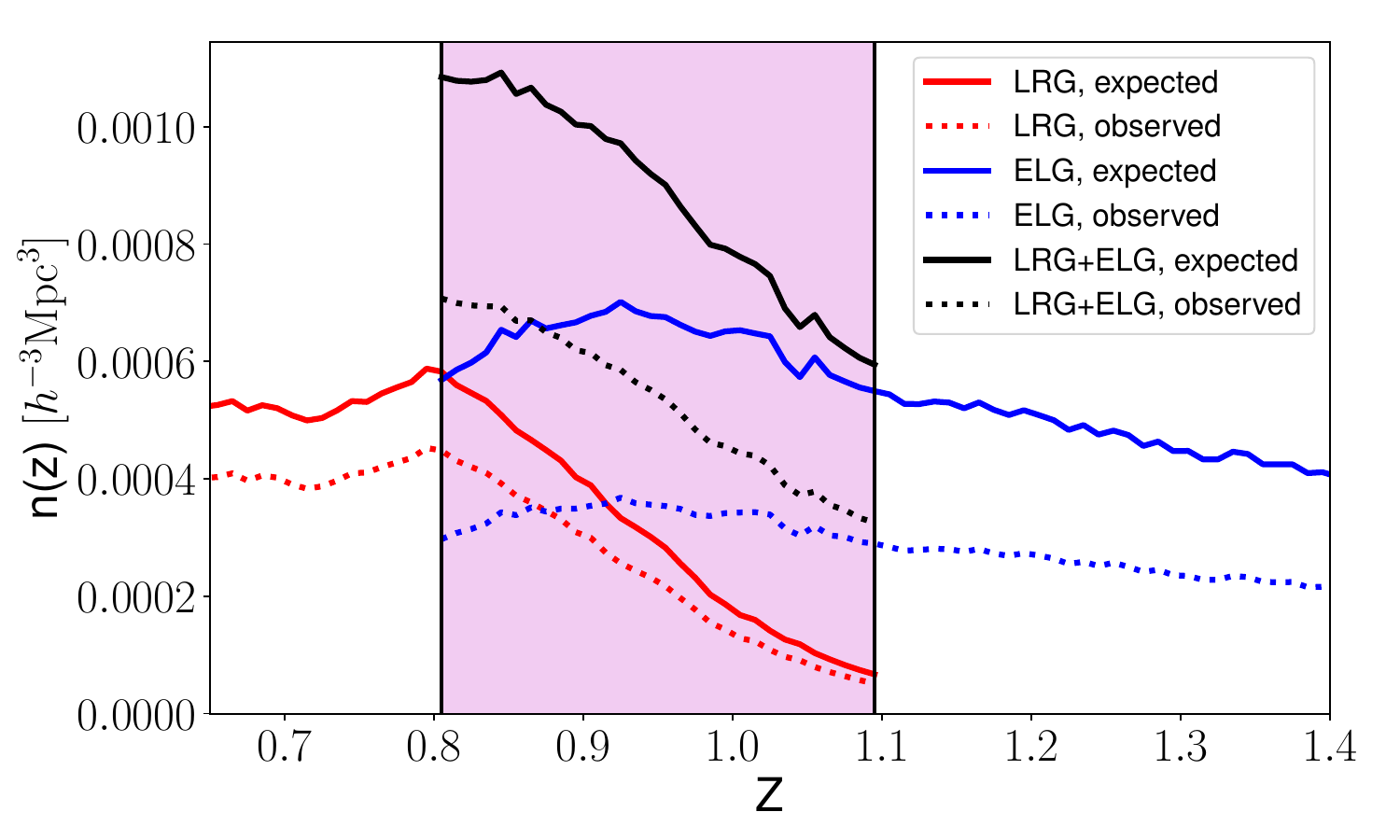}
    \includegraphics[width=0.45\textwidth]{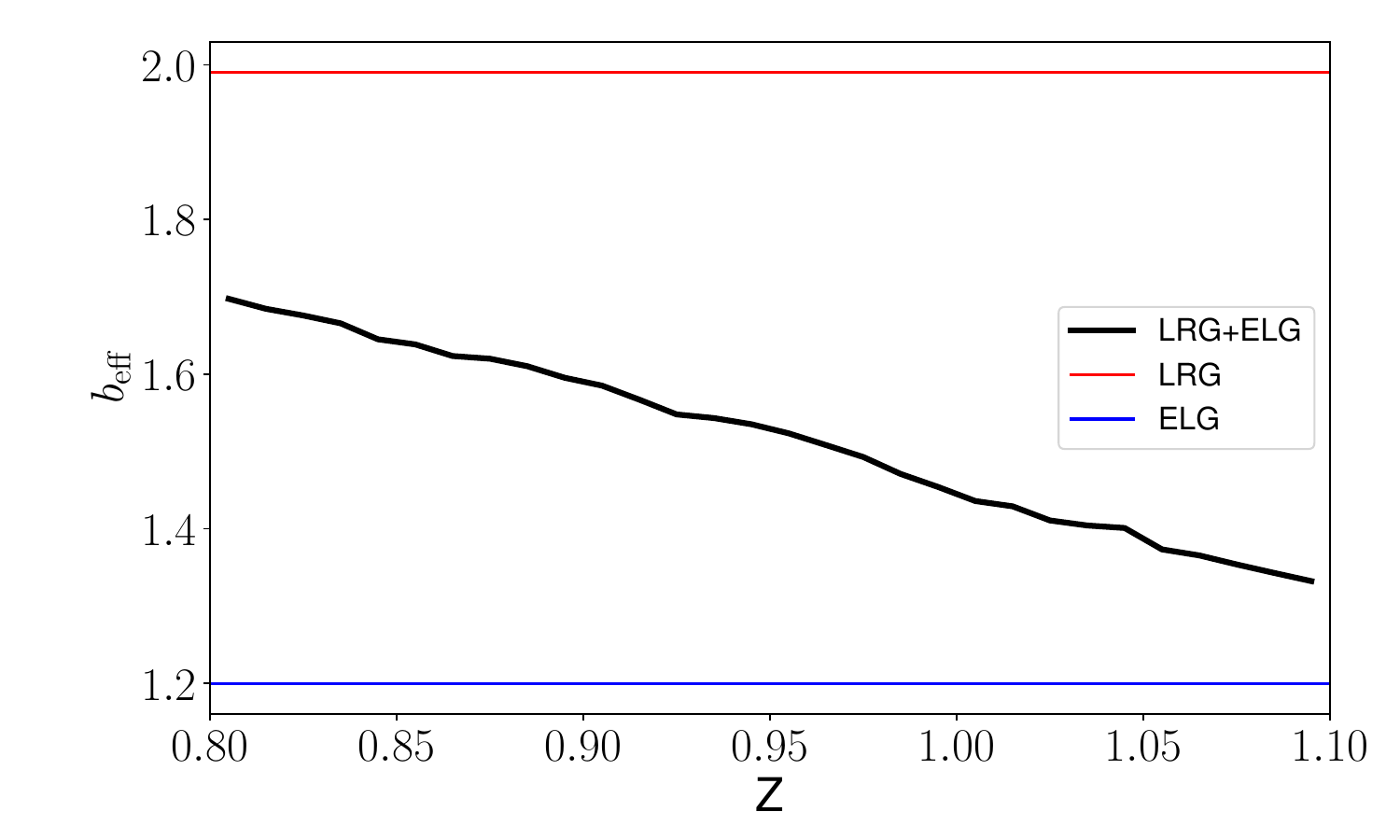}
    \caption{The left panel shows comoving number densities $n(z)$ of the relevant tracers. The shaded region corresponds to the overlapping range of the redshift we consider. The dotted lines correspond to the raw observed number density, whereas solid lines are corrected for incompleteness in target assignment.
    The black lines correspond to the effective number density of the combined catalog after each tracer is weighted according to \cref{eq:neffz}. The right panel shows the effective bias of the combined tracer (black) as a function of redshift. The horizontal lines represent $b$ for LRGs (red) and ELGs (blue) that were assumed constant (following the convention of \cite{DESI2024.III.KP4}) when calculating the effective bias for the combined tracer. In practice, galaxy bias evolves with redshift, such that the amplitude of the overall clustering stays about constant.  }
    \label{fig:red_disty1}
\end{figure*}

Combining the tracers into a unified catalog before the BAO analysis can have additional advantages compared to combining the BAO analysis results from multiple tracers.
First, in case a tracer suffers from high noise, its BAO constraint would be weak with a highly non-Gaussian distribution. 
This is the case for the DR1 ELG sample at $0.8<z<1.1$ due to its low fiber assignment completeness (35.3\%).
The sample represents less than 1/4 of the final DESI ELG sample. \cref{fig:red_disty1} shows the raw observed number density (dotted lines) and the expected number density corrected for completeness (solid lines); the difference between the two lines is much greater for ELGs than for LRGs, due to their low completeness.
Besides, as we can see on the right panel of \cref{fig:red_disty1}, the clustering amplitude of ELGs is also lower.
Both of these lower the signal-to-noise ratio of the BAO for ELGs, making the analysis suffer more from non-Gaussianity in the likelihood and less robust compared to other tracers.
Second, the combined tracer clustering includes information from the auto-clustering statistics of the single tracers as well as the cross-clustering statistics between the tracers.
The combination of tracers into a unified catalog before the clustering measurements and BAO analysis, therefore, enables a more robust (i.e., not impacted by non-Gaussianity of the single tracer ELG analysis) and complete (i.e., includes auto as well as the cross-clustering information with ELG) integration of the ELG information.
Third, the lower shot noise of the combined tracer can potentially improve the reconstruction efficiency beyond the reconstruction of the individual tracers, especially for ELGs.
This is why we expect the combination of the LRG and ELG catalogs over the overlapping redshifts to be beneficial to BAO analyses.
Whereas we also have QSOs in this redshift bin, due to its very low sample density, the gain by including quasars in this combined tracer is expected to be marginal.
Therefore, in this paper, we focus on the combined tracer of LRGs and ELGs. 
Based on the results presented in this paper, DESI DR1 BAO \cite{DESI2024.III.KP4} adopted the combined tracer \lrgelg\ as the baseline for this redshift bin.

The structure of the paper is as follows: in \cref{section:Dataset} we briefly summarize the DESI DR1 dataset and the properties of the large-scale structure catalogs, focusing on the ELGs and LRGs $0.8<z<1.1$.
In \cref{sec:Mocks} we describe the DESI DR1 mocks we used for the tests in this paper.
In \cref{sec:Methods} we present how we construct the combined catalog, measure and analyze the clustering statistics, and perform the BAO analysis.
In \cref{sec:Results} we assess the gain of the combined catalog using mocks as well as using DR1 data and test for systematics in the BAO measurements using the overlapping tracers.
Finally, in \cref{sec:Conclusion} we summarize the benefits of the combination of overlapping tracers and the prospects for future data releases.

\section{DESI Data Release 1 (DR1)}
\label{section:Dataset}
The DESI instrument \citep{DESI2016b.Instr,DESI2022.KP1.Instr} is located on the Mayall Telescope at Kitt Peak, Arizona.
It can simultaneously observe up to 5000 targets using robotic positioners that precisely place optical fibers on the focal plane at the celestial coordinates of the targets \citep{FocalPlane.Silber.2023, FBA.Raichoor.2024,Corrector.Miller.2023,FiberSystem.Poppett.2024,SurveyOps.Schlafly.2023}.
The collected light is then carried out to one of the 10 spectrographs available. 
A set of targets at a specific sky position is collected through observations of `tiles'~\citep{TS.Pipeline.Myers.2023}.
DESI observing strategy is separated into two categories depending on the observing conditions.
BGS observations are scheduled during ``bright'' time (with moonlight or during twilight), while QSO, LRG, and ELG targets are prioritized during ``dark'' time when the sky background is low.
During these observations, the tracers compete for fiber assignment based on a predefined priority system designed to optimize science goals.

The galaxies observed in the DESI Data Release 1 (DR1, \citep{DESI2024.I.DR1}) were collected during the main survey operations starting from May 14, 2021, after a period of survey validation \citep{DESI2023a.KP1.SV}, through to June 14, 2022.
During this period, 2744 dark time tiles and 2275 bright time tiles were observed covering a surface area of order 7,500~deg$^2$, just over half of the expected final coverage of 14,200~deg$^2$.
The observed data are processed by the DESI spectroscopic pipeline \citep{Spectro.Pipeline.Guy.2023} daily for immediate quality checks.
The redshift catalogs used for this analysis and released with DESI DR1 are obtained from a spectroscopic reduction run with a fixed pipeline version internally denoted as ``iron''.
From the redshifts and parent target catalogs, large-scale structure catalogs and two-point function measurements were made following the procedure described in \citep{DESI2024.II.KP3, KP3s15-Ross}.
Along with the data, random sample catalogs (``randoms''), designed to account for the survey geometry, were also produced (cf. \cite{TS.Pipeline.Myers.2023, KP3s15-Ross,DESI2024.II.KP3} for the detailed methodology).
The redshift distributions are matched between data and randoms in each region separately (North or South Galactic cap --- NGC/SGC).

One of the key science goals of DESI is to measure the BAO scale from redshift $\sim 0.1$ to redshift $\sim 3.5$.
To cover such a wide range of redshifts, DESI relies on different types of galaxies, which we will briefly introduce here.
At the lowest redshift ($0.1<z<0.4$), DESI observes a r-band magnitude-limited sample BGS, which has a significantly higher density compared to past surveys in this redshift range~\cite{2001MNRAS.328.1039C,2009ApJS..182..543A,2011MNRAS.413..971D}.
At higher redshift ($0.4<z<1.1$), DESI observes LRGs, similar to the Baryon Oscillation Spectroscopic Survey (BOSS, \citep{Reid2015:1509.06529v2}).
At even higher redshift ($0.8<z<1.6$), DESI observes ELGs, which make up the largest sample in DESI. ELGs suffer from low completeness (fiber assignment completeness is only 35.3\% for DR1), due to the observing strategy of DESI, which favors quasars and LRGs during dark time.
To go beyond redshift $1.6$, DESI targets quasars.
There are two different QSO selections: a homogeneous QSO selection up to redshift $2.1$ is designed for the study of the BAO feature in the QSO auto-correlation \cite{DESI2024.III.KP4}, and another QSO selection up to redshift $3.5$ is designed for the BAO study of the Lyman-$\alpha$ forest \cite{DESI2024.IV.KP6}.

\begin{table*}
\centering
\begin{tabular}{|l|c|c|c|c|c|}
\hline
Tracer &  $N_{\rm tracer}$ & $z_{\rm eff}$  & $P_0(k=0.14)$ & $V_{\rm eff}$ (Gpc$^3$) \\  \hline
LRG   & 859,824 & 0.92 & $\sim 8.4\times10^3$ & 5.0 \\
ELG  & 1,016,340 & 0.95 & $\sim 2.6\times10^3$ & 2.0 \\
LRG+ELG  & 1,876,164 & 0.93 & $\sim 6\times10^3$ & $\sim 6.5$ \\
\hline
\end{tabular}
\caption{Statistics of DESI DR1 LRG and ELG samples over $0.8<z<1.1$ used in this paper as well as the overlapping tracer \lrgelg.
Note that these samples exactly correspond to \lrgth, \elgo, and \lrgth+\elgo\ of DR1 \cite{DESI2024.III.KP4}.
The effective redshift $z_{\rm eff}$ and the effective volume $V_{\rm eff}$ are taken from \citep{DESI2024.III.KP4} (also \cref{eq:zeff,eq:veff}).
The $V_{\rm eff}$ of \lrgelg\ is about 1.3 times that of LRGs, implying a gain of $\sim 1.14$, compared to LRG alone.}
\label{tab:Y1data}
\end{table*}

In this paper, we only focus on the LRGs, ELGs, and their combination (\lrgelg) in the redshift range $0.8<z<1.1$. Note that these correspond to \lrgth, \elgo\ and \lrgth+\elgo\ in \cite{DESI2024:2404.03002v1}~\footnote{DESI DR1 analysis splits LRGs ($0.4<z<1.1$) into 3 redshift bins, \lrgo, \lrgt, and \lrgth\ and ELGs ($0.8<z<1.6$) into 2 redshift bins, \elgo\ and \elgt. }, respectively.
\cref{tab:Y1data} provides some details on the properties of each sample, taken from \cite{DESI2024.III.KP4}.
Here $z_{\rm eff}$ represents the redshift at which the BAO fit parameters can be converted into physical distances and is calculated weighting by the square of the weighted number density of randoms (with the weights---including FKP\footnote{We follow the prescription from \cite{FKP1994}, which balances the cosmic-variance-dominated regions (high-density regions) and the shot noise-dominated regions (low-density regions) to minimize the variance.}), $n_{\rm ran}(z)$:
\begin{equation}
z_{\rm eff} =\frac{\int r^2dr zn^2_{\rm ran}(z)}{\int r^2dr n^2_{\rm ran}(z)},
\label{eq:zeff}
\end{equation}
where $r$ is the comoving distance to the redshift $z$.
The effective volume estimate is obtained for each redshift bin via 
\begin{equation}
    V_{\rm eff} = 
    \int \left[\frac{\bar{n}_{\rm tracer}(z)P_0(k=0.14)}{1+\bar{n}_{\rm tracer}(z)P_0(k=0.14)}\right]^2 dV(z)
    \label{eq:veff}
\end{equation}
where $P_0(k=0.14)$ are taken from \cref{tab:Y1data} and represent the amplitude of the observed power spectrum at the wave mode $k=0.14\ihMpc$ that is considered most relevant for the BAO information.\footnote{Values for $P_0$ are taken from \cite{KP3s5-Pinon}.
Specifically, the ones that remove the effect of angular separations less than 0.05 degrees. $k=0.14 \ihMpc$ was chosen to maximize the trade-off between area and number density at a fixed total number of objects for $\bar{n}P = 1$ \cite{FontRibera2014}.}
The comoving number density as a function of redshift $\bar{n}_{\rm tracer}(z)$ for each sample is shown in the left panel of \cref{fig:red_disty1}.
The right panel of \cref{fig:red_disty1} shows the galaxy bias (right panel) of individual LRGs (red), ELGs (blue), and the weighted combination for \lrgelg\ (black) in this redshift range (shaded region).

\section{DR1 Mock catalogs}
\label{sec:Mocks}

We utilize mock catalogs for multiple purposes.
First, to test the realistic gain given the survey realism.
Second, to identify any potential systematics when using the combined tracer.
Third, to understand the expected range of consistency between the single-tracer and combined-tracer BAO measurements.
The details of mocks are presented in various DESI DR1 papers \cite{DESI2024.II.KP3, DESI2024.III.KP4}; here, we summarize the key details once again to make this paper self-contained.

\subsection{\abacussecond\ DR1 mocks}
\label{sec:Abacusmocks}

The \abacus\ simulations are high-resolution gravity-only N-body simulations~\citep{Abacus-code}.
We use a set of 25 simulation boxes from the {\sc AbacusSummit} suite~\citep{AbacusSummit}, each with a volume of $\rm (2h^{-1} Gpc)^3$ and $6912^3$ particles.
These {\sc AbacusSummit} simulations assumed the Planck 2018 $\Lambda$CDM cosmology, specifically the mean estimates of the Planck TT,TE,EE+lowE+lensing likelihood chains: $\Omega_c h^2 = 0.1200$, $\Omega_b h^2 = 0.02237$, $\sigma_8 = 0.811355$, $n_s = 0.9649$, $h = 0.6736$, and $w = -1$~\citep{Planck2018}. 
This is also the fiducial cosmology used throughout the DESI DR1 analysis.

The \abacus\ halos\footnote{Identified with the {\sc CompaSO} halo finder~\citep{CompaSO-halo-finder}. We refer to the full simulation suite as \abacus{}, and to the specific realizations used in this work as \abacussecond{} DR1 mocks for example.} are populated with galaxies using a flexible halo occupation distribution model (HOD) \citep{abacushod} which has been fitted to the galaxy two-point correlation functions of the final Early Data Release (EDR)~\cite{DESI2023b.KP1.EDR,EDR_HOD_LRGQSO2023,EDR_HOD_ELG2023} that included correction for all the systematics and included a detailed model for DESI focal plane effects.
More details about the production and utilization of the mocks are provided in \citep{DESI2024.III.KP4}.
From these simulations, we create mock catalogs designed to mimic the survey realism, called ``cutsky'' \abacussecond{}\footnote{\abacussecond{} is the refined second iteration that came after {\tt Abacus-1} mock galaxy catalogs \cite{DESI2024.II.KP3}. We do not use {\tt Abacus-1} in this work.} mocks, after applying the DESI footprint and survey selection function, including realistic redshift failures and targeting masks as described in \cite{DESI2024.II.KP3}.
As a caveat, the LRG DR1 mocks are produced using the simulation output at $z=0.8$, while the ELG DR1 mocks were produced using the simulation output at $z=0.950$ \cite{DESI2024.II.KP3}.
In comparison, the actual effective redshifts of the two DR1 samples (\lrgth, \elgo ) are very close: 0.92 and 0.95, respectively (\cref{tab:Y1data}).
Despite the offset in the output redshifts of the mocks, the cross-correlations between the two tracers reasonably agree between the mocks and the data on the BAO scale, as will be shown in \cref{sec:Results} (\cref{fig:mock_correlation}).
We, therefore, ignore the effect of this redshift offset. 

\begin{figure*}
    \centering    
    \includegraphics[width=0.8\textwidth]{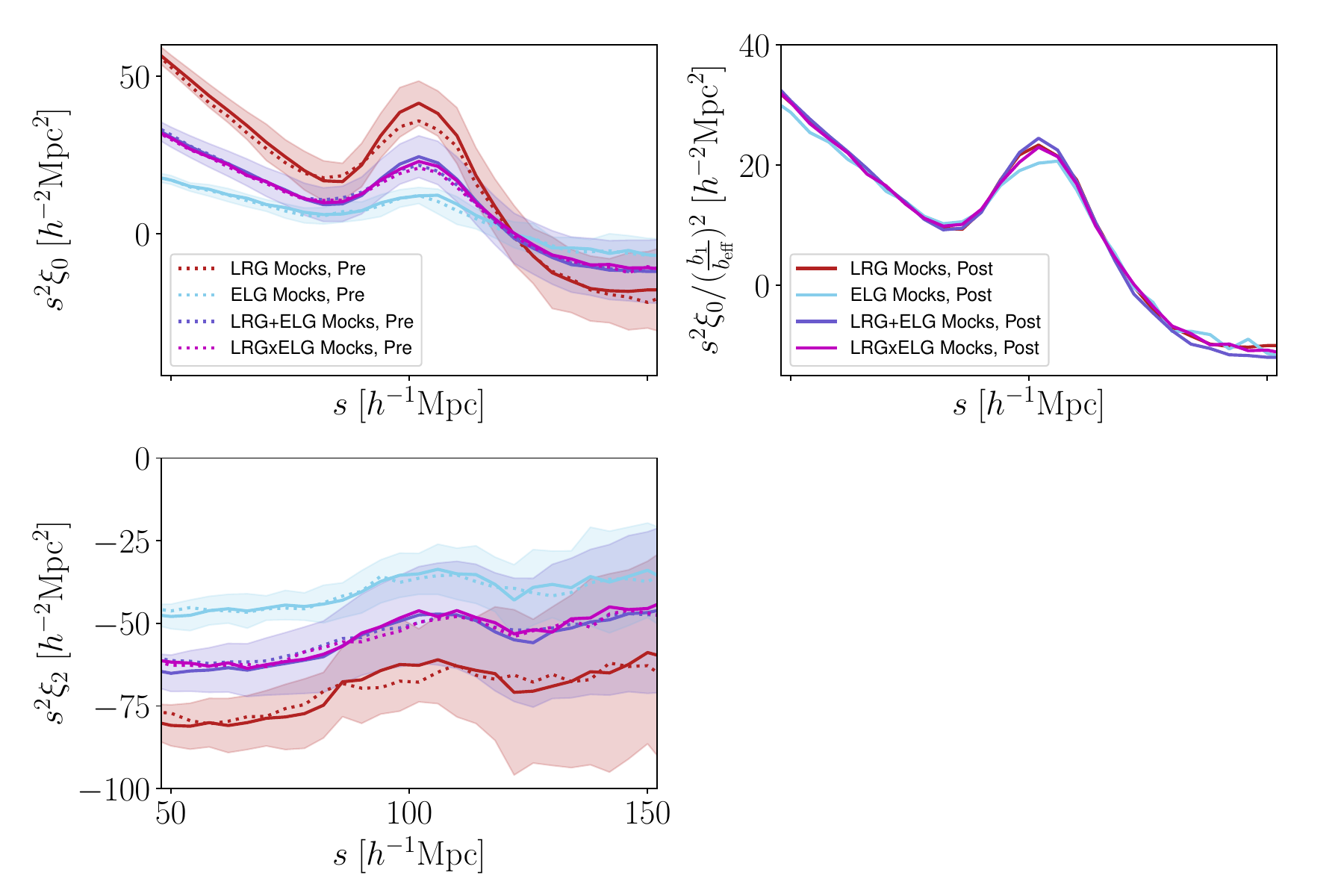}
    \caption{
    Mean of the 25 \abacussecond\ DR1 realizations for the BAO fitting range $48 < s < 152\,h^{-1}\mathrm{Mpc}$. 
    The monopole (top row) and quadrupole (bottom row) are shown before reconstruction (left, dotted) 
    and after standard reconstruction (right, solid). Shaded regions show the $1\sigma$ scatter among 
    the 25 realizations for the post-reconstruction case. In the top-right panel, the post-reconstruction 
    monopoles are rescaled to account for bias differences between tracers, to facilitate comparison of 
    the BAO feature. The combined tracer exhibits a slightly stronger BAO signal than the single LRG tracer, 
    and the LRG$\times$ELG cross-correlation shows a clear BAO feature, seemingly more distinct than that of ELG alone.
    }
    \label{fig:xi_mocks}
\end{figure*}

As the final step, fiber assignment in the mocks is modeled using the DESI fiber assignment pipeline, which replicates the actual survey tiling, priority, and collision logic.
It is important to simulate this effect as it might bias our observation and so our BAO measurements.
For details, see \cite{KP3s15-Ross, KP3s11-Sikandar}.
This is called the `{\bf altmtl}' (the Alternate Merged Target Ledgers  \citep{KP3s7-Lasker}) Abacus DR1 mocks. 
Even though they are more computationally expensive, they are the most realistic simulations of the DR1 data as they accurately reproduce DESI fiber assignments.
\cref{fig:xi_mocks} shows the clustering amplitude of the DR1 mocks for \lrg, \elg, their cross correlation \lrgxelg\ and their combination \lrgelg.

\subsection{{\ezmock\ DR1}}
\label{sec:EZmocks}

The second type of mocks is based on the effective Zel’dovich approximation mocks~\citep{Chuang:2014vfa}, which we call \ezmocks.
\ezmocks\ are almost as accurate as \abacus\ mocks on linear scales but less computationally expensive.
We use 1000 realizations of DR1 \ezmocks\ that are calibrated to produce two-point clustering statistics of \abacussecond\ DR1, while constructed from a cubic volume of ${\rm (6h^{-1} Gpc)^3}$, as described in \cite{DESI2024.III.KP4}.
When applying the survey realisms to \ezmocks, instead of applying and simulating \textsc{fiberassign}, we use the ‘fast-fiberassign (FFA)’ method that emulates the fiber assignment process by generating and averaging realizations from the average targeting probability as a function of number of overlapping tiles and local (angular) clustering, learned from the data.
This `{\bf FFA}' method approximately reproduces the fiber-collisions pairwise incompleteness and is much faster than `{\bf altmtl}'.
We use these DR1 \ezmocks\ only for the tests of the gain in BAO precision (\cref{sec:impactbao}), which requires a greater number of independent simulations and needs to avoid a potential box replication effect \footnote{To reach the full survey volume, the simulation box must be periodically replicated. This artificial tiling can introduce spurious large-scale correlations or suppress variance on scales approaching or exceeding the box size.} \cite{DESI2024.II.KP3}.
The rest of the mock tests in this paper are performed using \abacussecond\ DR1 mocks.

\section{Methods}
\label{sec:Methods}

\subsection{Nomenclature and fiducial cosmology}

Our analysis requires an assumption of the fiducial cosmology for transforming the observational coordinates to the comoving coordinates and calculating the two-point correlation function, as well as in the BAO fitting.
Whereas we use the Planck 2018 $\Lambda$CDM cosmology as the DESI fiducial cosmology (\cref{sec:Abacusmocks}), the robustness against the choice of fiducial cosmology is extensively demonstrated in \cite{KP4s9-Perez-Fernandez}.

Throughout this paper, we follow the definition of $\apar,\aperp$ from the DESI DR1 BAO analysis \cite{DESI2024.III.KP4}.:
\begin{align}    
    \aperp &= \frac{D_A(z)r_d^{\rm fid}}{D_A^{\rm fid}(z)r_d}\label{eq:aperp}\;  \\
  \mbox{and\;\;}  \apar &= \frac{H^{\rm fid}(z)r_d^{\rm fid}}{H(z)r_d}\; ,
    \label{eq:apar}
\end{align}
where $r_d$ is the sound horizon at the drag epoch, $D_A(z)$ is the angular diameter distance, $H(z)$ is the Hubble parameter, and $\apar$ and $\aperp$ are the dilations along and across the line of sight, respectively.
Quantities with the superscript ``fid'' are measured in the fiducial cosmology that we introduced above.
Because $D_A(z)$ also depends on $H(z)$, the two $\alpha$ parameters are correlated at some level.
It is useful in some cases to express the 2-D dilation in terms of the isotropic $\aiso$ and anisotropic $\alap$ distortion parameters defined as  
\begin{align}
    \aiso &= \left(\apar \aperp^2 \right)^{1/3} \; \label{eq:aiso}\\
    \mbox{and \;\;}
    \alap &= \frac{\apar}{\aperp}\;,
    \label{eq:aap}
\end{align}
where $\aiso$ measures the spherically averaged BAO scale, and $\alap$\footnote{AP refers to the Alcock-Paczynski effect \cite{Alcock} i.e anisotropic distortion of the BAO feature.} captures the anisotropy in the BAO feature.

\subsection{Creation and weighting of the combined catalogs}
\label{sec:creation}

\def\ti{Tracer$_1$}
\def\tj{Tracer$_2$}
\def\tij{Tracer$_{1+2}$}
\def\wfkp{w_{\rm FKP}}
\def\wfkpc{w_{\rm FKPeff}}
\def\beff{b_{\rm eff}} 

The size and location of the BAO feature have been demonstrated to be approximately tracer-independent, especially after reconstruction~\citep{Mehta2011:1104.1178v1,Padmanabhan2008:0812.2905v3}.
Therefore, the most naive approach to combining different BAO tracers is to concatenate the catalogs.
However, given the different galaxy biases and sampling noise of the two tracers, an optimal approach would be to weight each tracer according to its signal-to-noise ratio when combining the catalogs.
In this paper, we propose and test an approximate optimal weight assuming a scale-independent and isotropic signal-to-noise ratio. 
Once combined, we treat this as a single catalog, where the information it provides should be equivalent to or better than (especially if there is a gain during reconstruction) the combined information from all three clustering measurements available: the two auto-clustering measurements of the individual tracers and their cross-clustering.
The cross-covariance among these three clustering statistics is naturally accounted for by treating them as a single catalog (at least within the approximation we use to derive the weight for the combination).

The ultimate information we want to derive from the combined tracer is the underlying matter density field (and the BAO feature in it) and the corresponding displacement field.
LRG and ELG trace the underlying matter density field in a biased way.
We construct an approximate maximum likelihood estimator of the matter density field, ignoring the anisotropies due to redshift-space distortions and scale-dependence of bias: 

\begin{equation}
\delta_{\rm ML} \simeq \frac{b_1 ({n}_1-\Bar{n}_1)  + b_2 ({n}_2-\Bar{n}_2) }{b_1^2 \Bar{n}_1 + b_2^2 \Bar{n}_2},  \label{eq:deltaML}
\end{equation}
where $b_1$ and $b_2$ refer to the linear galaxy bias parameters for LRGs and ELGs, ${n}_1$ and $n_2$ are the observed number densities of LRG and ELG at a given location/pixel and $\Bar{n}_1$ and $\Bar{n}_2$ are the average number densities expected at given location/pixel, which are typically traced with randoms.
The derivation for \cref{eq:deltaML} is presented in \cref{construct_field}.

Based on the estimator, we choose to weight each tracer with its estimated linear galaxy bias before merging the two catalogs.
Then, the overdensity field of this combined catalog can be written as:
\begin{equation}
     \delta_{\rm Combined} = \frac{b_1 ({n}_1-\Bar{n}_1) + b_2 ({n}_2-\Bar{n}_2)}{b_1 \Bar{n}_1 + b_2 \Bar{n}_2}, 
     \label{eq:deltacomb}
\end{equation}
and the effective bias of this combined field is:
\begin{equation}
b_{\rm eff} =\frac{b_1^2 \Bar{n}_1 + b_2^2 \Bar{n}_2}{b_1 \Bar{n}_1 + b_2 \Bar{n}_2}.\label{eq:beff}
\end{equation}
When comparing \cref{eq:deltacomb,eq:beff} with \cref{eq:deltaML}, one can show
\begin{equation}
\delta_{\rm ML} = \frac{\delta_{\rm Combined}}{b_{\rm eff}}. 
\end{equation}

For simplicity, we ignore the position/redshift-dependence on $b_{\rm eff}$ when deriving the displacement field during reconstruction (\cref{Density field reconstruction}).
But we account for the redshift-dependence of $b_{\rm eff}$ when updating the FKP weight for the combined tracer (\cref{sec:combined-fkp-weights}).
The procedure for constructing the combined catalog can be summarized as follows:

\begin{enumerate}
\item Estimate $\bar{n}_{\rm eff}(z)$ of the combined galaxy density field, assuming that each tracer is weighted with its estimated linear galaxy bias, and compute the new FKP weight based on it (\cref{sec:combined-fkp-weights}).
This can be done without altering the clustering catalogs.
\item In the clustering catalogs, weight LRGs with $b_1$ and ELGs with $b_2$, i.e., multiply the existing clustering weight by the corresponding bias for each galaxy.
Weight the random particles in the same way.
\item Renormalize the randoms so that the data-to-randoms weighted sum ratio is the same for both tracers.
See \cref{sec:reweight}.
\item Concatenate the two galaxy catalogs with the new weights (with biases, the new FKP weight).
Do the same to the randoms (with the additional renormalization).
\end{enumerate}

In the following subsections, we present practical details of each step for interested readers.

\subsubsection{Single-tracer weights for galaxies and randoms}

As we mentioned in \cref{section:Dataset}, each data and random in the DESI catalogs is associated with a set of weights; $w_t$ ("{\tt WEIGHT}" column in the catalog file) corrects for the variations in the selection function and the FKP weight, $\wfkp$ ("{\tt WEIGHT\_FKP}" column), is designed to construct a minimum variance power spectrum estimator given the redshift-dependent number density.
Each source is then multiplied by the net weight $W_i$:

\begin{align}
    W &= w_t \times \wfkp .
    \label{eq:weightgen}
\end{align}
The FKP weight is defined as 
 \begin{equation}
     \wfkp(z, n_{\rm tile}) = \frac{1}{1+\Bar{n}(z)\times \comntile\times P_0}\;,
     \label{eq:fkp}
\end{equation}
where $\comntile$\footnote{For details of $\comntile$, please refer to \cite{DESI2024.II.KP3}.} is the mean tile completeness at a given (discrete) number of overlapping tiles $n_{\rm tile}$, $\Bar{n}(z)$ is the expected comoving number density (solid lines in \cref{fig:red_disty1}), $\Bar{n}(z)\times \comntile$, is the actual, observed comoving number density (dashed lines in \cref{fig:red_disty1}), and \( P_0 \) represents an approximate value of the power spectrum monopole at \( k \approx 0.15\,h\,\mathrm{Mpc}^{-1} \).
It is used as a reference amplitude in the FKP weighting scheme, and is set to \( 10{,}000\,h^{-3}\,\mathrm{Mpc}^3 \) for Luminous Red Galaxies (LRGs) and \( 4{,}000\,h^{-3}\,\mathrm{Mpc}^3 \) for Emission Line Galaxies (ELGs) in DESI DR1~\cite{DESI2024.II.KP3}.

Since we want to treat the combined catalog as a single tracer after weighing each tracer with its linear bias, we need to redefine effective FKP weights for the combined sample.

\subsubsection{Updating the FKP weights}
\label{sec:combined-fkp-weights}

Here we are going to describe how we derive the combined quantities leading to the effective FKP weights: 
\begin{enumerate}
    \item First we split the complete redshift range of the two tracers $0.4<z<1.6$ (with an overlap between $0.8<z<1.1$. See \cref{fig:red_disty1}) in bins of $dz=0.01$.
    We then construct the effective redshift-dependent bias $b_{\rm eff}(z)$ from 
    \begin{equation}
    b_{\rm eff}(z) = \frac{b_L^2 \Bar{n}_L(z) + b_E^2 \Bar{n}_E(z)}{b_L \Bar{n}_L(z) + b_E \Bar{n}_E(z)},  
    \label{eq:beffz} 
    \end{equation}
    where $\Bar{n}_L$, $\Bar{n}_E$ stand for the expected comoving number density, and $b_L$, $b_E$ for the linear bias assumed for LRG and ELG, respectively.
    The extra factor of $b_L$ and $b_E$ in the numerator reflects each tracer being weighed with its own $b$. 
    \item Now that we have the effective bias, the next step is to derive the effective comoving density distribution $\Bar{n}_{\rm eff}(z)$ as follows:
    \begin{equation}
    \Bar{n}_{\rm eff}(z) = \frac{b_L\Bar{n}_L(z)+ b_E\Bar{n}_E(z)}{\beff(z)}\; .  
    \label{eq:neffz}
    \end{equation}

    \item Based on the mean $\beff = 1.6$ when averaged over all sources between $0.8<z<1.1$, we choose $P_0 = 6000~h^{-3} {\rm Mpc}^{3}$ for \lrgelg.  Given the formula of the FKP weights, we derive the new effective FKP weight:
    \begin{equation}
    \wfkpc(z, n_{\rm tile})  = \frac{1}{1+\Bar{n}_{\rm eff}(z) \times \comntile\times P_0}\; .
    \label{eq:fkpnew}
    \end{equation}
    $\comntile$ is position-dependent in a way that is specific to the type of the tracers so that $\wfkpc(z, n_{\rm tile})$ would be calculated differently between LRG targets versus ELG targets within the concatenated catalog, despite the common $\Bar{n}_{\rm eff}(z)$: $\wfkpcl$ and $\wfkpce$.
    Therefore, although combined, we are weighing different tracers in the same location slightly differently depending on their tiling completeness\footnote{Ideally, we would use  $1+[b_L^2 \bar{n}_L(x)+b_E^2\bar{n}_E(x)]P_{m,0}$ in the denominator of \cref{eq:fkpnew}, where $\bar{n}_i(x)=\bar{n}_i(z)\times \comntilei$ for each tracer.}.
 
\end{enumerate}

\subsubsection{Re-weighting and concatenating the catalogs}\label{sec:reweight}

Before combining, we first update the weight ($w_t$, hereafter $t$ is the tracer label) by multiplying it with the estimated linear bias for each tracer $b_t$.
Then, with the new FKP weight (\cref{eq:fkpnew}), the weight assigned for data and randoms inside this combined catalog is:
\begin{subequations}
\begin{align}
    W_t &= b_t \times w_t \times \wfkpc \mbox{\; for LRG ($t=L$) and ELG ($t=E$) data} \\
    W_{L,r} &= b_L \times w_{L,r} \times \wfkpc \mbox{\; for LRG randoms} \\
    W_{E,r} &= b_E \times w_{E,r} \times \wfkpc \times \frac{\sum w_E \wfkpc}{\sum w_{E,r} \wfkpc} \times \frac{\sum w_{L,r} \wfkpc}{\sum w_L \wfkpc} \mbox{\; for ELG randoms}.
\end{align}
\label{eq:weff}
\end{subequations}
Here $w_t$, $w_{L,r}$ and $w_{E,r}$ corresponds to the WEIGHT in the original catalogs. Note that we renormalize the weights for ELG randoms to match the LRG (weighted) data-to-random ratio\footnote{Symmetric weight renormalization $W_{t,r} = b_t \times w_{t,r} \times \wfkpc \times (\sum w_t \wfkpc) / (\sum w_{t,r} \wfkpc)$ for both randoms must give equivalent results. Still, we prioritize exact documentation of the DESI DR1 combined catalog production.}.
This relative normalization ensures that when the combined random catalog is normalized to match the weighted sum of the combined tracer catalog, the weighted sum of randoms for each tracer also matches that of the corresponding tracer.
Without this step, the combined randoms might misrepresent the combined catalog's selection function due to the tracers' different structures.
We then merge all these sources into a single catalog.

\subsection{Two-point correlation function}
Following the pipeline defined in \citep{DESI2024.III.KP4, KP4s4-Paillas}, the correlation functions are computed using the Landy-Szalay estimator \citep{Landy1993}
\begin{equation}
    \xi(s,\mu) = \frac{DD(s,\mu) - 2DR(s,\mu) + RR(s,\mu)}{RR(s,\mu)}\; ,
\end{equation}
with $\mu \in [-1,1]$ being the cosine of the angle between the galaxy pair and the line of sight, $D,R$ refer respectively to points from the data and random catalogs, and $s$ is the pair separation (here we use bins with a width of $4 \hMpc$).
To compute the two-point correlation we use the publicly available code \textsc{pycorr}\footnote{\url{https://github.com/cosmodesi/pycorr}} \cite{pycorr} based on the \textsc{Corrfunc}\footnote{\url{https://github.com/manodeep/Corrfunc}} engine \cite{corrfunc-1,corrfunc-2}.
For our analyses, we convert the output to Legendre multipoles (specifically the monopole ($l=0$) and the quadrupole ($l=2$)).
To extract the maximum out of the clustering measurement, we weigh each galaxy by a combination of weights, as described in \cref{sec:creation} (\cref{eq:weff} for the combined tracer, analogous to \cref{eq:weightgen} for regular single tracers).

Since the data reduction produces catalogs for each of the galactic caps, we choose to compute the correlation functions for NGC and SGC separately.
Accordingly, we perform the catalog combination separately for NGC and SGC.
Then the combination is performed by summing the pair counts computed in each region independently.
This allows us to compare the consistency of the two regions and test for possible systematics.
In some cases, an insufficient number of randoms can degrade the computation of the two-point statistics.
To avoid this, we concatenate multiple random catalogs (typically 18) so that the number of randoms is more than 50 times the number of data galaxies.

\subsection{Covariance matrices}
\label{sec:cov_mat}

\begin{table}[hbtp]
\centering
\resizebox{\columnwidth}{!}{%
\begin{tabular}{l|l|l}
\hline\hline
{\bf Name} & {\bf Tracer} & {\bf Notes}  \\ \hline
\rascalc-LRG & LRG & Based on DR1\ LRG clustering (pre and post)\\
\rascalc-ELG & ELG & Based on DR1\ ELG clustering (pre and post) \\
\rascalc-\cross & \lrgxelg\ & Based on DR1\ LRG, \lrgxelg\ and ELG clustering (only post-recon) \\ 
\rascalc-{\rm comb} & \lrgelg\ & Based on DR1\ \lrgelg\  clustering (pre and post) \\
\hline
\rascalc-LRG$_{\rm EZ}$ & LRG & Based on \ezmocks\ LRG clustering (only post-recon) \\
\rascalc-{\rm comb}$_{\rm EZ}$ & \lrgelg\ & Based on \ezmocks\ \lrgelg\  clustering (only post-recon) \\ 
\hline \hline
\end{tabular}%
}
\caption{Summary of semi-analytic covariance matrices used in this work.}
\label{tab:covariance}
\end{table}

In this work, we used semi-analytical covariance matrices for the 2-point correlation function multipoles produced with the \rascalc{} code\footnote{\url{https://github.com/oliverphilcox/RascalC}} \cite{rascal,rascal-jackknife,rascalC,2023MNRAS.524.3894R,KP4s7-Rashkovetskyi}.
It computes the covariance matrix terms based on an empirical 2-point function (but not 3- and connected 4-point functions) and applies a shot-noise rescaling to mimic non-Gaussian effects.
We refer the readers to \cite{KP4s7-Rashkovetskyi} for further details on the methodology.

The covariance matrix for the combined tracer was produced similarly to other single tracers, using the combined data and random catalogs and tuning the shot-noise to the jackknife covariance\footnote{Scripts available at \url{https://github.com/cosmodesi/RascalC-scripts/tree/DESI2024/DESI/Y1/comb}.}.
Therefore, the LRG, ELG and \lrgelg\ \rascalc{} covariance matrices in this paper are the same as in the main DESI 2024 galaxy BAO analysis \cite{DESI2024.III.KP4}.
The covariance matrix of the cross-correlation function (\lrgxelg) was produced based on data correlation functions and random catalogs according to Appendix A of \cite{KP4s7-Rashkovetskyi}\footnote{Scripts available at \url{https://github.com/cosmodesi/RascalC-scripts/tree/DESI2024/DESI/Y1/cross}.}, using the shot-noise rescaling values calibrated with jackknives on LRG and ELG auto-correlations.
We also produced LRG and \lrgelg{} covariance matrices using the average \ezmocks{} clustering and shot-noise rescaling tuned to the \ezmocks{} sample covariance\footnote{Scripts available at \url{https://github.com/cosmodesi/RascalC-scripts/tree/DESI2024/DESI/Y1/EZmocks/average/post}.}.
This is important because the covariance of \ezmocks{} is ``smaller'' than DR1 covariance by 22-25\% \citep{KP4s6-Forero-Sanchez,DESI2024.II.KP3,DESI2024.V.KP5}, probably due to approximate higher-point clustering and fiber assignment modeling.
We use the \ezmocks{} covariances exclusively for BAO fits to \ezmocks{} in \cref{sec:impactbao,tab:keybaofitsmocks1}.
\cref{tab:covariance} summarizes the covariances we use in this paper.

\subsection{BAO fitting}

The BAO fitting procedure for the DESI collaboration is based on the extensive tests that are summarized in \citep{KP4s2-Chen}.
The model consists of a combination of a physically motivated model from quasi-linear theory and a parameterized model to marginalize over non-linearities that may otherwise affect our measurements of the BAO scale.
For interested readers, \citep{DESI2024.III.KP4, KP4s2-Chen} explain the fitting model for different choices of the reconstruction convention, how the power spectrum multipoles are transformed into configuration space multipoles, the broadband modeling and its marginalization, etc.
We show the parameters of interest in \cref{tab:priors}.

To summarize, we use a mixture of flat ($\aiso, \alap$) and Gaussian priors (damping terms).
To isolate the BAO feature, the range of galaxy separation is restricted to $48<s<152 \hMpc$.
The fit is performed with the publicly available code on two-point correlation functions of combined galactic caps (NGC+SGC) using the \textsc{desilike}\footnote{\url{https://github.com/cosmodesi/desilike}}.
Fitting methods available are MCMC sampling (e.g. \textsc{emcee} \citep{2013PASP..125..306F}, \textsc{pocoMC} \citep{karamanis2022accelerating}) or posterior profiling (\textsc{MINUIT} \citep{minuit}).
The software offers the possibility to fit in several parameter spaces.
In this paper, we chose to present our results in $\aiso-\alap$ or $\apar -\aperp$.

\begin{table}
    \scriptsize
    \centering
    \resizebox{0.8\columnwidth}{!}{
        \begin{tabular}{|l|c|c|}
            \hline
            Parameter &  $\xi(r)$ prior & Description \\ \hline
            $\aiso$ & [0.8, 1.2] & Isotropic BAO dilation \\
            $\alap^*$ & [0.8, 1.2] & Anisotropic (AP) BAO dilation \\
            $\Sigma{\perp}$ &$\mathcal{N}(\Sigma^{\mathrm{fid}}_{\perp}, 1.0)$ & Transverse BAO damping [$\hMpc$]\\
            $\Sigma{\parallel}$  &$\mathcal{N}(\Sigma^{\mathrm{fid}}_{\parallel}, 2.0)$ & Line-of-sight BAO damping [$\hMpc$]\\
            $\Sigma_s$ & $\mathcal{N}(2.0,2.0)$ & Finger of God damping [$\hMpc$] \\
            Fitting range & [48, 152]$\hMpc$ & Measurement bin edges \\
            Data binning & $4\hMpc$ & Measurement bin width \\
            \hline
        \end{tabular}%
    }
    \caption{The free parameters and their priors in configuration-space analyses.
    $\mathcal{N}(\mu, \sigma)$ refers to a normal distribution of mean $\mu$ and standard deviation $\sigma$, $[x_{1}, x_{2}]$ to a flat distribution between $x_{1}$ and $x_{2}$ inclusive.
    Parameters with superscript `$*$' are fixed to the following values when only a 1D fit is performed: $\alap=1$.}
    \label{tab:priors}  
\end{table}

For the priors on the damping parameters, the central values denoted by the superscript `fid' are tracer-specific and they are shown in \cref{tab:priors2}.
These fiducial values are motivated by a combination of theoretical calculations, measurements of the cross-correlation between the initial and post-reconstruction density fields, and fits to mock catalogs.
Finally, the total number of free parameters for our 1D and 2D fits is 7 and 13, respectively, for $\xi(r)$.
The corresponding numbers of degrees of freedom for our 1D and 2D fits are 19 and 39.

\subsection{Density field reconstruction and specific configuration}
\label{Density field reconstruction}

We apply the density field reconstruction technique 
\citep{Eisenstein2007:astro-ph/0604362v1} on the catalogs considered in this paper, single and combined tracers, to partially recover the BAO feature that has been degraded due to structure growth and redshift-space distortions (RSD).
We follow the DESI DR1 default reconstruction setup in \cite{DESI2024.III.KP4}, which was determined based on a set of extensive tests in \cite{KP4s4-Paillas,KP4s3-Chen}.

To summarize, we use \textsc{pyrecon},\footnote{\url{https://github.com/cosmodesi/pyrecon}} a \textsc{Python} package developed by the DESI collaboration, and adopt \texttt{IterativeFFTReconstruction} (hereafter, `IFFT') that implements the iterative procedure described in \citep{Burden2015:1504.02591v2}, with the \texttt{RecSym} convention.
\footnote{\texttt{RecSym} is a choice to recover the large-scale anisotropy due to redshift-space distortions in the process of reconstruction.}
The overdensity field is smoothed by a Gaussian kernel of width $15 \hMpc$ for all of the tracers we consider in this paper.
For reconstruction, we use the linear bias in \cref{tab:priors2} and the growth rate in \cref{tab:Y1data}.
The value of linear bias, $b_{\rm eff} = 1.6$, is computed from \cref{eq:beffz} where the formula is integrated over the redshift range.

\begin{table}
    \centering
    \resizebox{\columnwidth}{!}{%
      \begin{tabular}{|l|c|c|c|c|c|c|c|}
        \hline
        \multirow{2}{*}{Tracer} & \multicolumn{2}{c|}{$\Sigma^{\mathrm{fid}}_{\perp} [\hMpc]$} & \multicolumn{2}{c|}{$\Sigma^{\mathrm{fid}}_{\parallel} [\hMpc]$} & \multicolumn{1}{c|}{\multirow{2}{*}{Linear bias}} & \multicolumn{1}{c|}{\multirow{2}{*}{Growth rate}} & \multicolumn{1}{c|}{\multirow{2}{*}{Reconstruction range}} \\
        \cline{2-5} 
         & Pre-recon & Post-recon & Pre-recon & Post-recon &  &  &  \\ \hline
        LRG     & 4.5 & 3.0 & 9.0 & 6.0 & 2.0 & 0.83 & 0.4 -- 1.1 \\
        ELG     & 4.5 & 3.0 & 8.5 & 6.0 & 1.2 & 0.90 & 0.8 -- 1.6 \\
        LRG+ELG & 4.5 & 3.0 & 9.0 & 6.0 & 1.6 & 0.87 & 0.8 -- 1.1 \\ \hline
      \end{tabular}%
    }
    \caption{The baseline setup for the density field reconstruction.
    The linear bias for the combined catalog is derived from equation \cref{eq:beffz}.
    While we use the entire redshift range of $0.4<z<1.1$ for LRGs and $0.8<z<1.6$ for ELGs, we restrict the reconstruction redshift range of \lrgelg\ to $0.8<z<1.1$ so that the input bias and growth rate are better representative of the overlapping sample.
    See \cref{Density field reconstruction}.}
    \label{tab:priors2}
\end{table}

\begin{figure*}
    \centering
    \includegraphics[width=\textwidth]{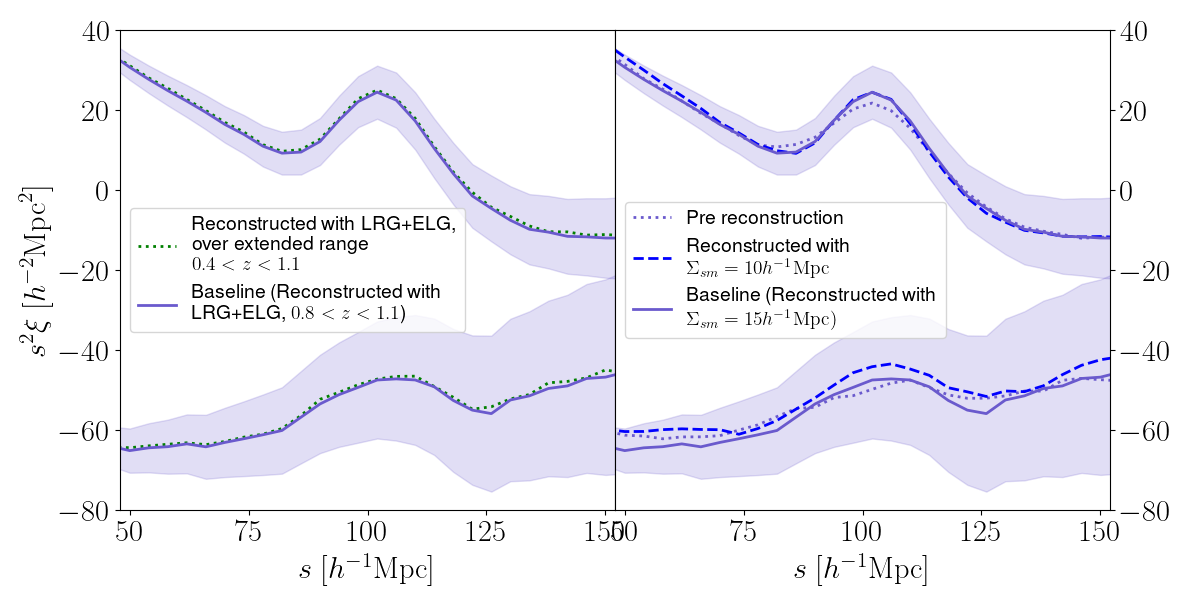}
    \caption{Correlation functions of the \lrgelg\ mock catalogs showing different reconstruction configurations in the BAO fitting range $48<s<152 \hMpc$.
    Here we show only 25 realizations of \abacussecond\ DR1 mocks.
    Left: we compare two cases, one reconstructed using the density field over $0.4<z<1.1$, in comparison to the other reconstructed using the density field over $0.8<z<1.1$ (our baseline).
    There is little difference between reconstructing with $0.4<z<1.1$ versus $0.8<z<1.1$.
    Right: we compare the impact of the smoothing scale.
    The gain of using a slightly smaller smoothing scale does not appear to be obvious for the BAO feature in the monopoles.
    Each curve represents the mean clustering amplitude of the realizations and the shaded region is the standard deviation of the baseline case.}
    \label{fig:combined_recon}
\end{figure*}

As a caveat, whereas the default reconstruction setup of the individual tracers is performed using the displacement field constructed over the full redshift range for LRG and ELG, i.e.,  $0.4<z<1.1$ for LRG and $0.8<z<1.6$ for ELG, the reconstruction of the combined catalog is performed using the displacement field estimated from $0.8<z<1.1$, i.e., the range where the two tracers overlap, as stated in \citep{DESI2024.III.KP4}.
The latter choice was made to input bias and growth rate that are better representative of the overlapping sample to the reconstruction pipeline, rather than using an effective bias and the growth rate averaged over a wide range of redshift (i.e. $0.4<z<1.1$ or $0.8<z<1.6$).
The downside of this choice is a potentially greater boundary effect at $z=0.8$ when estimating the displacement field across the boundaries.
The left panel of \cref{fig:combined_recon} shows the mean correlation functions of 25 \lrgelg\ mocks over $0.8<z<1.1$ using the displacement field constructed from $0.4<z<1.1$ (green-dotted, the convention used for LRGs) and from $0.8<z<1.1$ (blue, our default choice for \lrgelg).
The figure shows that the choice of the reconstruction redshift range between the two options has little impact on the reconstructed correlation function. Table \ref{tab:recon_configs_bao_25} shows that there is no impact on the BAO precision and negligible impact on the best fits.
Therefore, the comparisons between LRG and the combined tracer in this paper are minimally affected by differences in reconstruction redshift ranges.

In the right panel, we also test the effect of the Gaussian smoothing kernel for reconstruction.
Given the effective number density of the combined tracer, greater than LRG and ELG,  we test a more aggressive kernel $10 \hMpc$ (dashed blue line): the gain of using a smaller smoothing kernel does not appear obvious in the BAO feature of the monopoles, which mainly determines the constraint on $\aiso$.
\cref{tab:recon_configs_bao_25} (the second versus the third rows) shows that there is about 6\%-7\% gain by using a smaller smoothing scale with negligible difference in the best fits, compared to the statistical error associated with this test.
Despite the small additional gain, we adopt $15 \hMpc$ that is the default for DR1 LRG and ELG samples for a more conservative comparison.

\begin{table}
\centering
\caption{Best-fit isotropic and anisotropic BAO dilation parameters for different reconstruction configurations tested in \cref{fig:combined_recon}, obtained from fits to the mean of \abacussecond\ DR1 realizations. We hold the covariance matrix fixed to the baseline case. Quoted uncertainties correspond to $1\sigma$ marginalized intervals. The baseline corresponds to reconstructing it with \lrgelg\ density field over $0.8<z<1.1$ with the smoothing scale of $\Sigma_{\rm sm}=15\hMpc$, while the ``extended redshift range'' was reconstructed using the density field of $0.4<z<1.1$.}
\label{tab:recon_configs_bao_25}
\begin{tabular}{lcc}
\hline\hline
Configuration & $q_{\rm iso}$ & $q_{\rm AP}$ \\
\hline
Pre-reconstruction
  & $0.9980 \pm 0.0110$
  & $0.9920 \pm 0.0380$ \\

Post-reconstruction (baseline, $\Sigma_{\rm sm}=15\hMpc$)
  & $0.9967 \pm 0.0085$
  & $1.0000 \pm 0.0290$ \\

Post-reconstruction ($\Sigma_{\rm sm}=10\hMpc$)
  & $0.9970 \pm 0.0080$
  & $0.9980 \pm 0.0270$ \\

Post-reconstruction (extended redshift range)
  & $0.9969 \pm 0.0085$
  & $1.0020 \pm 0.0290$ \\

Post-reconstruction (LRG-only displacement field)
  & $0.9957 \pm 0.0082$
  & $1.0000 \pm 0.0280$ \\
\hline\hline
\end{tabular}

\end{table}

For the cross-correlation \lrgxelg\ of LRG and ELG, we cross-correlate after each of LRG and ELG is separately reconstructed.
One could also use the cross-correlation of the two tracers after reconstructing using the common, combined displacement field, especially given the potentially noisy displacement field from ELG alone.
For a more conservative systematic/consistency check, we, however, decided to cross-correlate the individually reconstructed two fields, which can be subject to more errors.

\section{Results}
\label{sec:Results}

In this section, we demonstrate that our method of combining tracers yields an unbiased BAO estimator and mildly enhances the precision compared to the single catalog.
After validating the method with the mocks, we apply it to the DR1 data, quantify the gain, and test for potential tracer-dependent systematics in the BAO measurements by comparing different tracers, including the cross-correlation \lrgxelg.

\subsection{Application to the DESI DR1 mock catalogs}

We first apply our pipeline on the 25 \abacussecond\ DR1 mocks and estimate the expected level of systematics and the gain from the combination of catalogs.

\begin{figure*}
    \centering
    \includegraphics[width=\textwidth]{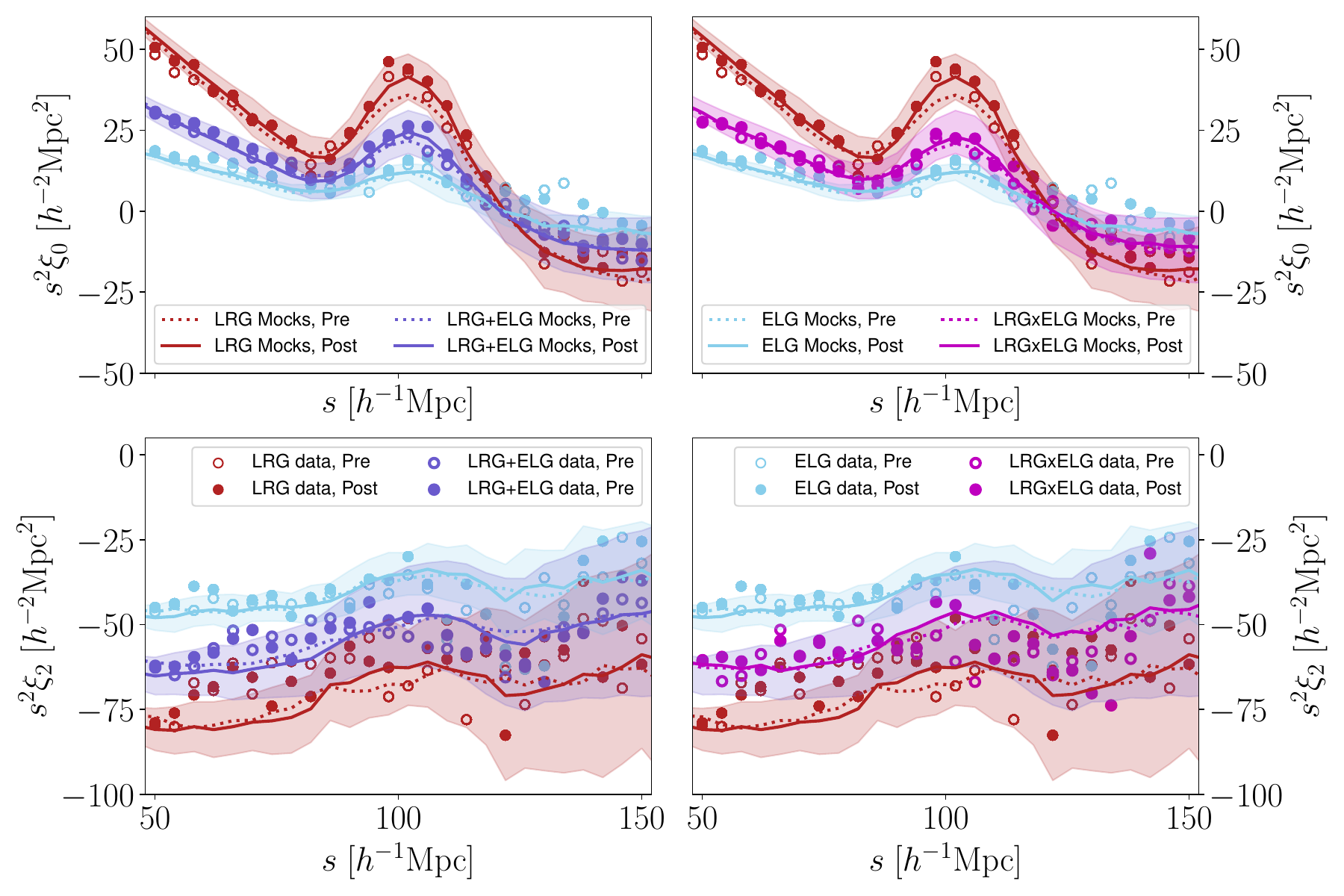}
    \caption{Comparison of the DR1 data (points) and \abacussecond\ DR1 mocks (lines, the mean of the 25 realizations).
    The left panel shows LRGs and ELGs, in comparison to \lrgelg. The right panel repeats the left panel but replaces \lrgelg\ with \lrgxelg.
    The shaded regions correspond to the standard deviation of the mocks.
    The top panel shows the monopole.
    The bottom panel is the quadrupole.
    The difference in the amplitudes is due to the different galaxy biases of each sample.}
    \label{fig:mock_correlation}
\end{figure*}

In \cref{fig:mock_correlation} we show the mean of the correlation function for the 25 realizations of pre- and post-reconstruction for LRGs, ELGs, \lrgelg{}, and \lrgxelg{}.
The amplitude of clustering is proportional to the product of the galaxy biases of the two samples, and we can estimate the galaxy bias of \lrgelg\ and \lrgxelg\ from LRG ($b_{\rm LRG} = 2$) and ELG ($b_{\rm ELG} = 1.2$)~\footnote{The default bias values for LRG and ELG are taken from \cite{DESI2024.III.KP4}.}.
One can see a clear BAO feature around the 100 $\hMpc$ with a slightly sharper peak after reconstruction (a sign that some of the non-linearities have been successfully corrected).
This is especially the case for LRG and \lrgelg.
The ELG reconstruction was likely hindered by its low completeness, which should be improved with future data releases of the DESI survey. 

\subsubsection{Effects of tracer combination on the reconstruction efficiency}

One of the advantages we are seeking by using the combined tracer is to leverage the lower shot noise of the combined tracer to potentially improve the reconstruction efficiency beyond the reconstruction of the individual tracers.
On the other hand, DESI DR1 ELG over $0.8-1.1$ suffers a much larger residual observational systematics \cite{DESI2024.II.KP3,KP3s2-Rosado} in addition to the higher shot noise associated with its low completeness, which could end up increasing undesired systematics when combined to estimate the displacement field.

\begin{figure*}
    \centering
    \includegraphics[width=0.5\textwidth]{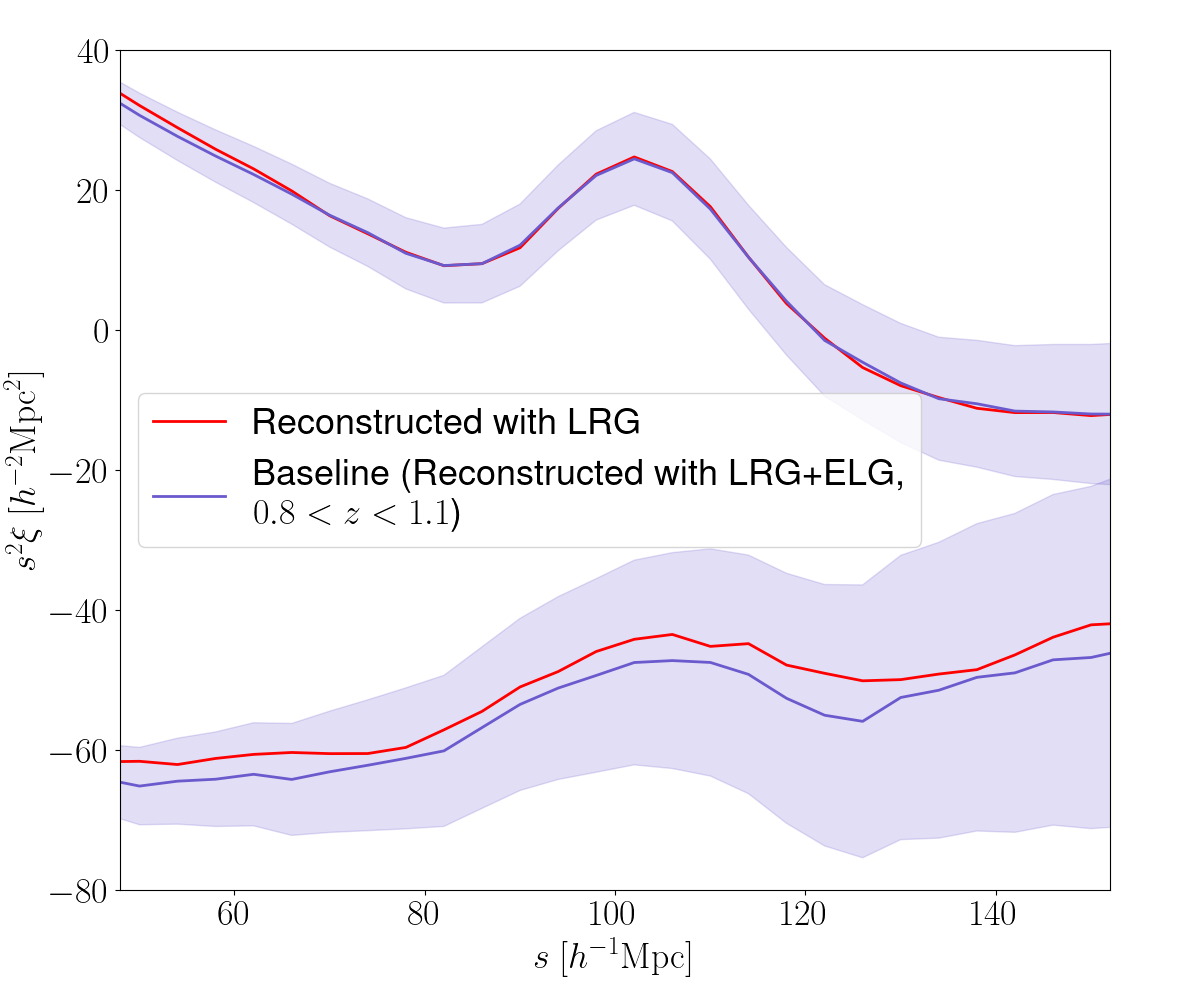}
    \caption{Comparison of correlation functions of the \lrgelg\ \abacussecond\ DR1 mocks, reconstructed with the displacement of \lrgelg\ (Baseline) or LRG only, in the BAO fitting range $48<s<152 \hMpc$.
    The shaded regions show the standard deviation of 25 mocks for the fiducial case.}
    \label{fig:combined_recon2}
\end{figure*}

Therefore, we first investigate the impact of ELGs in the displacement field calculation.
In \cref{fig:combined_recon2}, we test the reconstruction of the combined catalog using the displacement calculated with LRG density fields (red line, i.e., without ELG), in comparison to the default case using the displacement fields from the combined catalog (purple line).
We find little difference in the BAO feature in the monopole, implying that most of the signal in the displacement field is coming from LRGs and that ELGs do not significantly impact reconstruction.
\cref{tab:recon_configs_bao_25} (the second and the last rows) shows that, indeed, there is no gain from using the combined field; even 3-4\% worse precision, which could be merely statistical fluctuations.
That is, the primary benefit of the combined tracer for DR1, relative to LRGs alone, appears mainly to be the stable and robust integration of the low signal-to-noise ELGs by constructing a unified catalog before the BAO analysis, without an additional gain in the reconstruction efficiency.
We will revisit this quantitatively by comparing the reconstruction efficiency of \lrgelg\ in terms of the BAO constraint in the following section.

\subsubsection{The impact on the BAO constraints}\label{sec:impactbao}

We perform a two-dimensional fit using the monopole and quadrupole of the simulations and constrain BAO scales in the basis of the isotropic $\aiso = \apar^{1/3} \aperp^{2/3}$ and anisotropic $\alpha_{\rm AP} = \apar /\aperp$, and equivalently, in the basis of the radial ($\alpha_{\parallel}$) and perpendicular ($\alpha_{\perp}$) directions. We estimate the precision in terms of ${\sigma^D_{\alpha}}$, the standard deviation of the $n_S=25$\footnote{$n_S$ refers to the number of realizations of \abacussecond\ DR1 mocks.} best fits, and $\langle {\sigma^L_{\alpha}} \rangle$, the average of the $n_S$ likelihood errors ${\sigma^L_{\alpha}}$, as presented in \cref{tab:keybaofitsmocks1}.
We also present the correlation coefficient between the isotropic and anisotropic parameters $r_{\rm off}$, and the chi-square per degrees of freedom $\chi^2 / {\rm dof}$.
In this section and throughout the paper, a gain in terms of precision is calculated by dividing the difference in precision (between \lrgelg\ and \lrg) by the precision of the \lrg\ case.\footnote{Likewise, the gain of reconstruction will be quoted as the difference between pre and post reconstruction divided by the pre-reconstruction precision.}.

\begin{table*}
\centering
\resizebox{\textwidth}{!}{%
\begin{tabular}{|lllrrrrrrrl|}
\hline
Tracer & Recon & $\alpha_{\rm iso}$ & $\sigma^D_{\alpha_{\rm iso}}$ & $\langle \sigma^L_{\alpha_{\rm iso}} \rangle$ & $\alpha_{\rm AP}$ & $\sigma^D_{\alpha_{\rm AP}}$ & $\langle \sigma^L_{\alpha_{\rm AP}} \rangle$ & $r_{\rm off}$ & $\chi^2 / {\rm dof}$ \\
\hline
LRG & Pre & 0.9989 & 0.0143 & 0.0133 & 1.0037 & 0.0322 & 0.0491 & 0.1278 & 52.0/39 \\
ELG & Pre & 1.0014 & 0.0322 & 0.0282 & 0.9840 & 0.0495 & 0.0755 & -0.2354 & 41.4/39 \\
LRG$\times$ELG & Pre & 1.0000 & 0.0159 & 0.0133 & 0.9916 & 0.0433 & 0.0463 & 0.0043 & 44.7/39 \\
LRG+ELG & Pre & 0.9998 & 0.0148 & 0.0120 & 0.9937 & 0.0374 & 0.0433 & 0.0803 & 39.4/39 \\
\hline
LRG & Post & 0.9956 & 0.0094 & 0.0097 & 0.9926 & 0.0240 & 0.0339 & 0.0388 & 42.1/39 \\
ELG & Post & 0.9977 & 0.0235 & 0.0250 & 1.0000 & 0.0492 & 0.0680 & -0.4545 & 40.9/39 \\
LRG$\times$ELG & Post & 0.9973 & 0.0136 & 0.0108 & 0.9979 & 0.0318 & 0.0359 & -0.1340 & 43.0/39 \\
LRG+ELG & Post & 0.9971 & 0.0108 & 0.0086 & 0.9983 & 0.0258 & 0.0294 & -0.0188 & 39.9/39 \\
\hline
LRG$_{\tt EZ}$ & Post & 0.9994 & 0.0100 & 0.0096 & 0.9996 & 0.0329 & 0.0325 &  & 42.7/39 \\
LRG+ELG$_{\tt EZ}$ & Post & 1.0001 & 0.0082 & 0.0082 & 1.0005 & 0.0270 & 0.0275 &  & 41.9/39 \\
\hline
\end{tabular}%
}
\caption{BAO results from the simulations for testing the overlapping tracer for $0.8<z<1.1$.
Fits performed with \textsc{desilike} and covariance from \cref{tab:covariance}.
The first eight rows are from 25 \abacussecond\ DR1 mocks, and the last two rows are 1000 DR1 \ezmocks.
${\sigma^D_{\alpha}}$ represents the standard deviation of the best fits and $\langle {\sigma^L_{\alpha}}\rangle$ represents the mean of the individual likelihood errors.
$\langle {\sigma^L_{\alpha}}\rangle$ is derived from the MCMC sampling for \abacussecond\ and from \texttt{MINUIT} \citep{minuit} for \ezmocks\ for a fast calculation.
}
\label{tab:keybaofitsmocks1}
\end{table*}

First, we inspect the efficiency of reconstruction between the single and the combined tracers.
\cref{tab:keybaofitsmocks1} shows that, as expected from \cref{fig:mock_correlation}, reconstruction improves the precision of our BAO measurements.
Both the standard deviation ${\sigma^D_{\alpha}}$ and the mean likelihood error $\langle {\sigma^L_{\alpha}} \rangle$ show improvement for $\aiso$ and $\alap$: by $\sim$ 30\% for \lrgelg\ for both BAO parameters, a level of improvement similar to LRG.
Since the improvement fraction in reconstruction is not larger for \lrgelg\ than for LRG alone, we again do not find an obvious gain during reconstruction by constructing the combined displacement field.
The cross statistics after reconstruction, \lrgxelg\, present a reconstruction improvement of $15-30\%$. 

We next estimate the net gain of constructing a combined catalog relative to the LRG catalog alone.
The effective volumes ($V_{\rm eff} = 6.5\;{\rm Gpc}^3$ for \lrgelg\ and $5 \;{\rm Gpc}^3$ for \lrg\ from \cref{tab:Y1data}) predicts about 14\% improvement between \lrgelg\ and \lrg\ (assuming no additional gain during reconstruction).
Given that cosmology constraints from \elg\ at this redshift bin were excluded from the DESI DR1 analysis \cite{DESI2024.III.KP4,DESI2024.V.KP5} due to their low signal-to-noise, this modest improvement makes a case for using the combined tracer. 

We want to confirm this gain by inspecting the BAO constraints, ${\sigma^D_{\alpha}}$ and/or $\langle {\sigma^L_{\alpha}} \rangle$ of simulations.
If the likelihood distribution of the parameter is perfectly Gaussian, two estimates of precision should match in the infinite sample limit.
With the 25 realizations, unfortunately, the $1\sigma$ relative error of the sample standard deviation estimate, ${\sigma^D_{\alpha}}$, is expected to be as large as 14\%~\footnote{$[2(n_S-1)]^{-0.5} \approx 14\%$ with $n_S=25$.}, which is the level of gain we want to prove.
Also, \abacussecond\ DR1 mocks potentially suffer the box replication effect due to it being constructed from a box size of $2 \ihGpc$. On the other hand, $\langle {\sigma^L_{\alpha}}\rangle$ reflects the property of the DR1 data through the DR1 covariance matrix, which may not perfectly resemble the standard deviation of the mocks in a way that depends on tracers.
For example, ${\sigma^D_{\alap}}$ and  $\langle {\sigma^L_{\alap}}\rangle$  of \abacussecond\ \lrg\ shows a difference of a factor of 1.4.
We therefore cannot be confident that the ${\sigma^D_{\alpha}}$ from \abacussecond\ DR1 should be sufficiently precise or consistent with $\langle {\sigma^L_{\alpha}} \rangle$.
\cref{tab:keybaofitsmocks1} indeed gives inconclusive implications about the gain, being different depending on ${\sigma^D_{\alpha}}$ and $\langle {\sigma^L_{\alpha}}\rangle$. 

To avoid all the aforementioned potential sources of discrepancies, i.e., to increase the statistical precision of this comparison without the box replication effect and to have a matching covariance matrix for the mock data, we adopt 1000 DR1 \ezmocks\ that were constructed from $6\hGpc$ simulation boxes and use the covariance calibrated for the \ezmocks\ (\cref{tab:covariance}). Note that the variance of DR1 is different from the variance of the \ezmock\ that we are about to show; the former is higher by 1.22-1.25 (i.e., 11-12\% larger values in $\langle {\sigma^L_{\alpha}} \rangle$) \cite{DESI2024.II.KP3,KP4s6-Forero-Sanchez} mainly due to the approximation in the fiber assignment modeling and higher-point clustering in \ezmocks{} and therefore the absolute precision would be different between \abacussecond\ or DR1 data versus \ezmocks{}, while we check the relative gain of the combined tracer.

The estimation of gains based on ${\sigma^D_{\alpha}}$ (last two rows of Table \ref{tab:keybaofitsmocks1}) now allows the statistical comparison at the level of 2\%~\footnote{As \lrgelg\ and \lrg\ are correlated, this 2\% does not directly translate to the precision of comparing the two constraint values.}.

The bottom two rows of  \cref{tab:keybaofitsmocks1} show highly consistent BAO measurements between ${\sigma^D_{\alpha}}$ and $\langle {\sigma^L_{\alpha}} \rangle$ for both \lrg\ and \lrgelg\ when using \ezmocks{}, because of the reduced sample variance due to the large sample size.
We find a gain of 15-18\%  (15\% from $\langle {\sigma^L_{\alpha}} \rangle$ and 18\% from ${\sigma^D_{\alpha}}$) in both $\aiso$ and $\alap$ with \lrgelg\ over \lrg.
This gain is well aligned with the $V_{\rm eff}$ difference, 14\%.

\begin{figure*}[h]
    \centering
    \includegraphics[width=1\textwidth]{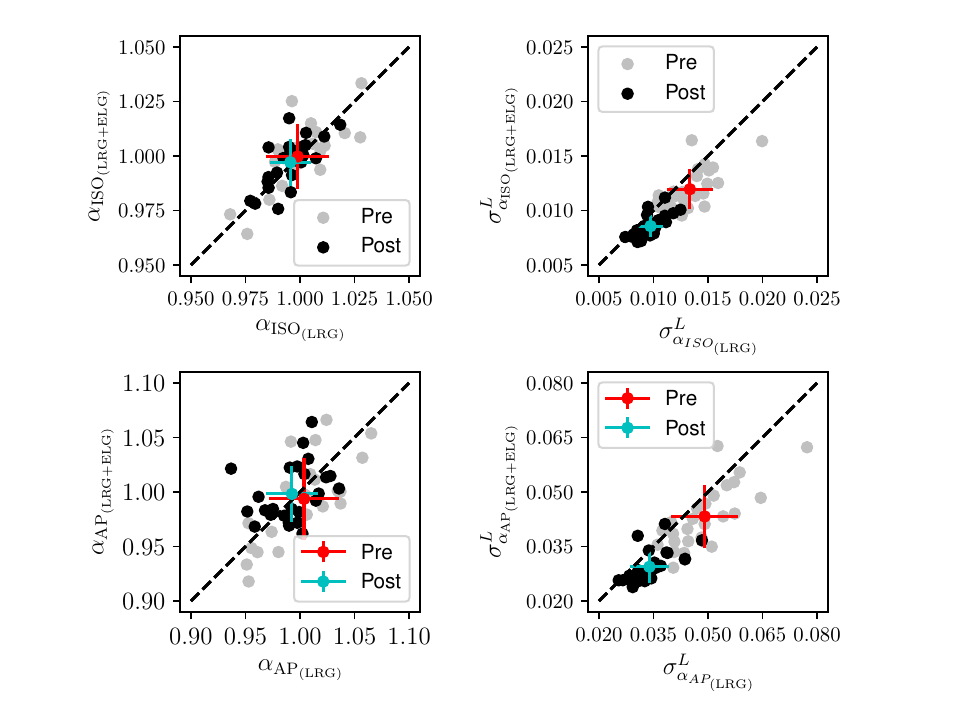}
    \caption{BAO constraints correlations between LRG and \lrgelg\ from DESI DR1 mocks.
    Left panels: the best fits are highly correlated, especially for $\aiso$.
    The red (pre-recon) and the cyan (post-recon) points and crosses show the average and the standard deviations of the 25 mocks.
    Right panels: the errors are also highly correlated, for both $\aiso$ and $\alap$.
    In the right panels, the scatter points tend to be below the $x=y$ (black dashed) line, meaning a gain in the likelihood error $\langle\sigma^L_{\alpha}\rangle$ for the combined tracers.}
\label{fig:mocks_prop2}
\end{figure*}

Going back to \abacussecond, the gain estimated from \ezmocks{}\ is more consistent with the gain estimated using \abacussecond\  $\langle {\sigma^L_{\alpha}}\rangle$: 11\% (13\%) for $\aiso$ ($\alap$). This is expected, as $\langle {\sigma^L_{\alpha}}\rangle$ depends on the covariance matrix used and would be more stable, whereas ${\sigma^D_{\aiso}}$ is subject to high sample variance of having 25 mocks and any box replication effect.
We therefore consider \abacussecond\ DR1 $\langle {\sigma^L_{\alpha}} \rangle$ as a more reliable measure of precision as a reference to check the DR1 data, than ${\sigma^D_{\aiso}}$ for the rest of the paper.
\cref{fig:mocks_prop2} shows the distribution of gains from the 25 DR1 mocks.

\subsubsection{Testing BAO systematic errors using the overlapping tracers}

We first test the bias in BAO measurement by comparing the average shift in the BAO scale parameters $\alpha$ from the mock best fits (with respect to the true $\alpha=1$ in our simulations) with respect to the statistical precision of the mocks.
The best-fit post-reconstruction BAO values of \abacussecond\ in \cref{tab:keybaofitsmocks1} demonstrate that \lrgelg\ as well as \lrgxelg\ give unbiased estimates of the BAO scales, with offsets from unity within a significance of $1.7~\sigma$, where the significance is defined in terms of the statistical precision of the 25 mocks \footnote{When quoting the significance $N\;\sigma$, $N = \frac{\langle \alpha \rangle -1}{\sigma_{\alpha}} \times \sqrt{n_S}$, where $n_S$ is the number of realizations and ${\sigma_{\alpha}}$ is from a smaller error between ${\sigma^D_{\alpha}}$ and $\langle {\sigma^L_{\alpha}}\rangle$.}.
This validates the use of \lrgelg\ as well as \lrgxelg\ as unbiased BAO tracers.
Note that the $2.3~\sigma$ offset from the unity in \lrg\ $\aiso$ was discussed in \cite{DESI2024.III.KP4}, and considered non-detection of systematics, as it does not pass the $3~\sigma$ threshold.

With multiple unbiased BAO tracers for this redshift bin, we can test the consistency among the BAO measurements.
In simulations, we can assure whether or not the BAO scales recovered are unbiased relative to the true BAO scale, as we have demonstrated above. With real data, however, we do not have an absolute reference to compare with to ensure that the observed BAO scales are unbiased. The existence of any residual systematics could be tested by checking the consistency among different BAO measurements that are potentially subject to different levels of systematics. As mentioned in \cref{sec:intro}, a disagreement between different BAO tracers in real data can potentially indicate a tracer-dependent source of systematics, either due to observational artifacts or due to a cosmological effect such as the relative velocity effect~\citep{2010PhRvD..82h3520T,barreira_baryon-cdm_2020,KP4s2-Chen}.
Since the tracers are correlated, tracing the same underlying matter distribution with a different clustering bias, this consistency test can benefit from the sample variance cancellation.
Using \abacussecond, we first test how much consistency is expected between the different but correlated tracers purely due to statistical fluctuations, in the absence of unknown systematics. 
As a reminder, these mocks include the fiber assignment effects {\tt `altmtl'}, the level of which is different between ELG and LRG, but do not include any other observational systematics.
The impact of the observational systematics, such as the imaging systematics and spectroscopic systematics, was shown to be negligible in \cite{DESI2024.III.KP4}.

\begin{table*}
\centering
\begin{tabular}{|c|c|c|c|c|c|}
\hline 
Estimates & Recon & $\langle \Delta\alpha \rangle_{\rm ISO} [\%]$ & $ \sigma(\Delta\alpha)_{\rm ISO} /5 [\%]$ & $\langle \Delta\alpha \rangle_{\rm AP} [\%]$ & $ \sigma(\Delta\alpha)_{\rm AP} /5 [\%]$ \\\hline 
$\alpha_E - \alpha_L$ & Post &
0.209 & 0.458 & 0.743 & 1.178\\
$(\alpha_\times - \alpha_L)$ & Post & 0.169 & 0.228 & 0.533 & 0.730\\
$(\alpha_+ - \alpha_L)$ & Post & 0.156 & 0.159 & 0.568 & 0.586\\
$(\alpha_+-1)^2-(\alpha_L-1)^2$ & Post & 0.0017 & 0.0028 & 0.0036 & 0.0276\\
\hline
$\alpha_E - \alpha_L$ & Pre &
0.252 & 0.554 & -1.973 & 0.985\\
$(\alpha_\times - \alpha_L)$ & Pre & 0.106 & 0.280 & -1.207 & 0.599 \\
$(\alpha_+ - \alpha_L)$ & Pre & 0.083 & 0.197 & -1.005 & 0.518 \\
$(\alpha_+-1)^2-(\alpha_L-1)^2$ & Pre & 0.0014 & 0.0053 & 0.0385 & 0.0334 \\
\hline
\end{tabular}
\caption{The expected level of consistency between BAO tracers estimated using DESI DR1 mocks of ELG, LRG, \lrgxelg,  and \lrgelg. $\alpha_E$ is the BAO constraint from ELG, $\alpha_L$ is for LRG, $\alpha_\times$ for \lrgxelg.
We derive the average and standard deviations of the difference of the matching initial condition using 25 BAO fits.
Due to the nonzero correlations, the sample variance is canceled to some extent, and the dispersions of the differences are smaller than the dispersions of individual tracer measurements.
Note that the standard deviations are divided by $\sqrt{25}$ to give the errors for the mean difference.
}
\label{tab:systematicsdiff}
\end{table*}

\begin{figure*}[h]
    \centering
    \includegraphics[width=0.9\textwidth]{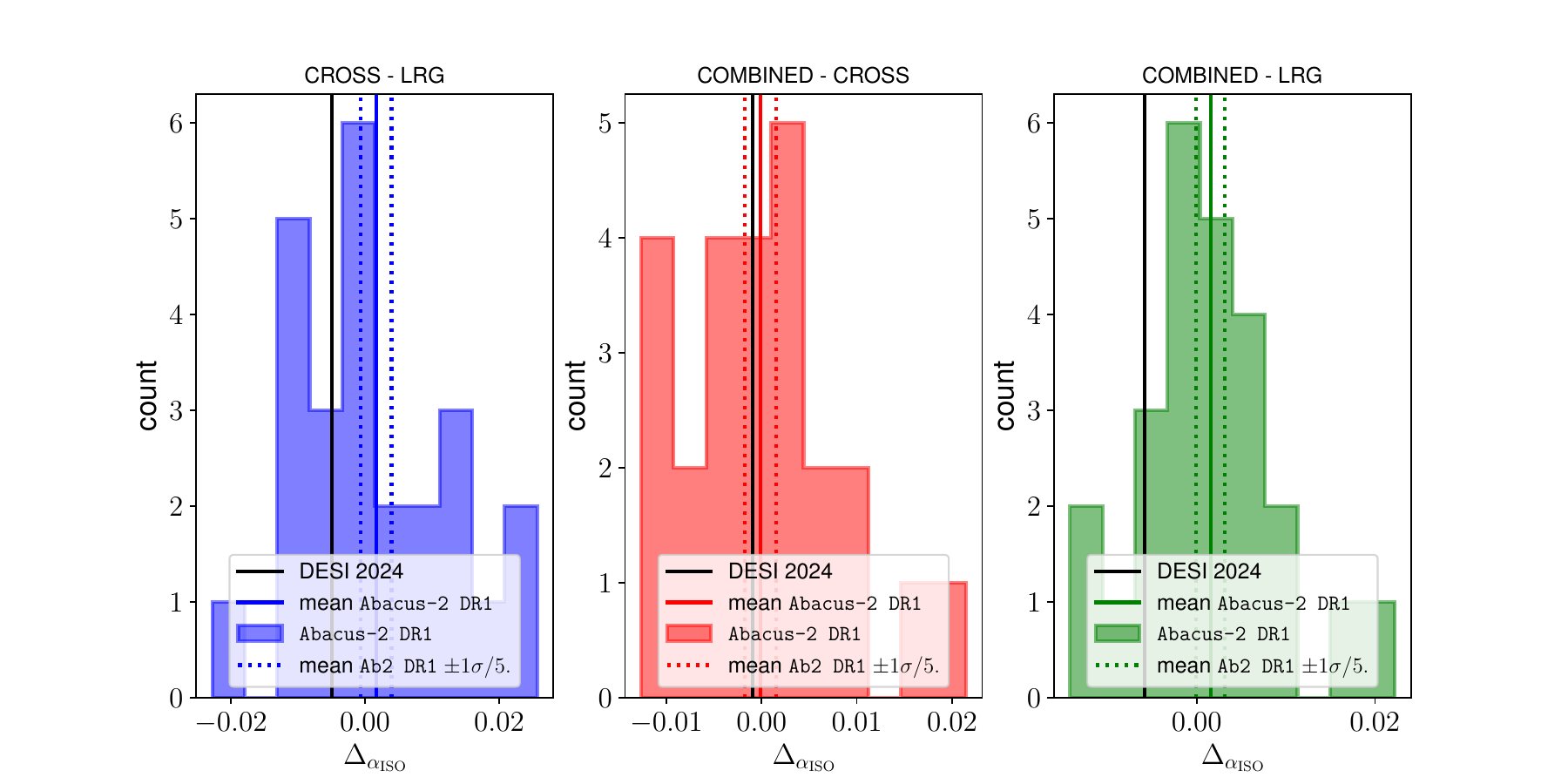}
    \includegraphics[width=0.9\textwidth]{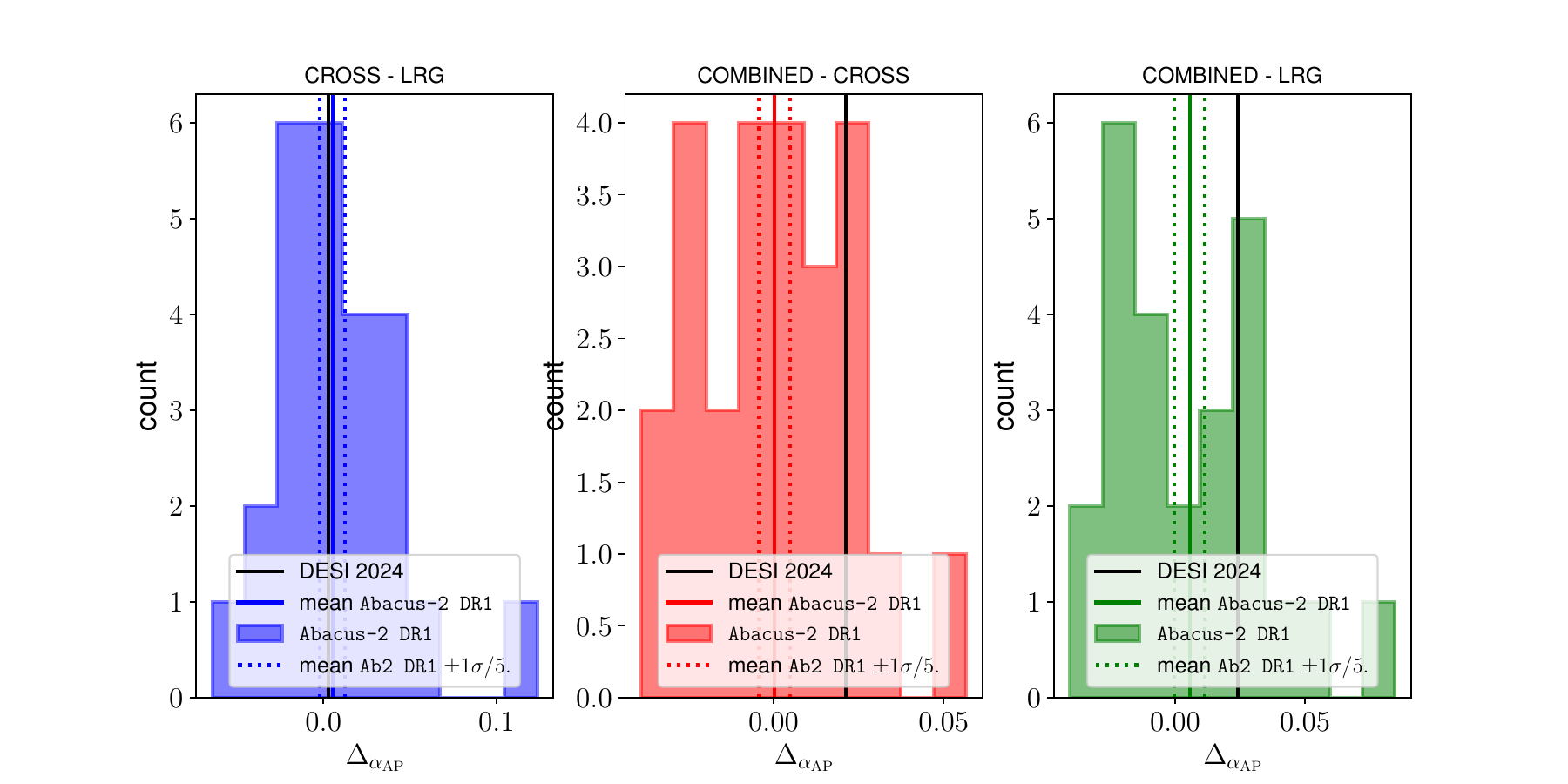}
    \caption{
    Histograms of the differences in $\alpha$ between LRG, LRG$\times$ELG and LRG+ELG using 25 \abacussecond\ DR1 mock realizations, compared to the corresponding differences measured from the DR1 data (vertical black solid lines).
    Solid colored lines indicate the mean of the mock distributions, while dashed colored lines show the expected $1\sigma$ uncertainty on the mean (i.e., the mock standard deviation divided by $\sqrt{25}$).
    As expected, the mock results are statistically consistent with zero, indicating no tracer-dependent systematics.
    The data (black solid lines) are compared with the full range of mock distributions to assess consistency with mocks that contain no tracer-dependent systematics; in all cases, the data lie within the expected mock scatter, indicating no significant systematic offset.}
    \label{fig:histogram_difference}
\end{figure*}

\cref{tab:systematicsdiff} presents the average and the dispersion of the differences between different BAO tracers.
The dispersions are divided by $\sqrt{n_S}$ with $n_S=25$ to represent the error associated with the average difference.
In \cref{fig:histogram_difference} we compare $\alpha_{\rm LRGxELG}- \alpha_{\rm LRG}$ (left panel in blue), $\alpha_{\rm {LRG+ELG}} - \alpha_{\rm LRGxELG}$ (middle panel in red) and $\alpha_{\rm {LRG+ELG}} - \alpha_{\rm LRG}$ (right panel in green) for $\aiso$ and $\alap$.
For every case, the differences are expected to be zero within the statistical precision; the mean of the mocks (solid colored vertical lines) agrees with zero difference within {$1\sigma$} of the statistical precision associated with the mean differences (vertical dashed lines, $1\sigma$ of the mean difference is equivalent to `$\sigma(\Delta \alpha))/5$' in \cref{tab:systematicsdiff}).
\cref{fig:histogram_difference} and \cref{tab:systematicsdiff} also imply that the purely statistical dispersions around the zero difference are typically around $\sim 1$\% (\lrg\ versus \lrgelg), implying that only relatively large tracer-dependent systematics would be detectable from our test.
The next section compares the differences measured from the DESI DR1 data with the expectation.

\subsection{Application to the DESI DR1 data catalog} 

\begin{figure*}
    \centering
    \includegraphics[width=\textwidth]{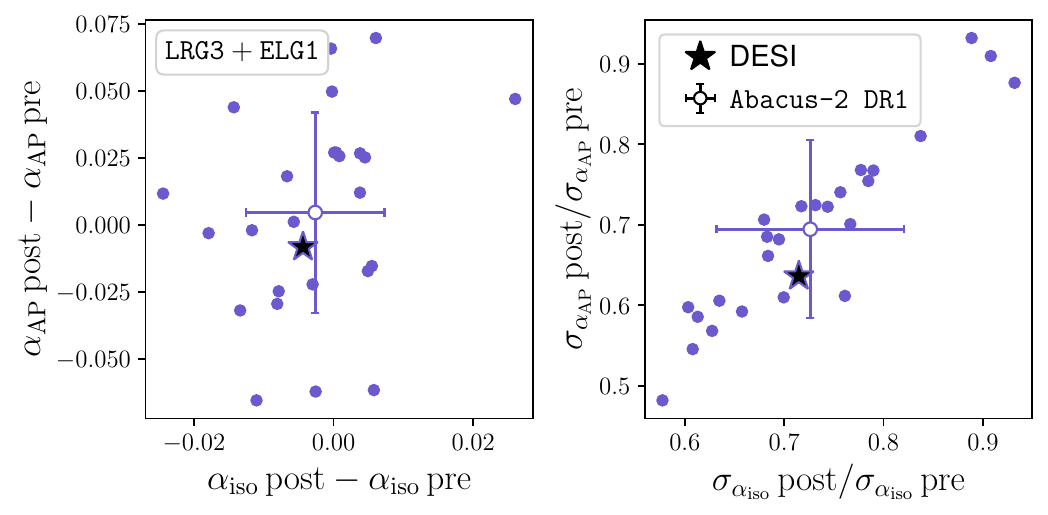}
    \caption{Pre- and post-reconstruction BAO measurements for \lrgelg, comparing \desidrone\ measurements (stars) with the 25 \abacussecond\ DR1 (the open points with the error bars are the means and the standard deviations around the means of the 25 mocks).
    The reconstruction efficiency of \desidrone\ \lrgelg\ is consistent with what is expected from the mocks.}
    \label{fig:mock_versus_data_scatter}
\end{figure*}

The last step of our analysis is to apply the combined tracer analysis on the DR1 data and investigate the gain and potential systematics.
In \cref{tab:Y1unblinded}, we present the results of BAO fits applied to the DESI DR1 data.
First, \cref{fig:mock_versus_data_scatter} repeats one of the sanity checks presented in \cite{DESI2024.III.KP4}: pre- and post-reconstruction BAO measurements for \lrgelg, comparing \desidrone\ measurements (stars) with the 25 \abacussecond\ DR1.
Therefore, the reconstruction efficiency of \desidrone\ \lrgelg\ is consistent with what is expected from the mocks.

\begin{figure*}
    \centering
    \includegraphics[width=\textwidth]{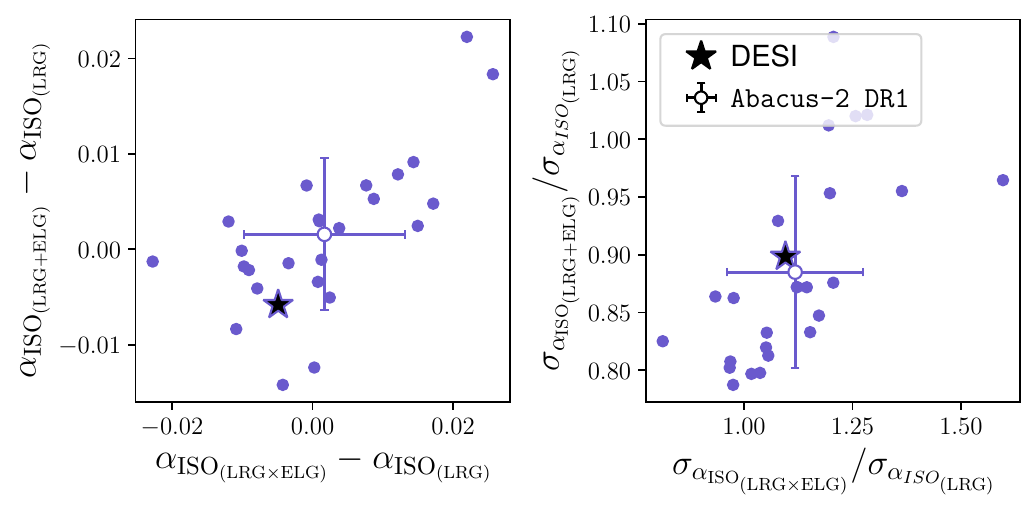}
    \includegraphics[width=\textwidth]{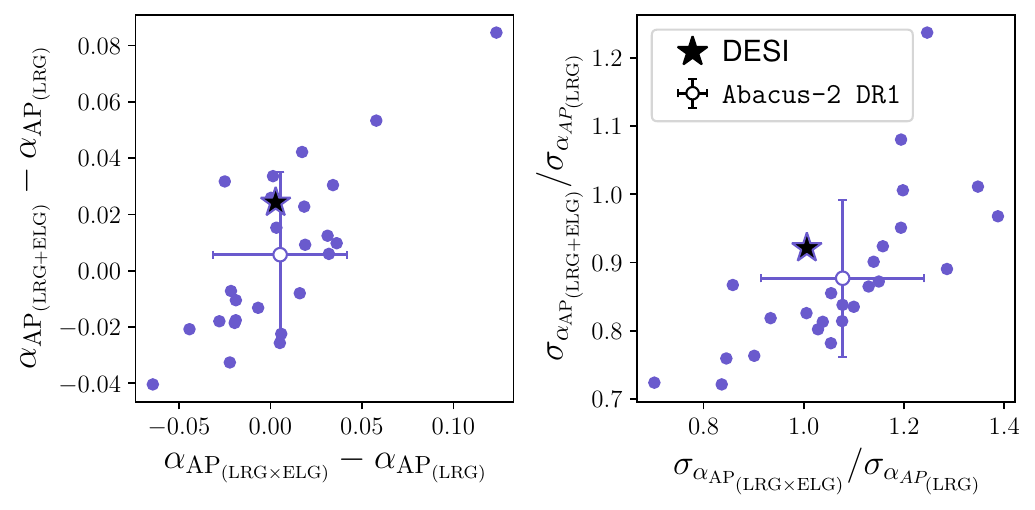}
    \caption{Post-reconstruction BAO measurements between LRG and \lrgelg\ as well as \lrgxelg, comparing \desidrone\ measurements (stars) with the 25 \abacussecond\ DR1 (the open points with the error bars are the means and the standard deviations around the means of the 25 mocks).
    The gain in precision of the combined tracer observed from the data (right panels) is consistent with the mocks, and the difference between the different BAO tracers (left panels) is consistent with no unknown systematics and no missing physics.}
    \label{fig:mock_versus_data_scatter2}
\end{figure*}

In terms of a gain in precision between \lrgelg\ and LRGs, \cref{tab:Y1unblinded} shows that, for $\aiso$ we find a gain of $5\%$ pre recon and $10\%$ post reconstruction, while the gain for $\alap$ is $\sim 2.2\%$ pre recon and $\sim 7.0\%$ post reconstruction. \cref{fig:mock_versus_data_scatter2} visualizes that post-reconstruction BAO measurements between LRG and \lrgelg\ (y-axis) as well as LRG and \lrgxelg\ (x-axis) fall within the distribution of the mocks.
Especially, the gain in precision of the combined tracer observed from the data (right panels) is consistent with the gain in terms of $\langle {\sigma^L_{\alpha}} \rangle$ of the \abacussecond\ mocks.

\cref{fig:histogram_difference} shows differences in $\alpha$'s from pairs of the BAO tracers, including \lrgxelg, comparing those of the mocks to the data.
All differences from DR1 data are within the range of the histograms, therefore consistent with zero offsets, and therefore no indication of a tracer-dependent systematics, whether due to different observational systematics or due to the relative velocity effect that would bias the BAO scales\footnote{Since we are testing a single realization, DR1, we compare the DR1 constraint with the entire width of the distribution, instead of the width divided by $\sqrt(25)$.}.
As a caveat, while \lrgxelg\ returns consistent precisions compared to the mocks, the data returns $\chi^2$ much higher than the average $\chi^2$ of the mocks in \cref{tab:keybaofitsmocks1}.
Although this implies $\xi_{\rm LRGxELG}$ may need a more careful inspection in terms of the goodness of the fits and its covariance matrix, we defer such a test to future study. We note that the relatively high $\chi^2$ values observed in the DR1 LRG×ELG measurement are reduced in DR2 (Sanders et al., in prep.), indicating improved consistency between the model and the data. For other cases, the reduced $\chi^2$, especially post-reconstruction, is reasonable and falls within the range covered by 25 \abacussecond DR1 mocks \citep{DESI2024.III.KP4}.

\begin{table}
    \centering
    \begin{tabular}{|l|c|c|c|c|c|r|}
    \hline
    Tracer & Redshift & Recon & $\sigma^L_{\alpha_{\rm iso}}$ & $\sigma^L_{\alpha_{\rm AP}}$ & $r_{\rm off}$ & $\chi^2 / {\rm dof}$ \\
    \hline
    {\tt LRG} & 0.8--1.1 & Post & 0.0090  & 0.0299 & -0.0597 & 46.2/39 \\
    {\lrgelg} & 0.8--1.1 & Post & 0.0081  & 0.0278 & -0.1019 & 33.1/39 \\
    {\lrgxelg}  & 0.8-1.1 & Post & 0.0099 & 0.0303 & -0.4203 & 63.5/39 \\
    {\tt ELG} & 0.8--1.1 & Post & 0.0200 &  & & 20.4/19 \\
    \hline
    {\tt LRG} & 0.8--1.1 & Pre & 0.0120  & 0.0446 & 0.2737 & 48.4/39 \\
    {\tt \lrgelg} & 0.8--1.1 & Pre & 0.0114  & 0.0436 & 0.2569 & 53.0/39 \\
    \lrgxelg{} & 0.8-1.1 & Pre & 0.0126 & 0.0473 & 0.2987 & 48.3/39 \\     
    {\tt ELG} & 0.8--1.1 & Pre & 0.0620 &  &  & 29.2/19 \\
    \hline
    \end{tabular}
    \caption{Standard deviations from the marginalized posteriors of the BAO scaling parameters from fits to the unblinded \desidrone\ correlation functions of LRG, \lrgelg, and \lrgxelg. $r_{\rm off}=C_{\aiso,\alap}/\sqrt{C_{\aiso,\aiso}C_{\alap,\alap}}$.
    LRG, ELG, and \lrgelg\ are consistent with what is presented in \cite{DESI2024.III.KP4}.}
    \label{tab:Y1unblinded}
\end{table}

\section{Conclusion}
\label{sec:Conclusion}

In this paper, we presented how the Dark Energy Spectroscopic Instrument (DESI) current and future BAO analyses can optimally combine overlapping tracers in the same redshift range.
We focused on the two tracers, DESI DR1 Luminous red galaxies (LRGs) and Emission line galaxies (ELGs) in the redshift range $0.8<z<1.1$ and investigated the impact of the combined tracer on the BAO constraints.

The combination of tracers into a unified catalog enables a more robust integration of the two tracers, including the information from the cross-correlation of the two tracers.
This paper presented a simple method of combining the two catalogs and tested the robustness of the combined catalog as a BAO tracer. In parallel, we tested if the lower shot noise of the combined tracer could potentially improve the reconstruction efficiency beyond the reconstruction of the individual tracers.
We also tested if the different tracer types in the same volume allow for tests of potential tracer-dependent systematic effects, either observational or physics missing in our model, such as the relative velocity effect. 

We find that due to the low completeness of DR1 ELGs, the combined \lrgelg\ tracer does not exhibit a clear gain in reconstruction efficiency compared to that of the dominant tracer, LRG.
Instead, it achieves a similar level of efficiency.
We expect future DESI data with higher ELG completeness near the final DESI dataset to yield improved displacement field estimates. 

We observe a gain of $\sim 10\%$ in the BAO precision with the combined tracer compared to using only LRGs, which appears to be consistent with the additional information from the ELG BAO feature.
Therefore, the main advantage of the combined tracer over LRGs alone in DR1 is its stable and complete integration of low–S/N ELG data through a unified pre-analysis catalog, naturally incorporating auto- and cross-clustering to maximize information extraction.
This benefit remains applicable to future data releases with a higher S/N ELG as well and is indeed utilized in \cite{DESI.DR2.BAO.cosmo,Y3BAO-Andrade,Y3BAO-Sanders}.

Using DR1 mocks, we demonstrate that the combined tracer \lrgelg\ constructed using our method, as well as the cross-statistics, \lrgxelg, are unbiased tracers of BAO post-reconstruction in the presence of the simulated survey realism.
Using these BAO tracers from the DR1 data in addition to LRG and ELGs, we tested for potential systematics on the BAO scale at the redshift range of $0.8-1.1$ by comparing the BAO measurement across different overlapping tracers and found results that are consistent with no tracer-dependency between different tracers.
That is, we detect neither the net effect of systematics nor the effect of relative velocity effects, within the statistical precision of DR1.
With DR2, we revisit and extend the set of tests to other tracers and redshift ranges. \citep{Y3BAO-Sanders}.

The method presented in this paper was integrated into the DESI DR1 BAO analysis to produce the \lrgelg\ at the redshift bin of $0.8-1.1$, which provided the most precise DESI BAO measurement of DR1 with a 0.81\% constraint on $D_V/r_d$ corresponding to a 9.1$\sigma$ detection of the isotropic BAO feature. 

\section*{Data Availability}

DESI Data Release 1 is available at \url{https://data.desi.lbl.gov/doc/releases/}.
Additional data and code to reproduce the figures are available at \supplementarymaterial{}.

\acknowledgments

DV and H-JS acknowledge support from the U.S. Department of Energy, Office of Science, Office of High Energy Physics under grant No. DE-SC0019091 and No. DE-SC0023241.
MR has been supported by U.S. Department of Energy grant DE-SC0013718 and by the Simons Foundation Investigator program.

This material is based upon work supported by the U.S. Department of Energy (DOE), Office of Science, Office of High-Energy Physics, under Contract No. DE–AC02–05CH11231, and by the National Energy Research Scientific Computing Center, a DOE Office of Science User Facility under the same contract. Additional support for DESI was provided by the U.S. National Science Foundation (NSF), Division of Astronomical Sciences under Contract No. AST-0950945 to the NSF’s National Optical-Infrared Astronomy Research Laboratory; the Science and Technology Facilities Council of the United Kingdom; the Gordon and Betty Moore Foundation; the Heising-Simons Foundation; the French Alternative Energies and Atomic Energy Commission (CEA); the National Council of Humanities, Science and Technology of Mexico (CONAHCYT); the Ministry of Science, Innovation and Universities of Spain (MICIU/AEI/10.13039/501100011033), and by the DESI Member Institutions: \url{https://www.desi.lbl.gov/collaborating-institutions}. Any opinions, findings, and conclusions or recommendations expressed in this material are those of the author(s) and do not necessarily reflect the views of the U. S. National Science Foundation, the U. S. Department of Energy, or any of the listed funding agencies.

The authors are honored to be permitted to conduct scientific research on I'oligam Du'ag (Kitt Peak), a mountain with particular significance to the Tohono O’odham Nation.

This work has used the following software packages: \textsc{astropy} \citep{astropy:2013, astropy:2018, astropy:2022}, \textsc{healpy} and \textsc{HEALPix}\footnote{\url{https://healpix.sourceforge.net}} \citep{Zonca2019, 2005ApJ...622..759G,healpy_8404216}, \textsc{Jupyter} \citep{2007CSE.....9c..21P, kluyver2016jupyter}, \textsc{matplotlib} \citep{Hunter:2007}, \textsc{numpy} \citep{numpy}, \textsc{pycorr}\footnote{\url{https://github.com/cosmodesi/pycorr}} \citep{pycorr,corrfunc-1,corrfunc-2}, \textsc{python} \citep{python}, \rascalc{}\footnote{\url{https://github.com/oliverphilcox/RascalC}} \citep{rascal,rascal-jackknife,rascalC,2023MNRAS.524.3894R,KP4s7-Rashkovetskyi}, \textsc{scipy} \citep{2020SciPy-NMeth, scipy_10155614}, and \textsc{scikit-learn} \citep{scikit-learn, sklearn_api, scikit-learn_10034229}.

This research has used NASA's Astrophysics Data System.
Software citation information has been aggregated using \texttt{\href{https://www.tomwagg.com/software-citation-station/}{The Software Citation Station}} \citep{software-citation-station-paper, software-citation-station-zenodo}.

\bibliographystyle{JHEP}
\bibliography{biblio,DESI_supporting_papers,software}

@ARTICLE{DESI2016a.Science,
       author = {{DESI Collaboration} and {Aghamousa}, Amir and {Aguilar}, Jessica and {Ahlen}, Steve and {Alam}, Shadab and {Allen}, Lori E. and {Allende Prieto}, Carlos and {Annis}, James and {Bailey}, Stephen and {Balland}, Christophe and {Ballester}, Otger and {Baltay}, Charles and {Beaufore}, Lucas and {Bebek}, Chris and {Beers}, Timothy C. and {Bell}, Eric F. and {Bernal}, Jos{\'e} Luis and {Besuner}, Robert and {Beutler}, Florian and {Blake}, Chris and {Bleuler}, Hannes and {Blomqvist}, Michael and {Blum}, Robert and {Bolton}, Adam S. and {Briceno}, Cesar and {Brooks}, David and {Brownstein}, Joel R. and {Buckley-Geer}, Elizabeth and {Burden}, Angela and {Burtin}, Etienne and {Busca}, Nicolas G. and {Cahn}, Robert N. and {Cai}, Yan-Chuan and {Cardiel-Sas}, Laia and {Carlberg}, Raymond G. and {Carton}, Pierre-Henri and {Casas}, Ricard and {Castander}, Francisco J. and {Cervantes-Cota}, Jorge L. and {Claybaugh}, Todd M. and {Close}, Madeline and {Coker}, Carl T. and {Cole}, Shaun and {Comparat}, Johan and {Cooper}, Andrew P. and {Cousinou}, M. -C. and {Crocce}, Martin and {Cuby}, Jean-Gabriel and {Cunningham}, Daniel P. and {Davis}, Tamara M. and {Dawson}, Kyle S. and {de la Macorra}, Axel and {De Vicente}, Juan and {Delubac}, Timoth{\'e}e and {Derwent}, Mark and {Dey}, Arjun and {Dhungana}, Govinda and {Ding}, Zhejie and {Doel}, Peter and {Duan}, Yutong T. and {Ealet}, Anne and {Edelstein}, Jerry and {Eftekharzadeh}, Sarah and {Eisenstein}, Daniel J. and {Elliott}, Ann and {Escoffier}, St{\'e}phanie and {Evatt}, Matthew and {Fagrelius}, Parker and {Fan}, Xiaohui and {Fanning}, Kevin and {Farahi}, Arya and {Farihi}, Jay and {Favole}, Ginevra and {Feng}, Yu and {Fernandez}, Enrique and {Findlay}, Joseph R. and {Finkbeiner}, Douglas P. and {Fitzpatrick}, Michael J. and {Flaugher}, Brenna and {Flender}, Samuel and {Font-Ribera}, Andreu and {Forero-Romero}, Jaime E. and {Fosalba}, Pablo and {Frenk}, Carlos S. and {Fumagalli}, Michele and {Gaensicke}, Boris T. and {Gallo}, Giuseppe and {Garcia-Bellido}, Juan and {Gaztanaga}, Enrique and {Pietro Gentile Fusillo}, Nicola and {Gerard}, Terry and {Gershkovich}, Irena and {Giannantonio}, Tommaso and {Gillet}, Denis and {Gonzalez-de-Rivera}, Guillermo and {Gonzalez-Perez}, Violeta and {Gott}, Shelby and {Graur}, Or and {Gutierrez}, Gaston and {Guy}, Julien and {Habib}, Salman and {Heetderks}, Henry and {Heetderks}, Ian and {Heitmann}, Katrin and {Hellwing}, Wojciech A. and {Herrera}, David A. and {Ho}, Shirley and {Holland}, Stephen and {Honscheid}, Klaus and {Huff}, Eric and {Hutchinson}, Timothy A. and {Huterer}, Dragan and {Hwang}, Ho Seong and {Illa Laguna}, Joseph Maria and {Ishikawa}, Yuzo and {Jacobs}, Dianna and {Jeffrey}, Niall and {Jelinsky}, Patrick and {Jennings}, Elise and {Jiang}, Linhua and {Jimenez}, Jorge and {Johnson}, Jennifer and {Joyce}, Richard and {Jullo}, Eric and {Juneau}, St{\'e}phanie and {Kama}, Sami and {Karcher}, Armin and {Karkar}, Sonia and {Kehoe}, Robert and {Kennamer}, Noble and {Kent}, Stephen and {Kilbinger}, Martin and {Kim}, Alex G. and {Kirkby}, David and {Kisner}, Theodore and {Kitanidis}, Ellie and {Kneib}, Jean-Paul and {Koposov}, Sergey and {Kovacs}, Eve and {Koyama}, Kazuya and {Kremin}, Anthony and {Kron}, Richard and {Kronig}, Luzius and {Kueter-Young}, Andrea and {Lacey}, Cedric G. and {Lafever}, Robin and {Lahav}, Ofer and {Lambert}, Andrew and {Lampton}, Michael and {Landriau}, Martin and {Lang}, Dustin and {Lauer}, Tod R. and {Le Goff}, Jean-Marc and {Le Guillou}, Laurent and {Le Van Suu}, Auguste and {Lee}, Jae Hyeon and {Lee}, Su-Jeong and {Leitner}, Daniela and {Lesser}, Michael and {Levi}, Michael E. and {L'Huillier}, Benjamin and {Li}, Baojiu and {Liang}, Ming and {Lin}, Huan and {Linder}, Eric and {Loebman}, Sarah R. and {Luki{\'c}}, Zarija and {Ma}, Jun and {MacCrann}, Niall and {Magneville}, Christophe and {Makarem}, Laleh and {Manera}, Marc and {Manser}, Christopher J. and {Marshall}, Robert and {Martini}, Paul and {Massey}, Richard and {Matheson}, Thomas and {McCauley}, Jeremy and {McDonald}, Patrick and {McGreer}, Ian D. and {Meisner}, Aaron and {Metcalfe}, Nigel and {Miller}, Timothy N. and {Miquel}, Ramon and {Moustakas}, John and {Myers}, Adam and {Naik}, Milind and {Newman}, Jeffrey A. and {Nichol}, Robert C. and {Nicola}, Andrina and {Nicolati da Costa}, Luiz and {Nie}, Jundan and {Niz}, Gustavo and {Norberg}, Peder and {Nord}, Brian and {Norman}, Dara and {Nugent}, Peter and {O'Brien}, Thomas and {Oh}, Minji and {Olsen}, Knut A.~G.},
        title = "{The DESI Experiment Part I: Science,Targeting, and Survey Design}",
      journal = {arXiv e-prints},
         year = 2016,
        month = oct,
          eid = {arXiv:1611.00036},
        pages = {arXiv:1611.00036},
          doi = {10.48550/arXiv.1611.00036},
archivePrefix = {arXiv},
       eprint = {1611.00036},
 primaryClass = {astro-ph.IM},
       adsurl = {https://ui.adsabs.harvard.edu/abs/2016arXiv161100036D}
}

@ARTICLE{DESI2022.KP1.Instr,
       author = {{DESI Collaboration} and {Abareshi}, B. and {Aguilar}, J. and {Ahlen}, S. and {Alam}, Shadab and {Alexander}, David M. and {Alfarsy}, R. and {Allen}, L. and {Allende Prieto}, C. and {Alves}, O. and {Ameel}, J. and {Armengaud}, E. and {Asorey}, J. and {Aviles}, Alejandro and {Bailey}, S. and {Balaguera-Antol{\'\i}nez}, A. and {Ballester}, O. and {Baltay}, C. and {Bault}, A. and {Beltran}, S.~F. and {Benavides}, B. and {BenZvi}, S. and {Berti}, A. and {Besuner}, R. and {Beutler}, Florian and {Bianchi}, D. and {Blake}, C. and {Blanc}, P. and {Blum}, R. and {Bolton}, A. and {Bose}, S. and {Bramall}, D. and {Brieden}, S. and {Brodzeller}, A. and {Brooks}, D. and {Brownewell}, C. and {Buckley-Geer}, E. and {Cahn}, R.~N. and {Cai}, Z. and {Canning}, R. and {Capasso}, R. and {Carnero Rosell}, A. and {Carton}, P. and {Casas}, R. and {Castander}, F.~J. and {Cervantes-Cota}, J.~L. and {Chabanier}, S. and {Chaussidon}, E. and {Chuang}, C. and {Circosta}, C. and {Cole}, S. and {Cooper}, A.~P. and {da Costa}, L. and {Cousinou}, M. -C. and {Cuceu}, A. and {Davis}, T.~M. and {Dawson}, K. and {de la Cruz-Noriega}, R. and {de la Macorra}, A. and {de Mattia}, A. and {Della Costa}, J. and {Demmer}, P. and {Derwent}, M. and {Dey}, A. and {Dey}, B. and {Dhungana}, G. and {Ding}, Z. and {Dobson}, C. and {Doel}, P. and {Donald-McCann}, J. and {Donaldson}, J. and {Douglass}, K. and {Duan}, Y. and {Dunlop}, P. and {Edelstein}, J. and {Eftekharzadeh}, S. and {Eisenstein}, D.~J. and {Enriquez-Vargas}, M. and {Escoffier}, S. and {Evatt}, M. and {Fagrelius}, P. and {Fan}, X. and {Fanning}, K. and {Fawcett}, V.~A. and {Ferraro}, S. and {Ereza}, J. and {Flaugher}, B. and {Font-Ribera}, A. and {Forero-Romero}, J.~E. and {Frenk}, C.~S. and {Fromenteau}, S. and {G{\"a}nsicke}, B.~T. and {Garcia-Quintero}, C. and {Garrison}, L. and {Gazta{\~n}aga}, E. and {Gerardi}, F. and {Gil-Mar{\'\i}n}, H. and {Gontcho a Gontcho}, S. and {Gonzalez-Morales}, Alma X. and {Gonzalez-de-Rivera}, G. and {Gonzalez-Perez}, V. and {Gordon}, C. and {Graur}, O. and {Green}, D. and {Grove}, C. and {Gruen}, D. and {Gutierrez}, G. and {Guy}, J. and {Hahn}, C. and {Harris}, S. and {Herrera}, D. and {Herrera-Alcantar}, Hiram K. and {Honscheid}, K. and {Howlett}, C. and {Huterer}, D. and {Ir{\v{s}}i{\v{c}}}, V. and {Ishak}, M. and {Jelinsky}, P. and {Jiang}, L. and {Jimenez}, J. and {Jing}, Y.~P. and {Joyce}, R. and {Jullo}, E. and {Juneau}, S. and {Kara{\c{c}}ayl{\i}}, N.~G. and {Karamanis}, M. and {Karcher}, A. and {Karim}, T. and {Kehoe}, R. and {Kent}, S. and {Kirkby}, D. and {Kisner}, T. and {Kitaura}, F. and {Koposov}, S.~E. and {Kov{\'a}cs}, A. and {Kremin}, A. and {Krolewski}, Alex and {L'Huillier}, B. and {Lahav}, O. and {Lambert}, A. and {Lamman}, C. and {Lan}, Ting-Wen and {Landriau}, M. and {Lane}, S. and {Lang}, D. and {Lange}, J.~U. and {Lasker}, J. and {Le Guillou}, L. and {Leauthaud}, A. and {Le Van Suu}, A. and {Levi}, Michael E. and {Li}, T.~S. and {Magneville}, C. and {Manera}, M. and {Manser}, Christopher J. and {Marshall}, B. and {Martini}, Paul and {McCollam}, W. and {McDonald}, P. and {Meisner}, Aaron M. and {Mena-Fern{\'a}ndez}, J. and {Meneses-Rizo}, J. and {Mezcua}, M. and {Miller}, T. and {Miquel}, R. and {Montero-Camacho}, P. and {Moon}, J. and {Moustakas}, J. and {Mueller}, E. and {Mu{\~n}oz-Guti{\'e}rrez}, Andrea and {Myers}, Adam D. and {Nadathur}, S. and {Najita}, J. and {Napolitano}, L. and {Neilsen}, E. and {Newman}, Jeffrey A. and {Nie}, J.~D. and {Ning}, Y. and {Niz}, G. and {Norberg}, P. and {Noriega}, Hern{\'a}n E. and {O'Brien}, T. and {Obuljen}, A. and {Palanque-Delabrouille}, N. and {Palmese}, A. and {Zhiwei}, P. and {Pappalardo}, D. and {PENG}, X. and {Percival}, W.~J. and {Perruchot}, S. and {Pogge}, R. and {Poppett}, C. and {Porredon}, A. and {Prada}, F. and {Prochaska}, J. and {Pucha}, R. and {P{\'e}rez-Fern{\'a}ndez}, A. and {P{\'e}rez-R{\`a}fols}, I. and {Rabinowitz}, D. and {Raichoor}, A. and {Ramirez-Solano}, S. and {Ram{\'\i}rez-P{\'e}rez}, C{\'e}sar and {Ravoux}, C. and {Reil}, K. and {Rezaie}, M. and {Rocher}, A. and {Rockosi}, C. and {Roe}, N.~A. and {Roodman}, A. and {Ross}, A.~J. and {Rossi}, G. and {Ruggeri}, R. and {Ruhlmann-Kleider}, V. and {Sabiu}, C.~G. and {Safonova}, S. and {Said}, K. and {Saintonge}, A. and {Salas Catonga}, Javier and {Samushia}, L. and {Sanchez}, E. and {Saulder}, C. and {Schaan}, E. and {Schlafly}, E. and {Schlegel}, D. and {Schmoll}, J. and {Scholte}, D. and {Schubnell}, M. and {Secroun}, A. and {Seo}, H. and {Serrano}, S. and {Sharples}, Ray M. and {Sholl}, Michael J. and {Silber}, Joseph Harry and {Silva}, D.~R. and {Sirk}, M. and {Siudek}, M. and {Smith}, A. and {Sprayberry}, D. and {Staten}, R. and {Stupak}, B. and {Tan}, T. and {Tarl{\'e}}, Gregory and {Tie}, Suk Sien and {Tojeiro}, R. and {Ure{\~n}a-L{\'o}pez}, L.~A. and {Valdes}, F. and {Valenzuela}, O. and {Valluri}, M. and {Vargas-Maga{\~n}a}, M. and {Verde}, L. and {Walther}, M. and {Wang}, B. and {Wang}, M.~S. and {Weaver}, B.~A. and {Weaverdyck}, C. and {Wechsler}, R. and {Wilson}, Michael J. and {Yang}, J. and {Yu}, Y. and {Yuan}, S. and {Y{\`e}che}, Christophe and {Zhang}, H. and {Zhang}, K. and {Zhao}, Cheng and {Zhou}, Rongpu and {Zhou}, Zhimin and {Zou}, H. and {Zou}, J. and {Zou}, S. and {Zu}, Y. and {DESI Collaboration}},
        title = "{Overview of the Instrumentation for the Dark Energy Spectroscopic Instrument}",
      journal = {\aj},
         year = 2022,
        month = nov,
       volume = {164},
       number = {5},
          eid = {207},
        pages = {207},
          doi = {10.3847/1538-3881/ac882b},
archivePrefix = {arXiv},
       eprint = {2205.10939},
 primaryClass = {astro-ph.IM},
       adsurl = {https://ui.adsabs.harvard.edu/abs/2022AJ....164..207A}
}

@ARTICLE{DESI2023a.KP1.SV,
       author = {{DESI Collaboration} and {Adame}, A.~G. and {Aguilar}, J. and {Ahlen}, S. and {Alam}, S. and {Aldering}, G. and {Alexander}, D.~M. and {Alfarsy}, R. and {Allende Prieto}, C. and {Alvarez}, M. and {Alves}, O. and {Anand}, A. and {Andrade-Oliveira}, F. and {Armengaud}, E. and {Asorey}, J. and {Avila}, S. and {Aviles}, A. and {Bailey}, S. and {Balaguera-Antol{\'\i}nez}, A. and {Ballester}, O. and {Baltay}, C. and {Bault}, A. and {Bautista}, J. and {Behera}, J. and {Beltran}, S.~F. and {BenZvi}, S. and {Beraldo e Silva}, L. and {Bermejo-Climent}, J.~R. and {Berti}, A. and {Besuner}, R. and {Beutler}, F. and {Bianchi}, D. and {Blake}, C. and {Blum}, R. and {Bolton}, A.~S. and {Brieden}, S. and {Brodzeller}, A. and {Brooks}, D. and {Brown}, Z. and {Buckley-Geer}, E. and {Burtin}, E. and {Cabayol-Garcia}, L. and {Cai}, Z. and {Canning}, R. and {Cardiel-Sas}, L. and {Carnero Rosell}, A. and {Castander}, F.~J. and {Cervantes-Cota}, J.~L. and {Chabanier}, S. and {Chaussidon}, E. and {Chaves-Montero}, J. and {Chen}, S. and {Chen}, X. and {Chuang}, C. and {Claybaugh}, T. and {Cole}, S. and {Cooper}, A.~P. and {Cuceu}, A. and {Davis}, T.~M. and {Dawson}, K. and {de Belsunce}, R. and {de la Cruz}, R. and {de la Macorra}, A. and {de Mattia}, A. and {Demina}, R. and {Demirbozan}, U. and {DeRose}, J. and {Dey}, A. and {Dey}, B. and {Dhungana}, G. and {Ding}, J. and {Ding}, Z. and {Doel}, P. and {Doshi}, R. and {Douglass}, K. and {Edge}, A. and {Eftekharzadeh}, S. and {Eisenstein}, D.~J. and {Elliott}, A. and {Escoffier}, S. and {Fagrelius}, P. and {Fan}, X. and {Fanning}, K. and {Fawcett}, V.~A. and {Ferraro}, S. and {Ereza}, J. and {Flaugher}, B. and {Font-Ribera}, A. and {Forero-S{\'a}nchez}, D. and {Forero-Romero}, J.~E. and {Frenk}, C.~S. and {G{\"a}nsicke}, B.~T. and {Garc{\'\i}a}, L. {\'A}. and {Garc{\'\i}a-Bellido}, J. and {Garcia-Quintero}, C. and {Garrison}, L.~H. and {Gil-Mar{\'\i}n}, H. and {Golden-Marx}, J. and {Gontcho A Gontcho}, S. and {Gonzalez-Morales}, A.~X. and {Gonzalez-Perez}, V. and {Gordon}, C. and {Graur}, O. and {Green}, D. and {Gruen}, D. and {Guy}, J. and {Hadzhiyska}, B. and {Hahn}, C. and {Han}, J.~J. and {Hanif}, M.~M.~S. and {Herrera-Alcantar}, H.~K. and {Honscheid}, K. and {Hou}, J. and {Howlett}, C. and {Huterer}, D. and {Ir{\v{s}}i{\v{c}}}, V. and {Ishak}, M. and {Jana}, A. and {Jiang}, L. and {Jimenez}, J. and {Jing}, Y.~P. and {Joudaki}, S. and {Jullo}, E. and {Joyce}, R. and {Juneau}, S. and {Kizhuprakkat}, N. and {Kara{\c{c}}ayl{\i}}, N.~G. and {Karim}, T. and {Kehoe}, R. and {Kent}, S. and {Khederlarian}, A. and {Kim}, S. and {Kirkby}, D. and {Kisner}, T. and {Kitaura}, F. and {Kneib}, J. and {Koposov}, S.~E. and {Kov{\'a}cs}, A. and {Kremin}, A. and {Krolewski}, A. and {L'Huillier}, B. and {Lahav}, O. and {Lambert}, A. and {Lamman}, C. and {Lan}, T. -W. and {Landriau}, M. and {Lang}, D. and {Lange}, J.~U. and {Lasker}, J. and {Le Guillou}, L. and {Leauthaud}, A. and {Levi}, M.~E. and {Li}, T.~S. and {Linder}, E. and {Lyons}, A. and {Magneville}, C. and {Manera}, M. and {Manser}, C.~J. and {Margala}, D. and {Martini}, P. and {McDonald}, P. and {Medina}, G.~E. and {Medina-Varela}, L. and {Meisner}, A. and {Mena-Fern{\'a}ndez}, J. and {Meneses-Rizo}, J. and {Mezcua}, M. and {Miquel}, R. and {Montero-Camacho}, P. and {Moon}, J. and {Moore}, S. and {Moustakas}, J. and {Mueller}, E. and {Mundet}, J. and {Mu{\~n}oz-Guti{\'e}rrez}, A. and {Myers}, A.~D. and {Nadathur}, S. and {Napolitano}, L. and {Neveux}, R. and {Newman}, J.~A. and {Nie}, J. and {Niz}, G. and {Norberg}, P. and {Noriega}, H.~E. and {Paillas}, E. and {Palanque-Delabrouille}, N. and {Palmese}, A. and {Zhiwei}, P. and {Parkinson}, D. and {Penmetsa}, S. and {Percival}, W.~J. and {P{\'e}rez-Fern{\'a}ndez}, A. and {P{\'e}rez-R{\`a}fols}, I. and {Pieri}, M. and {Poppett}, C. and {Porredon}, A. and {Prada}, F. and {Pucha}, R. and {Raichoor}, A. and {Ram{\'\i}rez-P{\'e}rez}, C. and {Ramirez-Solano}, S. and {Rashkovetskyi}, M. and {Ravoux}, C. and {Rocher}, A. and {Rockosi}, C. and {Ross}, A.~J. and {Rossi}, G. and {Ruggeri}, R. and {Ruhlmann-Kleider}, V. and {Sabiu}, C.~G. and {Said}, K. and {Saintonge}, A. and {Samushia}, L. and {Sanchez}, E. and {Saulder}, C. and {Schaan}, E. and {Schlafly}, E.~F. and {Schlegel}, D. and {Scholte}, D. and {Schubnell}, M. and {Seo}, H. and {Shafieloo}, A. and {Sharples}, R. and {Sheu}, W. and {Silber}, J. and {Sinigaglia}, F. and {Siudek}, M. and {Slepian}, Z. and {Smith}, A. and {Sprayberry}, D. and {Stephey}, L. and {Su{\'a}rez-P{\'e}rez}, J. and {Sun}, Z. and {Tan}, T. and {Tarl{\'e}}, G. and {Tojeiro}, R. and {Ure{\~n}a-L{\'o}pez}, L.~A. and {Vaisakh}, R. and {Valcin}, D. and {Valdes}, F. and {Valluri}, M. and {Vargas-Maga{\~n}a}, M. and {Variu}, A. and {Verde}, L. and {Walther}, M. and {Wang}, B. and {Wang}, M.~S. and {Weaver}, B.~A. and {Weaverdyck}, N. and {Wechsler}, R.~H. and {White}, M. and {Xie}, Y. and {Yang}, J. and {Y{\`e}che}, C. and {Yu}, J. and {Yuan}, S. and {Zhang}, H. and {Zhang}, Z. and {Zhao}, C. and {Zheng}, Z. and {Zhou}, R. and {Zhou}, Z. and {Zou}, H. and {Zou}, S. and {Zu}, Y. and {DESI Collaboration}},
        title = "{Validation of the Scientific Program for the Dark Energy Spectroscopic Instrument}",
      journal = {\aj},
         year = 2024,
        month = feb,
       volume = {167},
       number = {2},
          eid = {62},
        pages = {62},
          doi = {10.3847/1538-3881/ad0b08},
archivePrefix = {arXiv},
       eprint = {2306.06307},
 primaryClass = {astro-ph.CO},
       adsurl = {https://ui.adsabs.harvard.edu/abs/2024AJ....167...62A}
}

@ARTICLE{DESI2023b.KP1.EDR,
       author = {{DESI Collaboration} and {Adame}, A.~G. and {Aguilar}, J. and {Ahlen}, S. and {Alam}, S. and {Aldering}, G. and {Alexander}, D.~M. and {Alfarsy}, R. and {Allende Prieto}, C. and {Alvarez}, M. and {Alves}, O. and {Anand}, A. and {Andrade-Oliveira}, F. and {Armengaud}, E. and {Asorey}, J. and {Avila}, S. and {Aviles}, A. and {Bailey}, S. and {Balaguera-Antol{\'\i}nez}, A. and {Ballester}, O. and {Baltay}, C. and {Bault}, A. and {Bautista}, J. and {Behera}, J. and {Beltran}, S.~F. and {BenZvi}, S. and {Beraldo e Silva}, L. and {Bermejo-Climent}, J.~R. and {Berti}, A. and {Besuner}, R. and {Beutler}, F. and {Bianchi}, D. and {Blake}, C. and {Blum}, R. and {Bolton}, A.~S. and {Brieden}, S. and {Brodzeller}, A. and {Brooks}, D. and {Brown}, Z. and {Buckley-Geer}, E. and {Burtin}, E. and {Cabayol-Garcia}, L. and {Cai}, Z. and {Canning}, R. and {Cardiel-Sas}, L. and {Carnero Rosell}, A. and {Castander}, F.~J. and {Cervantes-Cota}, J.~L. and {Chabanier}, S. and {Chaussidon}, E. and {Chaves-Montero}, J. and {Chen}, S. and {Chen}, X. and {Chuang}, C. and {Claybaugh}, T. and {Cole}, S. and {Cooper}, A.~P. and {Cuceu}, A. and {Davis}, T.~M. and {Dawson}, K. and {de Belsunce}, R. and {de la Cruz}, R. and {de la Macorra}, A. and {Della Costa}, J. and {de Mattia}, A. and {Demina}, R. and {Demirbozan}, U. and {DeRose}, J. and {Dey}, A. and {Dey}, B. and {Dhungana}, G. and {Ding}, J. and {Ding}, Z. and {Doel}, P. and {Doshi}, R. and {Douglass}, K. and {Edge}, A. and {Eftekharzadeh}, S. and {Eisenstein}, D.~J. and {Elliott}, A. and {Ereza}, J. and {Escoffier}, S. and {Fagrelius}, P. and {Fan}, X. and {Fanning}, K. and {Fawcett}, V.~A. and {Ferraro}, S. and {Flaugher}, B. and {Font-Ribera}, A. and {Forero-Romero}, J.~E. and {Forero-S{\'a}nchez}, D. and {Frenk}, C.~S. and {G{\"a}nsicke}, B.~T. and {Garc{\'\i}a}, L. {\'A}. and {Garc{\'\i}a-Bellido}, J. and {Garcia-Quintero}, C. and {Garrison}, L.~H. and {Gil-Mar{\'\i}n}, H. and {Golden-Marx}, J. and {Gontcho A Gontcho}, S. and {Gonzalez-Morales}, A.~X. and {Gonzalez-Perez}, V. and {Gordon}, C. and {Graur}, O. and {Green}, D. and {Gruen}, D. and {Guy}, J. and {Hadzhiyska}, B. and {Hahn}, C. and {Han}, J.~J. and {Hanif}, M.~M.~S. and {Herrera-Alcantar}, H.~K. and {Honscheid}, K. and {Hou}, J. and {Howlett}, C. and {Huterer}, D. and {Ir{\v{s}}i{\v{c}}}, V. and {Ishak}, M. and {Jacques}, A. and {Jana}, A. and {Jiang}, L. and {Jimenez}, J. and {Jing}, Y.~P. and {Joudaki}, S. and {Joyce}, R. and {Jullo}, E. and {Juneau}, S. and {Kara{\c{c}}ayl{\i}}, N.~G. and {Karim}, T. and {Kehoe}, R. and {Kent}, S. and {Khederlarian}, A. and {Kim}, S. and {Kirkby}, D. and {Kisner}, T. and {Kitaura}, F. and {Kizhuprakkat}, N. and {Kneib}, J. and {Koposov}, S.~E. and {Kov{\'a}cs}, A. and {Kremin}, A. and {Krolewski}, A. and {L'Huillier}, B. and {Lahav}, O. and {Lambert}, A. and {Lamman}, C. and {Lan}, T. -W. and {Landriau}, M. and {Lang}, D. and {Lange}, J.~U. and {Lasker}, J. and {Leauthaud}, A. and {Le Guillou}, L. and {Levi}, M.~E. and {Li}, T.~S. and {Linder}, E. and {Lyons}, A. and {Magneville}, C. and {Manera}, M. and {Manser}, C.~J. and {Margala}, D. and {Martini}, P. and {McDonald}, P. and {Medina}, G.~E. and {Medina-Varela}, L. and {Meisner}, A. and {Mena-Fern{\'a}ndez}, J. and {Meneses-Rizo}, J. and {Mezcua}, M. and {Miquel}, R. and {Montero-Camacho}, P. and {Moon}, J. and {Moore}, S. and {Moustakas}, J. and {Mueller}, E. and {Mundet}, J. and {Mu{\~n}oz-Guti{\'e}rrez}, A. and {Myers}, A.~D. and {Nadathur}, S. and {Napolitano}, L. and {Neveux}, R. and {Newman}, J.~A. and {Nie}, J. and {Nikutta}, R. and {Niz}, G. and {Norberg}, P. and {Noriega}, H.~E. and {Paillas}, E. and {Palanque-Delabrouille}, N. and {Palmese}, A. and {Pan}, Z. and {Parkinson}, D. and {Penmetsa}, S. and {Percival}, W.~J. and {P{\'e}rez-Fern{\'a}ndez}, A. and {P{\'e}rez-R{\`a}fols}, I. and {Pieri}, M. and {Poppett}, C. and {Porredon}, A. and {Pothier}, S. and {Prada}, F. and {Pucha}, R. and {Raichoor}, A. and {Ram{\'\i}rez-P{\'e}rez}, C. and {Ramirez-Solano}, S. and {Rashkovetskyi}, M. and {Ravoux}, C. and {Rocher}, A. and {Rockosi}, C. and {Ross}, A.~J. and {Rossi}, G. and {Ruggeri}, R. and {Ruhlmann-Kleider}, V. and {Sabiu}, C.~G. and {Said}, K. and {Saintonge}, A. and {Samushia}, L. and {Sanchez}, E. and {Saulder}, C. and {Schaan}, E. and {Schlafly}, E.~F. and {Schlegel}, D. and {Scholte}, D. and {Schubnell}, M. and {Seo}, H. and {Shafieloo}, A. and {Sharples}, R. and {Sheu}, W. and {Silber}, J. and {Sinigaglia}, F. and {Siudek}, M. and {Slepian}, Z. and {Smith}, A. and {Soumagnac}, M.~T. and {Sprayberry}, D. and {Stephey}, L. and {Su{\'a}rez-P{\'e}rez}, J. and {Sun}, Z. and {Tan}, T. and {Tarl{\'e}}, G. and {Tojeiro}, R. and {Ure{\~n}a-L{\'o}pez}, L.~A. and {Vaisakh}, R. and {Valcin}, D. and {Valdes}, F. and {Valluri}, M. and {Vargas-Maga{\~n}a}, M. and {Variu}, A. and {Verde}, L. and {Walther}, M. and {Wang}, B. and {Wang}, M.~S. and {Weaver}, B.~A. and {Weaverdyck}, N. and {Wechsler}, R.~H. and {White}, M. and {Xie}, Y. and {Yang}, J. and {Y{\`e}che}, C. and {Yu}, J. and {Yuan}, S. and {Zhang}, H. and {Zhang}, Z. and {Zhao}, C. and {Zheng}, Z. and {Zhou}, R. and {Zhou}, Z. and {Zou}, H. and {Zou}, S. and {Zu}, Y.},
        title = "{The Early Data Release of the Dark Energy Spectroscopic Instrument}",
      journal = {\aj},
         year = 2024,
        month = aug,
       volume = {168},
       number = {2},
          eid = {58},
        pages = {58},
          doi = {10.3847/1538-3881/ad3217},
archivePrefix = {arXiv},
       eprint = {2306.06308},
 primaryClass = {astro-ph.CO},
       adsurl = {https://ui.adsabs.harvard.edu/abs/2024AJ....168...58D}
}

@ARTICLE{DESI2024.I.DR1,
       author = {{DESI Collaboration} and {Abdul-Karim}, M  and {Adame}, A.~G. and {Aguado}, D. and {Aguilar}, J. and {Ahlen}, S. and {Alam}, S. and {Aldering}, G. and {Alexander}, D.~M. and {Alfarsy}, R. and {Allen}, L. and {Allende Prieto}, C. and {Alves}, O. and {Anand}, A. and {Andrade}, U. and {Armengaud}, E. and {Avila}, S. and {Aviles}, A. and {Awan}, H. and {Bailey}, S. and {Baleato Lizancos}, A. and {Ballester}, O. and {Bault}, A. and {Bautista}, J. and {BenZvi}, S. and {Beraldo e Silva}, L. and {Bermejo-Climent}, J.~R. and {Beutler}, F. and {Bianchi}, D. and {Blake}, C. and {Blum}, R. and {Bolton}, A.~S. and {Bonici}, M. and {Brieden}, S. and {Brodzeller}, A. and {Brooks}, D. and {Buckley-Geer}, E. and {Burtin}, E. and {Canning}, R. and {Carnero Rosell}, A. and {Carr}, A. and {Carrilho}, P. and {Casas}, L. and {Castander}, F.~J. and {Cereskaite}, R. and {Cervantes-Cota}, J.~L. and {Chaussidon}, E. and {Chaves-Montero}, J. and {Chen}, S. and {Chen}, X. and {Claybaugh}, T. and {Cole}, S. and {Cooper}, A.~P. and {Cousinou}, M. -C. and {Cuceu}, A. and {Davis}, T.~M. and {Dawson}, K.~S. and {de Belsunce}, R. and {de la Cruz}, R. and {de la Macorra}, A. and {de Mattia}, A. and {Deiosso}, N. and {Della Costa}, J. and {Demina}, R. and {Demirbozan}, U. and {DeRose}, J. and {Dey}, A. and {Dey}, B. and {Ding}, J. and {Ding}, Z. and {Doel}, P. and {Douglass}, K. and {Dowicz}, M. and {Ebina}, H. and {Edelstein}, J. and {Eisenstein}, D.~J. and {Elbers}, W. and {Emas}, N. and {Escoffier}, S. and {Fagrelius}, P. and {Fan}, X. and {Fanning}, K. and {Fawcett}, V.~A. and {Fern{\'a}ndez-Garc{\'\i}a}, E. and {Ferraro}, S. and {Findlay}, N. and {Font-Ribera}, A. and {Forero-Romero}, J.~E. and {Forero-S{\'a}nchez}, D. and {Frenk}, C.~S. and {G{\"a}nsicke}, B.~T. and {Galbany}, L. and {Garc{\'\i}a-Bellido}, J. and {Garcia-Quintero}, C. and {Garrison}, L.~H. and {Gazta{\~n}aga}, E. and {Gil-Mar{\'\i}n}, H. and {Gnedin}, O.~Y. and {Gontcho}, S. Gontcho A and {Gonzalez-Morales}, A.~X. and {Gonzalez-Perez}, V. and {Gordon}, C. and {Graur}, O. and {Green}, D. and {Gruen}, D. and {Gsponer}, R. and {Guandalin}, C. and {Gutierrez}, G. and {Guy}, J. and {Hahn}, C. and {Han}, J.~J. and {Han}, J. and {He}, S. and {Herrera-Alcantar}, H.~K. and {Honscheid}, K. and {Hou}, J. and {Howlett}, C. and {Huterer}, D. and {Ir{\v{s}}i{\v{c}}}, V. and {Ishak}, M. and {Jacques}, A. and {Jimenez}, J. and {Jing}, Y.~P. and {Joachimi}, B. and {Joudaki}, S. and {Joyce}, R. and {Jullo}, E. and {Juneau}, S. and {Kara{\c{c}}ayl{\i}}, N.~G. and {Karim}, T. and {Kehoe}, R. and {Kent}, S. and {Khederlarian}, A. and {Kirkby}, D. and {Kisner}, T. and {Kitaura}, F. -S. and {Kizhuprakkat}, N. and {Kong}, H. and {Koposov}, S.~E. and {Kremin}, A. and {Krolewski}, A. and {Lahav}, O. and {Lai}, Y. and {Lamman}, C. and {Lan}, T. -W. and {Landriau}, M. and {Lang}, D. and {Lange}, J.~U. and {Lasker}, J. and {Le Goff}, J.~M. and {Le Guillou}, L. and {Leauthaud}, A. and {Levi}, M.~E. and {Li}, S. and {Li}, T.~S. and {Lodha}, K. and {Lokken}, M. and {Luo}, Y. and {Magneville}, C. and {Manera}, M. and {Manser}, C.~J. and {Margala}, D. and {Martini}, P. and {Maus}, M. and {McCullough}, J. and {McDonald}, P. and {Medina}, G.~E. and {Medina-Varela}, L. and {Meisner}, A. and {Mena-Fern{\'a}ndez}, J. and {Menegas}, A. and {Mezcua}, M. and {Miquel}, R. and {Montero-Camacho}, P. and {Moon}, J. and {Moustakas}, J. and {Mu{\~n}oz-Guti{\'e}rrez}, A. and {Mu{\~n}oz-Santos}, D. and {Myers}, A.~D. and {Myles}, J. and {Nadathur}, S. and {Najita}, J. and {Napolitano}, L. and {Newman}, J.~A. and {Nikakhtar}, F. and {Nikutta}, R. and {Niz}, G. and {Noriega}, H.~E. and {Padmanabhan}, N. and {Paillas}, E. and {Palanque-Delabrouille}, N. and {Palmese}, A. and {Pan}, J. and {Pan}, Z. and {Parkinson}, D. and {Peacock}, J.~A.  and {Percival}, W.~J. and {P{\'e}rez-Fern{\'a}ndez}, A. and {P{\'e}rez-R{\`a}fols}, I. and {Peterson}, P.},
        title = "{Data Release 1 of the Dark Energy Spectroscopic Instrument}",
      journal = {arXiv e-prints},
         year = 2025,
        month = mar,
          eid = {arXiv:2503.14745},
        pages = {arXiv:2503.14745},
          doi = {10.48550/arXiv.2503.14745},
archivePrefix = {arXiv},
       eprint = {2503.14745},
 primaryClass = {astro-ph.CO},
       adsurl = {https://ui.adsabs.harvard.edu/abs/2025arXiv250314745D}
}

@ARTICLE{DESI2024.II.KP3,
       author = {{DESI Collaboration} and {Adame}, A.~G. and {Aguilar}, J. and {Ahlen}, S. and {Alam}, S. and {Alexander}, D.~M. and {Alvarez}, M. and {Alves}, O. and {Anand}, A. and {Andrade}, U. and {Armengaud}, E. and {Avila}, S. and {Aviles}, A. and {Awan}, H. and {Bailey}, S. and {Baltay}, C. and {Bault}, A. and {Behera}, J. and {BenZvi}, S. and {Beutler}, F. and {Bianchi}, D. and {Blake}, C. and {Blum}, R. and {Brieden}, S. and {Brodzeller}, A. and {Brooks}, D. and {Brown}, Z. and {Buckley-Geer}, E. and {Burtin}, E. and {Calderon}, R. and {Canning}, R. and {Carnero Rosell}, A. and {Cereskaite}, R. and {Cervantes-Cota}, J.~L. and {Chabanier}, S. and {Chaussidon}, E. and {Chaves-Montero}, J. and {Chen}, S. and {Chen}, X. and {Claybaugh}, T. and {Cole}, S. and {Cuceu}, A. and {Davis}, T.~M. and {Dawson}, K. and {de la Macorra}, A. and {de Mattia}, A. and {Deiosso}, N. and {Demina}, R. and {Dey}, A. and {Dey}, B. and {Ding}, Z. and {Doel}, P. and {Edelstein}, J. and {Eftekharzadeh}, S. and {Eisenstein}, D.~J. and {Elliott}, A. and {Fagrelius}, P. and {Fanning}, K. and {Ferraro}, S. and {Ereza}, J. and {Findlay}, N. and {Flaugher}, B. and {Font-Ribera}, A. and {Forero-S{\'a}nchez}, D. and {Forero-Romero}, J.~E. and {Frenk}, C.~S. and {Garcia-Quintero}, C. and {Gazta{\~n}aga}, E. and {Gil-Mar{\'\i}n}, H. and {Gontcho}, S. Gontcho A. and {Gonzalez-Morales}, A.~X. and {Gonzalez-Perez}, V. and {Gordon}, C. and {Green}, D. and {Gruen}, D. and {Gsponer}, R. and {Gutierrez}, G. and {Guy}, J. and {Hadzhiyska}, B. and {Hahn}, C. and {Hanif}, M.~M.~S. and {Herrera-Alcantar}, H.~K. and {Honscheid}, K. and {Hou}, J. and {Howlett}, C. and {Huterer}, D. and {Ir{\v{s}}i{\v{c}}}, V. and {Ishak}, M. and {Juneau}, S. and {Kara{\c{c}}ayl{\i}}, N.~G. and {Kehoe}, R. and {Kent}, S. and {Kirkby}, D. and {Kitaura}, F. -S. and {Kong}, H. and {Kremin}, A. and {Krolewski}, A. and {Lai}, Y. and {Lan}, T. -W. and {Landriau}, M. and {Lang}, D. and {Lasker}, J. and {Le Goff}, J.~M. and {Le Guillou}, L. and {Leauthaud}, A. and {Levi}, M.~E. and {Li}, T.~S. and {Lodha}, K. and {Magneville}, C. and {Manera}, M. and {Margala}, D. and {Martini}, P. and {Maus}, M. and {McDonald}, P. and {Medina-Varela}, L. and {Meisner}, A. and {Mena-Fern{\'a}ndez}, J. and {Miquel}, R. and {Moon}, J. and {Moore}, S. and {Moustakas}, J. and {Mudur}, N. and {Mueller}, E. and {Mu{\~n}oz-Guti{\'e}rrez}, A. and {Myers}, A.~D. and {Nadathur}, S. and {Napolitano}, L. and {Neveux}, R. and {Newman}, J.~A. and {Nguyen}, N.~M. and {Nie}, J. and {Niz}, G. and {Noriega}, H.~E. and {Padmanabhan}, N. and {Paillas}, E. and {Palanque-Delabrouille}, N. and {Pan}, J. and {Penmetsa}, S. and {Percival}, W.~J. and {Pieri}, M.~M. and {Pinon}, M. and {Poppett}, C. and {Porredon}, A. and {Prada}, F. and {P{\'e}rez-Fern{\'a}ndez}, A. and {P{\'e}rez-R{\`a}fols}, I. and {Rabinowitz}, D. and {Raichoor}, A. and {Ram{\'\i}rez-P{\'e}rez}, C. and {Ramirez-Solano}, S. and {Rashkovetskyi}, M. and {Ravoux}, C. and {Rezaie}, M. and {Rich}, J. and {Rocher}, A. and {Rockosi}, C. and {Roe}, N.~A. and {Rosado-Marin}, A. and {Ross}, A.~J. and {Rossi}, G. and {Ruggeri}, R. and {Ruhlmann-Kleider}, V. and {Samushia}, L. and {Sanchez}, E. and {Saulder}, C. and {Schlafly}, E.~F. and {Schlegel}, D. and {Scholte}, D. and {Schubnell}, M. and {Seo}, H. and {Sharples}, R. and {Silber}, J. and {Slosar}, A. and {Smith}, A. and {Sprayberry}, D. and {Tan}, T. and {Tarl{\'e}}, G. and {Trusov}, S. and {Vaisakh}, R. and {Valcin}, D. and {Valdes}, F. and {Vargas-Maga{\~n}a}, M. and {Verde}, L. and {Walther}, M. and {Wang}, B. and {Wang}, M.~S. and {Weaver}, B.~A. and {Weaverdyck}, N. and {Wechsler}, R.~H. and {Weinberg}, D.~H. and {White}, M. and {Wilson}, M.~J. and {Yu}, J. and {Yu}, Y. and {Yuan}, S. and {Y{\`e}che}, C. and {Zaborowski}, E.~A. and {Zarrouk}, P. and {Zhang}, H. and {Zhao}, C. and {Zhao}, R.},
        title = "{DESI 2024 II: sample definitions, characteristics, and two-point clustering statistics}",
      journal = {\jcap},
         year = 2025,
        month = jul,
       volume = {2025},
       number = {7},
          eid = {017},
        pages = {017},
          doi = {10.1088/1475-7516/2025/07/017},
archivePrefix = {arXiv},
       eprint = {2411.12020},
 primaryClass = {astro-ph.CO},
       adsurl = {https://ui.adsabs.harvard.edu/abs/2025JCAP...07..017A}
}

@ARTICLE{DESI2024.III.KP4,
       author = {{DESI Collaboration} and {Adame}, A.~G. and {Aguilar}, J. and {Ahlen}, S. and {Alam}, S. and {Alexander}, D.~M. and {Alvarez}, M. and {Alves}, O. and {Anand}, A. and {Andrade}, U. and {Armengaud}, E. and {Avila}, S. and {Aviles}, A. and {Awan}, H. and {Bailey}, S. and {Baltay}, C. and {Bault}, A. and {Behera}, J. and {BenZvi}, S. and {Beutler}, F. and {Bianchi}, D. and {Blake}, C. and {Blum}, R. and {Brieden}, S. and {Brodzeller}, A. and {Brooks}, D. and {Buckley-Geer}, E. and {Burtin}, E. and {Calderon}, R. and {Canning}, R. and {Carnero Rosell}, A. and {Cereskaite}, R. and {Cervantes-Cota}, J.~L. and {Chabanier}, S. and {Chaussidon}, E. and {Chaves-Montero}, J. and {Chen}, S. and {Chen}, X. and {Claybaugh}, T. and {Cole}, S. and {Cuceu}, A. and {Davis}, T.~M. and {Dawson}, K. and {de la Macorra}, A. and {de Mattia}, A. and {Deiosso}, N. and {Dey}, A. and {Dey}, B. and {Ding}, Z. and {Doel}, P. and {Edelstein}, J. and {Eftekharzadeh}, S. and {Eisenstein}, D.~J. and {Elliott}, A. and {Fagrelius}, P. and {Fanning}, K. and {Ferraro}, S. and {Ereza}, J. and {Findlay}, N. and {Flaugher}, B. and {Font-Ribera}, A. and {Forero-S{\'a}nchez}, D. and {Forero-Romero}, J.~E. and {Garcia-Quintero}, C. and {Gazta{\~n}aga}, E. and {Gil-Mar{\'\i}n}, H. and {Gontcho a Gontcho}, S. and {Gonzalez-Morales}, A.~X. and {Gonzalez-Perez}, V. and {Gordon}, C. and {Green}, D. and {Gruen}, D. and {Gsponer}, R. and {Gutierrez}, G. and {Guy}, J. and {Hadzhiyska}, B. and {Hahn}, C. and {Hanif}, M.~M.~S. and {Herrera-Alcantar}, H.~K. and {Honscheid}, K. and {Howlett}, C. and {Huterer}, D. and {Ir{\v{s}}i{\v{c}}}, V. and {Ishak}, M. and {Juneau}, S. and {Kara{\c{c}}ayl{\i}}, N.~G. and {Kehoe}, R. and {Kent}, S. and {Kirkby}, D. and {Kong}, H. and {Kremin}, A. and {Krolewski}, A. and {Lai}, Y. and {Lan}, T. -W. and {Landriau}, M. and {Lang}, D. and {Lasker}, J. and {Le Goff}, J.~M. and {Le Guillou}, L. and {Leauthaud}, A. and {Levi}, M.~E. and {Li}, T.~S. and {Linder}, E. and {Lodha}, K. and {Magneville}, C. and {Manera}, M. and {Margala}, D. and {Martini}, P. and {Maus}, M. and {McDonald}, P. and {Medina-Varela}, L. and {Meisner}, A. and {Mena-Fern{\'a}ndez}, J. and {Miquel}, R. and {Moon}, J. and {Moore}, S. and {Moustakas}, J. and {Mueller}, E. and {Mu{\~n}oz-Guti{\'e}rrez}, A. and {Myers}, A.~D. and {Nadathur}, S. and {Napolitano}, L. and {Neveux}, R. and {Newman}, J.~A. and {Nguyen}, N.~M. and {Nie}, J. and {Niz}, G. and {Noriega}, H.~E. and {Padmanabhan}, N. and {Paillas}, E. and {Palanque-Delabrouille}, N. and {Pan}, J. and {Penmetsa}, S. and {Percival}, W.~J. and {Pieri}, M.~M. and {Pinon}, M. and {Poppett}, C. and {Porredon}, A. and {Prada}, F. and {P{\'e}rez-Fern{\'a}ndez}, A. and {P{\'e}rez-R{\`a}fols}, I. and {Rabinowitz}, D. and {Raichoor}, A. and {Ram{\'\i}rez-P{\'e}rez}, C. and {Ramirez-Solano}, S. and {Rashkovetskyi}, M. and {Ravoux}, C. and {Rezaie}, M. and {Rich}, J. and {Rocher}, A. and {Rockosi}, C. and {Roe}, N.~A. and {Rosado-Marin}, A. and {Ross}, A.~J. and {Rossi}, G. and {Ruggeri}, R. and {Ruhlmann-Kleider}, V. and {Samushia}, L. and {Sanchez}, E. and {Saulder}, C. and {Schlafly}, E.~F. and {Schlegel}, D. and {Schubnell}, M. and {Seo}, H. and {Sharples}, R. and {Silber}, J. and {Slosar}, A. and {Smith}, A. and {Sprayberry}, D. and {Swanson}, J. and {Tan}, T. and {Tarl{\'e}}, G. and {Trusov}, S. and {Vaisakh}, R. and {Valcin}, D. and {Valdes}, F. and {Vargas-Maga{\~n}a}, M. and {Verde}, L. and {Walther}, M. and {Wang}, B. and {Wang}, M.~S. and {Weaver}, B.~A. and {Weaverdyck}, N. and {Wechsler}, R.~H. and {Weinberg}, D.~H. and {White}, M. and {Wilson}, M.~J. and {Yu}, J. and {Yu}, Y. and {Yuan}, S. and {Y{\`e}che}, C. and {Zaborowski}, E.~A. and {Zarrouk}, P. and {Zhang}, H. and {Zhao}, C. and {Zhao}, R. and {Zhou}, R. and {Zou}, H. and {DESI Collaboration}},
        title = "{DESI 2024 III: baryon acoustic oscillations from galaxies and quasars}",
      journal = {\jcap},
         year = 2025,
        month = apr,
       volume = {2025},
       number = {4},
          eid = {012},
        pages = {012},
          doi = {10.1088/1475-7516/2025/04/012},
archivePrefix = {arXiv},
       eprint = {2404.03000},
 primaryClass = {astro-ph.CO},
       adsurl = {https://ui.adsabs.harvard.edu/abs/2025JCAP...04..012A}
}

@ARTICLE{DESI2024.IV.KP6,
       author = {{DESI Collaboration} and {Adame}, A.~G. and {Aguilar}, J. and {Ahlen}, S. and {Alam}, S. and {Alexander}, D.~M. and {Alvarez}, M. and {Alves}, O. and {Anand}, A. and {Andrade}, U. and {Armengaud}, E. and {Avila}, S. and {Aviles}, A. and {Awan}, H. and {Bailey}, S. and {Baltay}, C. and {Bault}, A. and {Bautista}, J. and {Behera}, J. and {BenZvi}, S. and {Beutler}, F. and {Bianchi}, D. and {Blake}, C. and {Blum}, R. and {Brieden}, S. and {Brodzeller}, A. and {Brooks}, D. and {Buckley-Geer}, E. and {Burtin}, E. and {Calderon}, R. and {Canning}, R. and {Carnero Rosell}, A. and {Cereskaite}, R. and {Cervantes-Cota}, J.~L. and {Chabanier}, S. and {Chaussidon}, E. and {Chaves-Montero}, J. and {Chen}, S. and {Chen}, X. and {Claybaugh}, T. and {Cole}, S. and {Cuceu}, A. and {Davis}, T.~M. and {Dawson}, K. and {de la Cruz}, R. and {de la Macorra}, A. and {de Mattia}, A. and {Deiosso}, N. and {Dey}, A. and {Dey}, B. and {Ding}, J. and {Ding}, Z. and {Doel}, P. and {Edelstein}, J. and {Eftekharzadeh}, S. and {Eisenstein}, D.~J. and {Elliott}, A. and {Fagrelius}, P. and {Fanning}, K. and {Ferraro}, S. and {Ereza}, J. and {Findlay}, N. and {Flaugher}, B. and {Font-Ribera}, A. and {Forero-S{\'a}nchez}, D. and {Forero-Romero}, J.~E. and {Garcia-Quintero}, C. and {Gazta{\~n}aga}, E. and {Gil-Mar{\'\i}n}, H. and {Gontcho}, S. Gontcho A. and {Gonzalez-Morales}, A.~X. and {Gonzalez-Perez}, V. and {Gordon}, C. and {Green}, D. and {Gruen}, D. and {Gsponer}, R. and {Gutierrez}, G. and {Guy}, J. and {Hadzhiyska}, B. and {Hahn}, C. and {Hanif}, M.~M.~S. and {Herrera-Alcantar}, H.~K. and {Honscheid}, K. and {Howlett}, C. and {Huterer}, D. and {Ir{\v{s}}i{\v{c}}}, V. and {Ishak}, M. and {Juneau}, S. and {Kara{\c{c}}ayl{\i}}, N.~G. and {Kehoe}, R. and {Kent}, S. and {Kirkby}, D. and {Kremin}, A. and {Krolewski}, A. and {Lai}, Y. and {Lan}, T. -W. and {Landriau}, M. and {Lang}, D. and {Lasker}, J. and {Le Goff}, J.~M. and {Le Guillou}, L. and {Leauthaud}, A. and {Levi}, M.~E. and {Li}, T.~S. and {Linder}, E. and {Lodha}, K. and {Magneville}, C. and {Manera}, M. and {Margala}, D. and {Martini}, P. and {Maus}, M. and {McDonald}, P. and {Medina-Varela}, L. and {Meisner}, A. and {Mena-Fern{\'a}ndez}, J. and {Miquel}, R. and {Moon}, J. and {Moore}, S. and {Moustakas}, J. and {Mueller}, E. and {Mu{\~n}oz-Guti{\'e}rrez}, A. and {Myers}, A.~D. and {Nadathur}, S. and {Napolitano}, L. and {Neveux}, R. and {Newman}, J.~A. and {Nguyen}, N.~M. and {Nie}, J. and {Niz}, G. and {Noriega}, H.~E. and {Padmanabhan}, N. and {Paillas}, E. and {Palanque-Delabrouille}, N. and {Pan}, J. and {Penmetsa}, S. and {Percival}, W.~J. and {Pieri}, M.~M. and {Pinon}, M. and {Poppett}, C. and {Porredon}, A. and {Prada}, F. and {P{\'e}rez-Fern{\'a}ndez}, A. and {P{\'e}rez-R{\`a}fols}, I. and {Rabinowitz}, D. and {Raichoor}, A. and {Ram{\'\i}rez-P{\'e}rez}, C. and {Ramirez-Solano}, S. and {Rashkovetskyi}, M. and {Ravoux}, C. and {Rezaie}, M. and {Rich}, J. and {Rocher}, A. and {Rockosi}, C. and {Roe}, N.~A. and {Rosado-Marin}, A. and {Ross}, A.~J. and {Rossi}, G. and {Ruggeri}, R. and {Ruhlmann-Kleider}, V. and {Samushia}, L. and {Sanchez}, E. and {Saulder}, C. and {Schlafly}, E.~F. and {Schlegel}, D. and {Schubnell}, M. and {Seo}, H. and {Sharples}, R. and {Silber}, J. and {Sinigaglia}, F. and {Slosar}, A. and {Smith}, A. and {Sprayberry}, D. and {Tan}, T. and {Tarl{\'e}}, G. and {Trusov}, S. and {Vaisakh}, R. and {Valcin}, D. and {Valdes}, F. and {Vargas-Maga{\~n}a}, M. and {Verde}, L. and {Walther}, M. and {Wang}, B. and {Wang}, M.~S. and {Weaver}, B.~A. and {Weaverdyck}, N. and {Wechsler}, R.~H. and {Weinberg}, D.~H. and {White}, M. and {Yu}, J. and {Yu}, Y. and {Yuan}, S. and {Y{\`e}che}, C. and {Zaborowski}, E.~A. and {Zarrouk}, P. and {Zhang}, H. and {Zhao}, C. and {Zhao}, R. and {Zhou}, R. and {Zou}, H. and {DESI Collaboration}},
        title = "{DESI 2024 IV: Baryon Acoustic Oscillations from the Lyman alpha forest}",
      journal = {\jcap},
         year = 2025,
        month = jan,
       volume = {2025},
       number = {1},
          eid = {124},
        pages = {124},
          doi = {10.1088/1475-7516/2025/01/124},
archivePrefix = {arXiv},
       eprint = {2404.03001},
 primaryClass = {astro-ph.CO},
       adsurl = {https://ui.adsabs.harvard.edu/abs/2025JCAP...01..124A}
}

@ARTICLE{DESI2024.V.KP5,
       author = {{DESI Collaboration} and {Adame}, A.~G. and {Aguilar}, J. and {Ahlen}, S. and {Alam}, S. and {Alexander}, D.~M. and {Alvarez}, M. and {Alves}, O. and {Anand}, A. and {Andrade}, U. and {Armengaud}, E. and {Avila}, S. and {Aviles}, A. and {Awan}, H. and {Bailey}, S. and {Baltay}, C. and {Bault}, A. and {Behera}, J. and {BenZvi}, S. and {Beutler}, F. and {Bianchi}, D. and {Blake}, C. and {Blum}, R. and {Brieden}, S. and {Brodzeller}, A. and {Brooks}, D. and {Buckley-Geer}, E. and {Burtin}, E. and {Calderon}, R. and {Canning}, R. and {Carnero Rosell}, A. and {Cereskaite}, R. and {Cervantes-Cota}, J.~L. and {Chabanier}, S. and {Chaussidon}, E. and {Chaves-Montero}, J. and {Chen}, S. and {Chen}, X. and {Claybaugh}, T. and {Cole}, S. and {Cuceu}, A. and {Davis}, T.~M. and {Dawson}, K. and {de la Macorra}, A. and {de Mattia}, A. and {Deiosso}, N. and {Dey}, A. and {Dey}, B. and {Ding}, Z. and {Doel}, P. and {Edelstein}, J. and {Eftekharzadeh}, S. and {Eisenstein}, D.~J. and {Elliott}, A. and {Fagrelius}, P. and {Fanning}, K. and {Ferraro}, S. and {Ereza}, J. and {Findlay}, N. and {Flaugher}, B. and {Font-Ribera}, A. and {Forero-S{\'a}nchez}, D. and {Forero-Romero}, J.~E. and {Garcia-Quintero}, C. and {Garrison}, L.~H. and {Gazta{\~n}aga}, E. and {Gil-Mar{\'\i}n}, H. and {Gontcho}, S. Gontcho A and {Gonzalez-Morales}, A.~X. and {Gonzalez-Perez}, V. and {Gordon}, C. and {Green}, D. and {Gruen}, D. and {Gsponer}, R. and {Gutierrez}, G. and {Guy}, J. and {Hadzhiyska}, B. and {Hahn}, C. and {Hanif}, M.~M. S and {Herrera-Alcantar}, H.~K. and {Honscheid}, K. and {Howlett}, C. and {Huterer}, D. and {Ir{\v{s}}i{\v{c}}}, V. and {Ishak}, M. and {Juneau}, S. and {Kara{\c{c}}ayl{\i}}, N.~G. and {Kehoe}, R. and {Kent}, S. and {Kirkby}, D. and {Kong}, H. and {Koposov}, S.~E. and {Kremin}, A. and {Krolewski}, A. and {Lai}, Y. and {Lan}, T. -W. and {Landriau}, M. and {Lang}, D. and {Lasker}, J. and {Le Goff}, J.~M. and {Le Guillou}, L. and {Leauthaud}, A. and {Levi}, M.~E. and {Li}, T.~S. and {Lodha}, K. and {Magneville}, C. and {Manera}, M. and {Margala}, D. and {Martini}, P. and {Maus}, M. and {McDonald}, P. and {Medina-Varela}, L. and {Meisner}, A. and {Mena-Fern{\'a}ndez}, J. and {Miquel}, R. and {Moon}, J. and {Moore}, S. and {Moustakas}, J. and {Mueller}, E. and {Mu{\~n}oz-Guti{\'e}rrez}, A. and {Myers}, A.~D. and {Nadathur}, S. and {Napolitano}, L. and {Neveux}, R. and {Newman}, J.~A. and {Nguyen}, N.~M. and {Nie}, J. and {Niz}, G. and {Noriega}, H.~E. and {Padmanabhan}, N. and {Paillas}, E. and {Palanque-Delabrouille}, N. and {Pan}, J. and {Penmetsa}, S. and {Percival}, W.~J. and {Pieri}, M.~M. and {Pinon}, M. and {Poppett}, C. and {Porredon}, A. and {Prada}, F. and {P{\'e}rez-Fern{\'a}ndez}, A. and {P{\'e}rez-R{\`a}fols}, I. and {Rabinowitz}, D. and {Raichoor}, A. and {Ram{\'\i}rez-P{\'e}rez}, C. and {Ramirez-Solano}, S. and {Rashkovetskyi}, M. and {Ravoux}, C. and {Rezaie}, M. and {Rich}, J. and {Rocher}, A. and {Rockosi}, C. and {Rodr{\'\i}guez-Mart{\'\i}nez}, F. and {Roe}, N.~A. and {Rosado-Marin}, A. and {Ross}, A.~J. and {Rossi}, G. and {Ruggeri}, R. and {Ruhlmann-Kleider}, V. and {Samushia}, L. and {Sanchez}, E. and {Saulder}, C. and {Schlafly}, E.~F. and {Schlegel}, D. and {Schubnell}, M. and {Seo}, H. and {Sharples}, R. and {Silber}, J. and {Slosar}, A. and {Smith}, A. and {Sprayberry}, D. and {Tan}, T. and {Tarl{\'e}}, G. and {Trusov}, S. and {Vaisakh}, R. and {Valcin}, D. and {Valdes}, F. and {Vargas-Maga{\~n}a}, M. and {Verde}, L. and {Walther}, M. and {Wang}, B. and {Wang}, M.~S. and {Weaver}, B.~A. and {Weaverdyck}, N. and {Wechsler}, R.~H. and {Weinberg}, D.~H. and {White}, M. and {Wilson}, M.~J. and {Yu}, J. and {Yu}, Y. and {Yuan}, S. and {Y{\`e}che}, C. and {Zaborowski}, E.~A. and {Zarrouk}, P. and {Zhang}, H. and {Zhao}, C. and {Zhao}, R. and {Zhou}, R. and {Zou}, H.},
        title = "{DESI 2024 V: Full-Shape Galaxy Clustering from Galaxies and Quasars}",
      journal = {arXiv e-prints},
         year = 2024,
        month = nov,
          eid = {arXiv:2411.12021},
        pages = {arXiv:2411.12021},
          doi = {10.48550/arXiv.2411.12021},
archivePrefix = {arXiv},
       eprint = {2411.12021},
 primaryClass = {astro-ph.CO},
       adsurl = {https://ui.adsabs.harvard.edu/abs/2024arXiv241112021D}
}

@ARTICLE{KP3s2-Rosado,
       author = {{Rosado-Mar{\'\i}n}, A.~J. and {Ross}, A.~J. and {Seo}, H. and {Rezaie}, M. and {Kong}, H. and {de Mattia}, A. and {Zhou}, R. and {Ahlen}, S. and {Bianchi}, D. and {Brooks}, D. and {Claybaugh}, T. and {de la Macorra}, A. and {Doel}, P. and {Fanning}, K. and {Ferraro}, S. and {Gontcho}, S. Gontcho A and {Gutierrez}, G. and {Hahn}, C. and {Juneau}, S. and {Kehoe}, R. and {Kremin}, A. and {Meisner}, A. and {Miquel}, R. and {Moustakas}, J. and {Newman}, J.~A. and {Palanque-Delabrouille}, N. and {Percival}, W.~J. and {Prada}, F. and {P{\'e}rez-R{\`a}fols}, I. and {Rossi}, G. and {Sanchez}, E. and {Schlegel}, D. and {Schubnell}, M. and {Sprayberry}, D. and {Vargas-Maga{\~n}a}, M. and {Weaver}, B.~A. and {Zou}, H. and {Ruggeri}, R. and {Krolewski}, A. and {Yu}, J. and {Raichoor}, A. and {Hanif}, M.~M. S},
        title = "{Mitigating Imaging Systematics for DESI 2024 Emission Line Galaxies and Beyond}",
      journal = {arXiv e-prints},
         year = 2024,
        month = nov,
          eid = {arXiv:2411.12024},
        pages = {arXiv:2411.12024},
          doi = {10.48550/arXiv.2411.12024},
archivePrefix = {arXiv},
       eprint = {2411.12024},
 primaryClass = {astro-ph.CO},
       adsurl = {https://ui.adsabs.harvard.edu/abs/2024arXiv241112024R}
}

@ARTICLE{KP3s5-Pinon,
       author = {{Pinon}, M. and {de Mattia}, A. and {McDonald}, P. and {Burtin}, E. and {Ruhlmann-Kleider}, V. and {White}, M. and {Bianchi}, D. and {Ross}, A.~J. and {Aguilar}, J. and {Ahlen}, S. and {Brooks}, D. and {Cahn}, R.~N. and {Chaussidon}, E. and {Claybaugh}, T. and {Cole}, S. and {de la Macorra}, A. and {Dey}, B. and {Doel}, P. and {Fanning}, K. and {Forero-Romero}, J.~E. and {Gazta{\~n}aga}, E. and {Gontcho A Gontcho}, S. and {Howlett}, C. and {Kirkby}, D. and {Kisner}, T. and {Kremin}, A. and {Lambert}, A. and {Landriau}, M. and {Lasker}, J. and {Le Guillou}, L. and {Levi}, M.~E. and {Manera}, M. and {Martini}, P. and {Meisner}, A. and {Miquel}, R. and {Moustakas}, J. and {Myers}, A.~D. and {Niz}, G. and {Palanque-Delabrouille}, N. and {Percival}, W.~J. and {Poppett}, C. and {Rossi}, G. and {Sanchez}, E. and {Schlegel}, D. and {Schubnell}, M. and {Seo}, H. and {Sprayberry}, D. and {Tarl{\'e}}, G. and {Vargas-Maga{\~n}a}, M. and {Weaver}, B.~A. and {Zarrouk}, P. and {Zhou}, R. and {Zou}, H.},
        title = "{Mitigation of DESI fiber assignment incompleteness effect on two-point clustering with small angular scale truncated estimators}",
      journal = {\jcap},
         year = 2025,
        month = jan,
       volume = {2025},
       number = {1},
          eid = {131},
        pages = {131},
          doi = {10.1088/1475-7516/2025/01/131},
archivePrefix = {arXiv},
       eprint = {2406.04804},
 primaryClass = {astro-ph.CO},
       adsurl = {https://ui.adsabs.harvard.edu/abs/2025JCAP...01..131P}
}

@ARTICLE{KP3s7-Lasker,
       author = {{Lasker}, J. and {Carnero Rosell}, A. and {Myers}, A.~D. and {Ross}, A.~J. and {Bianchi}, D. and {Hanif}, M.~M.~S. and {Kehoe}, R. and {de Mattia}, A. and {Napolitano}, L. and {Percival}, W.~J. and {Staten}, R. and {Aguilar}, J. and {Ahlen}, S. and {Bigwood}, L. and {Brooks}, D. and {Claybaugh}, T. and {Cole}, S. and {de la Macorra}, A. and {Ding}, Z. and {Doel}, P. and {Fanning}, K. and {Forero-Romero}, J.~E. and {Gazta{\~n}aga}, E. and {Gontcho A Gontcho}, S. and {Gutierrez}, G. and {Honscheid}, K. and {Howlett}, C. and {Juneau}, S. and {Kremin}, A. and {Landriau}, M. and {Le Guillou}, L. and {Levi}, M.~E. and {Manera}, M. and {Meisner}, A. and {Miquel}, R. and {Moustakas}, J. and {Mueller}, E. and {Nie}, J. and {Niz}, G. and {Oh}, M. and {Palanque-Delabrouille}, N. and {Poppett}, C. and {Prada}, F. and {Rezaie}, M. and {Rossi}, G. and {Sanchez}, E. and {Schlegel}, D. and {Schubnell}, M. and {Seo}, H. and {Sprayberry}, D. and {Tarl{\'e}}, G. and {Vargas-Maga{\~n}a}, M. and {Weaver}, B.~A. and {Wilson}, Michael J. and {Zheng}, Y. and {DESI Collaboration}},
        title = "{Production of alternate realizations of DESI fiber assignment for unbiased clustering measurement in data and simulations}",
      journal = {\jcap},
         year = 2025,
        month = jan,
       volume = {2025},
       number = {1},
          eid = {127},
        pages = {127},
          doi = {10.1088/1475-7516/2025/01/127},
archivePrefix = {arXiv},
       eprint = {2404.03006},
 primaryClass = {astro-ph.CO},
       adsurl = {https://ui.adsabs.harvard.edu/abs/2025JCAP...01..127L}
}

@ARTICLE{KP3s11-Sikandar,
	title = "{Fast Fiber Assign: Emulating fiber assignment effects for realistic DESI catalogs}",
         author = {{M.~M.~S~Hanif et al.}},
            year = 2025,
        journal = {in preparation}
}

@ARTICLE{KP3s15-Ross,
       author = {{Ross}, A.~J. and {Aguilar}, J. and {Ahlen}, S. and {Alam}, S. and {Anand}, A. and {Bailey}, S. and {Bianchi}, D. and {Brieden}, S. and {Brooks}, D. and {Burtin}, E. and {Carnero Rosell}, A. and {Chaussidon}, E. and {Claybaugh}, T. and {Cole}, S. and {Dawson}, K. and {de la Macorra}, A. and {de Mattia}, A. and {Dey}, A. and {Dey}, B. and {Doel}, P. and {Fanning}, K. and {Ferraro}, S. and {Ereza}, J. and {Font-Ribera}, A. and {Forero-Romero}, J.~E. and {Gazta{\~n}aga}, E. and {Gil-Mar{\'\i}n}, H. and {Gontcho A Gontcho}, S. and {Gonzalez-Morales}, A.~X. and {Guy}, J. and {Hahn}, C. and {Heydenreich}, S. and {Honscheid}, K. and {Howlett}, C. and {Ishak}, M. and {Karim}, T. and {Kirkby}, D. and {Kisner}, T. and {Kong}, H. and {Kremin}, A. and {Krolewski}, A. and {Lambert}, A. and {Landriau}, M. and {Lasker}, J. and {Guillou}, L.~L. and {Levi}, M.~E. and {Manera}, M. and {Martini}, P. and {McDonald}, P. and {Meisner}, A. and {Miquel}, R. and {Moon}, J. and {Moustakas}, J. and {Mu{\~n}oz-Guti{\'e}rrez}, A. and {Myers}, A.~D. and {Nadathur}, S. and {Napolitano}, L. and {Newman}, J.~A. and {Nie}, J. and {Niz}, G. and {Palanque-Delabrouille}, N. and {Percival}, W.~J. and {Poppett}, C. and {Prada}, F. and {Raichoor}, A. and {Ravoux}, C. and {Rezaie}, M. and {Rosado-Marin}, A. and {Rossi}, G. and {Samushia}, L. and {Sanchez}, E. and {Schlafly}, E.~F. and {Schlegel}, D. and {Seo}, H. and {Smith}, A. and {Sprayberry}, D. and {Tarl{\'e}}, G. and {Valcin}, D. and {Vargas-Maga{\~n}a}, M. and {Weaver}, B.~A. and {Wilson}, M.~J. and {Yu}, J. and {Zarrouk}, P. and {Zhao}, C. and {Zhou}, R. and {Zou}, H.},
        title = "{The construction of large-scale structure catalogs for the Dark Energy Spectroscopic Instrument}",
      journal = {\jcap},
         year = 2025,
        month = jan,
       volume = {2025},
       number = {1},
          eid = {125},
        pages = {125},
          doi = {10.1088/1475-7516/2025/01/125},
archivePrefix = {arXiv},
       eprint = {2405.16593},
 primaryClass = {astro-ph.CO},
       adsurl = {https://ui.adsabs.harvard.edu/abs/2025JCAP...01..125R}
}

@ARTICLE{2023MNRAS.524.3894R,
       author = {{Rashkovetskyi}, Michael and {Eisenstein}, Daniel J. and {Aguilar}, Jessica Nicole and {Brooks}, David and {Claybaugh}, Todd and {Cole}, Shaun and {Dawson}, Kyle and {de la Macorra}, Axel and {Doel}, Peter and {Fanning}, Kevin and {Font-Ribera}, Andreu and {Forero-Romero}, Jaime E. and {Gontcho A Gontcho}, Satya and {Hahn}, ChangHoon and {Honscheid}, Klaus and {Kehoe}, Robert and {Kisner}, Theodore and {Landriau}, Martin and {Levi}, Michael and {Manera}, Marc and {Miquel}, Ramon and {Moon}, Jeongin and {Nadathur}, Seshadri and {Nie}, Jundan and {Poppett}, Claire and {Ross}, Ashley J. and {Rossi}, Graziano and {Sanchez}, Eusebio and {Saulder}, Christoph and {Schubnell}, Michael and {Seo}, Hee-Jong and {Tarle}, Gregory and {Valcin}, David and {Weaver}, Benjamin Alan and {Zhao}, Cheng and {Zhou}, Zhimin and {Zou}, Hu},
        title = "{Validation of semi-analytical, semi-empirical covariance matrices for two-point correlation function for early DESI data}",
      journal = {\mnras},
         year = 2023,
        month = sep,
       volume = {524},
       number = {3},
        pages = {3894-3911},
          doi = {10.1093/mnras/stad2078},
archivePrefix = {arXiv},
       eprint = {2306.06320},
 primaryClass = {astro-ph.CO},
       adsurl = {https://ui.adsabs.harvard.edu/abs/2023MNRAS.524.3894R}
}

@ARTICLE{KP4s2-Chen,
       author = {{Chen}, S. -F. and {Howlett}, C. and {White}, M. and {McDonald}, P. and {Ross}, A.~J. and {Seo}, H. -J. and {Padmanabhan}, N. and {Aguilar}, J. and {Ahlen}, S. and {Alam}, S. and {Alves}, O. and {Andrade}, U. and {Blum}, R. and {Brooks}, D. and {Chen}, X. and {Cole}, S. and {Dawson}, K. and {de la Macorra}, A. and {Dey}, A. and {Ding}, Z. and {Doel}, P. and {Ferraro}, S. and {Font-Ribera}, A. and {Forero-S{\'a}nchez}, D. and {Forero-Romero}, J.~E. and {Garcia-Quintero}, C. and {Gazta{\~n}aga}, E. and {Gontcho}, S.~G.~A. and {Hanif}, M.~M.~S. and {Honscheid}, K. and {Kisner}, T. and {Kremin}, A. and {Lambert}, A. and {Landriau}, M. and {Levi}, M.~E. and {Manera}, M. and {Meisner}, A. and {Mena-Fern{\'a}ndez}, J. and {Miquel}, R. and {Munoz-Gutierrez}, A. and {Paillas}, E. and {Palanque-Delabrouille}, N. and {Percival}, W.~J. and {P{\'e}rez-Fern{\'a}ndez}, A. and {Prada}, F. and {Rashkovetskyi}, M. and {Rezaie}, M. and {Rosado-Marin}, A. and {Rossi}, G. and {Ruggeri}, R. and {Sanchez}, E. and {Schlegel}, D. and {Silber}, J. and {Tarl{\'e}}, G. and {Vargas-Maga{\~n}a}, M. and {Weaver}, B.~A. and {Yu}, J. and {Yuan}, S. and {Zhou}, R. and {Zhou}, Z.},
        title = "{Baryon acoustic oscillation theory and modelling systematics for the DESI 2024 results}",
      journal = {\mnras},
         year = 2024,
        month = oct,
       volume = {534},
       number = {1},
        pages = {544-574},
          doi = {10.1093/mnras/stae2090},
archivePrefix = {arXiv},
       eprint = {2402.14070},
 primaryClass = {astro-ph.CO},
       adsurl = {https://ui.adsabs.harvard.edu/abs/2024MNRAS.534..544C}
}

@ARTICLE{KP4s3-Chen,
       author = {{Chen}, X. and {Ding}, Z. and {Paillas}, E. and {Nadathur}, S. and {Seo}, H. and {Chen}, S. and {Padmanabhan}, N. and {White}, M. and {de Mattia}, A. and {McDonald}, P. and {Ross}, A.~J. and {Variu}, A. and {Carnero Rosell}, A. and {Hadzhiyska}, B. and {Hanif}, M.~M. S and {Forero-S{\'a}nchez}, D. and {Ahlen}, S. and {Alves}, O. and {Andrade}, U. and {BenZvi}, S. and {Bianchi}, D. and {Brooks}, D. and {Chaussidon}, E. and {Claybaugh}, T. and {de la Macorra}, A. and {Dey}, Biprateep and {Fanning}, K. and {Ferraro}, S. and {Font-Ribera}, A. and {Forero-Romero}, J.~E. and {Garcia-Quintero}, C. and {Gazta{\~n}aga}, E. and {Gontcho}, S. Gontcho A and {Gutierrez}, G. and {Hahn}, C. and {Honscheid}, K. and {Juneau}, S. and {Kehoe}, R. and {Kirkby}, D. and {Kisner}, T. and {Kremin}, A. and {Levi}, M.~E. and {Meisner}, A. and {Mena-Fern{\'a}ndez}, J. and {Miquel}, R. and {Moustakas}, J. and {Mu{\~n}oz-Guti{\'e}rrez}, A. and {Nikakhtar}, F. and {Palanque-Delabrouille}, N. and {Percival}, W.~J. and {Prada}, F. and {P{\'e}rez-R{\`a}fols}, I. and {Rashkovetskyi}, M. and {Rossi}, G. and {Ruggeri}, R. and {Sanchez}, E. and {Saulder}, C. and {Schlegel}, D. and {Schubnell}, M. and {Smith}, A. and {Sprayberry}, D. and {Tarl{\'e}}, G. and {Valcin}, D. and {Vargas-Maga{\~n}a}, M. and {Weaver}, B.~A. and {Yuan}, S. and {Zhou}, R.},
        title = "{Extensive analysis of reconstruction algorithms for DESI 2024 baryon acoustic oscillations}",
      journal = {arXiv e-prints},
         year = 2024,
        month = nov,
          eid = {arXiv:2411.19738},
        pages = {arXiv:2411.19738},
          doi = {10.48550/arXiv.2411.19738},
archivePrefix = {arXiv},
       eprint = {2411.19738},
 primaryClass = {astro-ph.CO},
       adsurl = {https://ui.adsabs.harvard.edu/abs/2024arXiv241119738C}
}

@ARTICLE{KP4s4-Paillas,
       author = {{Paillas}, E. and {Ding}, Z. and {Chen}, X. and {Seo}, H. and {Padmanabhan}, N. and {de Mattia}, A. and {Ross}, A.~J. and {Nadathur}, S. and {Howlett}, C. and {Aguilar}, J. and {Ahlen}, S. and {Alves}, O. and {Andrade}, U. and {Brooks}, D. and {Buckley-Geer}, E. and {Burtin}, E. and {Chen}, S. and {Claybaugh}, T. and {Cole}, S. and {Dawson}, K. and {de la Macorra}, A. and {Dey}, Arjun and {Doel}, P. and {Fanning}, K. and {Ferraro}, S. and {Forero-Romero}, J.~E. and {Garcia-Quintero}, C. and {Gazta{\~n}aga}, E. and {Gil-Mar{\'\i}n}, H. and {Gontcho}, S. Gontcho A. and {Gutierrez}, G. and {Hahn}, C. and {Hanif}, M.~M.~S. and {Honscheid}, K. and {Ishak}, M. and {Kehoe}, R. and {Kremin}, A. and {Landriau}, M. and {Le Guillou}, L. and {Levi}, M.~E. and {Manera}, M. and {Martini}, P. and {Medina-Varela}, L. and {Meisner}, A. and {Mena-Fern{\'a}ndez}, J. and {Miquel}, R. and {Moustakas}, J. and {Mueller}, E. and {Mu{\~n}oz-Guti{\'e}rrez}, A. and {Myers}, A.~D. and {Newman}, J.~A. and {Nie}, J. and {Niz}, G. and {Palanque-Delabrouille}, N. and {Percival}, W.~J. and {Poppett}, C. and {Prada}, F. and {P{\'e}rez-Fern{\'a}ndez}, A. and {Rashkovetskyi}, M. and {Rezaie}, M. and {Rosado-Marin}, A. and {Rossi}, G. and {Ruggeri}, R. and {Sanchez}, E. and {Saulder}, C. and {Schlafly}, E.~F. and {Schlegel}, D. and {Schubnell}, M. and {Sprayberry}, D. and {Tarl{\'e}}, G. and {Valcin}, D. and {Vargas-Maga{\~n}a}, M. and {Yu}, J. and {Yuan}, S. and {Zhou}, R. and {Zou}, H.},
        title = "{Optimal reconstruction of baryon acoustic oscillations for DESI 2024}",
      journal = {\jcap},
         year = 2025,
        month = jan,
       volume = {2025},
       number = {1},
          eid = {142},
        pages = {142},
          doi = {10.1088/1475-7516/2025/01/142},
archivePrefix = {arXiv},
       eprint = {2404.03005},
 primaryClass = {astro-ph.CO},
       adsurl = {https://ui.adsabs.harvard.edu/abs/2025JCAP...01..142P}
}

@ARTICLE{KP4s6-Forero-Sanchez,
       author = {{Forero-S{\'a}nchez}, Daniel and {Rashkovetskyi}, Michael and {Alves}, Ot{\'a}vio and {de Mattia}, Arnaud and {Nadathur}, Seshadri and {Zarrouk}, Pauline and {Gil-Mar{\'\i}n}, H{\'e}ctor and {Ding}, Zhejie and {Yu}, Jiaxi and {Andrade}, Uendert and {Chen}, Xinyi and {Garcia-Quintero}, Cristhian and {Mena-Fern{\'a}ndez}, Juan and {Ahlen}, Steven and {Bianchi}, Davide and {Brooks}, David and {Burtin}, Etienne and {Chaussidon}, Edmond and {Claybaugh}, Todd and {Cole}, Shaun and {de la Macorra}, Axel and {Enriquez Vargas}, Miguel and {Gazta{\~n}aga}, Enrique and {Gutierrez}, Gaston and {Honscheid}, Klaus and {Howlett}, Cullan and {Kisner}, Theodore and {Landriau}, Martin and {Le Guillou}, Laurent and {Levi}, Michael and {Miquel}, Ramon and {Moustakas}, John and {Palanque-Delabrouille}, Nathalie and {Percival}, Will and {P{\'e}rez-R{\`a}fols}, Ignasi and {Ross}, Ashley J. and {Rossi}, Graziano and {Sanchez}, Eusebio and {Schlegel}, David and {Schubnell}, Michael and {Seo}, Hee-Jong and {Sprayberry}, David and {Tarl{\'e}}, Gregory and {Vargas Magana}, Mariana and {Weaver}, Benjamin Alan and {Zou}, Hu},
        title = "{Analytical and EZmock covariance validation for the DESI 2024 results}",
      journal = {arXiv e-prints},
         year = 2024,
        month = nov,
          eid = {arXiv:2411.12027},
        pages = {arXiv:2411.12027},
          doi = {10.48550/arXiv.2411.12027},
archivePrefix = {arXiv},
       eprint = {2411.12027},
 primaryClass = {astro-ph.CO},
       adsurl = {https://ui.adsabs.harvard.edu/abs/2024arXiv241112027F}
}

@ARTICLE{KP4s7-Rashkovetskyi,
       author = {{Rashkovetskyi}, M. and {Forero-S{\'a}nchez}, D. and {de Mattia}, A. and {Eisenstein}, D.~J. and {Padmanabhan}, N. and {Seo}, H. and {Ross}, A.~J. and {Aguilar}, J. and {Ahlen}, S. and {Alves}, O. and {Andrade}, U. and {Brooks}, D. and {Burtin}, E. and {Chen}, X. and {Claybaugh}, T. and {Cole}, S. and {de la Macorra}, A. and {Ding}, Z. and {Doel}, P. and {Fanning}, K. and {Ferraro}, S. and {Font-Ribera}, A. and {Forero-Romero}, J.~E. and {Garcia-Quintero}, C. and {Gil-Mar{\'\i}n}, H. and {Gontcho A Gontcho}, S. and {Gonzalez-Morales}, A.~X. and {Gutierrez}, G. and {Honscheid}, K. and {Howlett}, C. and {Juneau}, S. and {Kremin}, A. and {Le Guillou}, L. and {Manera}, M. and {Medina-Varela}, L. and {Mena-Fern{\'a}ndez}, J. and {Miquel}, R. and {Mueller}, E. and {Mu{\~n}oz-Guti{\'e}rrez}, A. and {Myers}, A.~D. and {Nie}, J. and {Niz}, G. and {Paillas}, E. and {Percival}, W.~J. and {Poppett}, C. and {P{\'e}rez-Fern{\'a}ndez}, A. and {Rezaie}, M. and {Rosado-Marin}, A. and {Rossi}, G. and {Ruggeri}, R. and {Sanchez}, E. and {Saulder}, C. and {Schlegel}, D. and {Schubnell}, M. and {Sprayberry}, D. and {Tarl{\'e}}, G. and {Weaver}, B.~A. and {Yu}, J. and {Zhao}, C. and {Zou}, H.},
        title = "{Semi-analytical covariance matrices for two-point correlation function for DESI 2024 data}",
      journal = {\jcap},
         year = 2025,
        month = jan,
       volume = {2025},
       number = {1},
          eid = {145},
        pages = {145},
          doi = {10.1088/1475-7516/2025/01/145},
archivePrefix = {arXiv},
       eprint = {2404.03007},
 primaryClass = {astro-ph.CO},
       adsurl = {https://ui.adsabs.harvard.edu/abs/2025JCAP...01..145R}
}

@ARTICLE{KP4s9-Perez-Fernandez,
       author = {{P{\'e}rez-Fern{\'a}ndez}, A. and {Medina-Varela}, L. and {Ruggeri}, R. and {Vargas-Maga{\~n}a}, M. and {Seo}, H. and {Padmanabhan}, N. and {Ishak}, M. and {Aguilar}, J. and {Ahlen}, S. and {Alam}, S. and {Alves}, O. and {Andrade}, U. and {Brieden}, S. and {Brooks}, D. and {Carnero Rosell}, A. and {Chen}, X. and {Claybaugh}, T. and {Cole}, S. and {Dawson}, K. and {de la Macorra}, A. and {de Mattia}, A. and {Dey}, Arjun and {Ding}, Z. and {Doel}, P. and {Fanning}, K. and {Garcia-Quintero}, C. and {Gazta{\~n}aga}, E. and {Gontcho}, S. Gontcho A. and {Gutierrez}, G. and {Honscheid}, K. and {Juneau}, S. and {Kirkby}, D. and {Kisner}, T. and {Lambert}, A. and {Landriau}, M. and {Lasker}, J. and {Le Guillou}, L. and {Manera}, M. and {Martini}, P. and {Meisner}, A. and {Mena-Fern{\'a}ndez}, J. and {Miquel}, R. and {Moustakas}, J. and {Myers}, A.~D. and {Nadathur}, S. and {Newman}, J.~A. and {Niz}, G. and {Paillas}, E. and {Palanque-Delabrouille}, N. and {Percival}, W.~J. and {Poppett}, C. and {Prada}, F. and {Rashkovetskyi}, M. and {Rocher}, A. and {Rossi}, G. and {Sanchez}, A. and {Sanchez}, E. and {Schubnell}, M. and {Sprayberry}, D. and {Tarl{\'e}}, G. and {Valcin}, D. and {Weaver}, B.~A. and {Yu}, J. and {Zou}, H.},
        title = "{Fiducial-cosmology-dependent systematics for the DESI 2024 BAO analysis}",
      journal = {\jcap},
         year = 2025,
        month = jan,
       volume = {2025},
       number = {1},
          eid = {144},
        pages = {144},
          doi = {10.1088/1475-7516/2025/01/144},
archivePrefix = {arXiv},
       eprint = {2406.06085},
 primaryClass = {astro-ph.CO},
       adsurl = {https://ui.adsabs.harvard.edu/abs/2025JCAP...01..144P}
}

@ARTICLE{KP4s10-Mena-Fernandez,
       author = {{Mena-Fern{\'a}ndez}, J. and {Garcia-Quintero}, C. and {Yuan}, S. and {Hadzhiyska}, B. and {Alves}, O. and {Rashkovetskyi}, M. and {Seo}, H. and {Padmanabhan}, N. and {Nadathur}, S. and {Howlett}, C. and {Alam}, S. and {Rocher}, A. and {Ross}, A.~J. and {Sanchez}, E. and {Ishak}, M. and {Aguilar}, J. and {Ahlen}, S. and {Andrade}, U. and {BenZvi}, S. and {Brooks}, D. and {Burtin}, E. and {Chen}, S. and {Chen}, X. and {Claybaugh}, T. and {Cole}, S. and {de la Macorra}, A. and {de Mattia}, A. and {Dey}, Arjun and {Dey}, B. and {Ding}, Z. and {Doel}, P. and {Fanning}, K. and {Forero-Romero}, J.~E. and {Gazta{\~n}aga}, E. and {Gil-Mar{\'\i}n}, H. and {Gontcho}, S. Gontcho A. and {Gutierrez}, G. and {Guy}, J. and {Hahn}, C. and {Honscheid}, K. and {Juneau}, S. and {Kremin}, A. and {Landriau}, M. and {Le Guillou}, L. and {Levi}, M.~E. and {Manera}, M. and {Martini}, P. and {Medina-Varela}, L. and {Meisner}, A. and {Miquel}, R. and {Moustakas}, J. and {Mueller}, E. and {Mu{\~n}oz-Guti{\'e}rrez}, A. and {Myers}, A.~D. and {Newman}, J.~A. and {Nie}, J. and {Niz}, G. and {Paillas}, E. and {Palanque-Delabrouille}, N. and {Percival}, W.~J. and {Poppett}, C. and {P{\'e}rez-Fern{\'a}ndez}, A. and {Rosado-Marin}, A. and {Rossi}, G. and {Ruggeri}, R. and {Saulder}, C. and {Schlegel}, D. and {Schubnell}, M. and {Sprayberry}, D. and {Tarl{\'e}}, G. and {Vargas-Maga{\~n}a}, M. and {Weaver}, B.~A. and {Yu}, J. and {Zhang}, H. and {Zou}, H.},
        title = "{HOD-dependent systematics for luminous red galaxies in the DESI 2024 BAO analysis}",
      journal = {\jcap},
         year = 2025,
        month = jan,
       volume = {2025},
       number = {1},
          eid = {133},
        pages = {133},
          doi = {10.1088/1475-7516/2025/01/133},
archivePrefix = {arXiv},
       eprint = {2404.03008},
 primaryClass = {astro-ph.CO},
       adsurl = {https://ui.adsabs.harvard.edu/abs/2025JCAP...01..133M}
}

@ARTICLE{KP4s11-Garcia-Quintero,
       author = {{Garcia-Quintero}, C. and {Mena-Fern{\'a}ndez}, J. and {Rocher}, A. and {Yuan}, S. and {Hadzhiyska}, B. and {Alves}, O. and {Rashkovetskyi}, M. and {Seo}, H. and {Padmanabhan}, N. and {Nadathur}, S. and {Howlett}, C. and {Ishak}, M. and {Medina-Varela}, L. and {McDonald}, P. and {Ross}, A.~J. and {Xie}, Y. and {Chen}, X. and {Bera}, A. and {Aguilar}, J. and {Ahlen}, S. and {Andrade}, U. and {BenZvi}, S. and {Brooks}, D. and {Burtin}, E. and {Chen}, S. and {Claybaugh}, T. and {Cole}, S. and {de la Macorra}, A. and {de Mattia}, A. and {Dey}, A. and {Dey}, B. and {Ding}, Z. and {Doel}, P. and {Fanning}, K. and {Forero-Romero}, J.~E. and {Gazta{\~n}aga}, E. and {Gil-Mar{\'\i}n}, H. and {Gontcho}, S. Gontcho A. and {Gutierrez}, G. and {Guy}, J. and {Hahn}, C. and {Honscheid}, K. and {Kremin}, A. and {Landriau}, M. and {Le Guillou}, L. and {Levi}, M.~E. and {Manera}, M. and {Martini}, P. and {Meisner}, A. and {Miquel}, R. and {Moustakas}, J. and {Mueller}, E. and {Mu{\~n}oz-Guti{\'e}rrez}, A. and {Myers}, A.~D. and {Newman}, J.~A. and {Nie}, J. and {Niz}, G. and {Paillas}, E. and {Palanque-Delabrouille}, N. and {Percival}, W.~J. and {Poppett}, C. and {P{\'e}rez-Fern{\'a}ndez}, A. and {Rosado-Marin}, A. and {Rossi}, G. and {Ruggeri}, R. and {Sanchez}, E. and {Schlegel}, D. and {Schubnell}, M. and {Sprayberry}, D. and {Tarl{\'e}}, G. and {Vargas-Maga{\~n}a}, M. and {Weaver}, B.~A. and {Yu}, J. and {Zhang}, H. and {Zhou}, R. and {Zou}, H.},
        title = "{HOD-dependent systematics in Emission Line Galaxies for the DESI 2024 BAO analysis}",
      journal = {\jcap},
         year = 2025,
        month = jan,
       volume = {2025},
       number = {1},
          eid = {132},
        pages = {132},
          doi = {10.1088/1475-7516/2025/01/132},
archivePrefix = {arXiv},
       eprint = {2404.03009},
 primaryClass = {astro-ph.CO},
       adsurl = {https://ui.adsabs.harvard.edu/abs/2025JCAP...01..132G}
}

@ARTICLE{Snowmass2013.Levi,
       author = {{Levi}, Michael and {Bebek}, Chris and {Beers}, Timothy and {Blum}, Robert and {Cahn}, Robert and {Eisenstein}, Daniel and {Flaugher}, Brenna and {Honscheid}, Klaus and {Kron}, Richard and {Lahav}, Ofer and {McDonald}, Patrick and {Roe}, Natalie and {Schlegel}, David and {representing the DESI collaboration}},
        title = "{The DESI Experiment, a whitepaper for Snowmass 2013}",
      journal = {arXiv e-prints},
         year = 2013,
        month = aug,
          eid = {arXiv:1308.0847},
        pages = {arXiv:1308.0847},
archivePrefix = {arXiv},
       eprint = {1308.0847},
 primaryClass = {astro-ph.CO},
       adsurl = {https://ui.adsabs.harvard.edu/abs/2013arXiv1308.0847L}
}

@ARTICLE{DESI2016b.Instr,
       author = {{DESI Collaboration} and {Aghamousa}, Amir and {Aguilar}, Jessica and {Ahlen}, Steve and {Alam}, Shadab and {Allen}, Lori E. and {Allende Prieto}, Carlos and {Annis}, James and {Bailey}, Stephen and {Balland}, Christophe and {Ballester}, Otger and {Baltay}, Charles and {Beaufore}, Lucas and {Bebek}, Chris and {Beers}, Timothy C. and {Bell}, Eric F. and {Bernal}, Jos{\'e} Luis and {Besuner}, Robert and {Beutler}, Florian and {Blake}, Chris and {Bleuler}, Hannes and {Blomqvist}, Michael and {Blum}, Robert and {Bolton}, Adam S. and {Briceno}, Cesar and {Brooks}, David and {Brownstein}, Joel R. and {Buckley-Geer}, Elizabeth and {Burden}, Angela and {Burtin}, Etienne and {Busca}, Nicolas G. and {Cahn}, Robert N. and {Cai}, Yan-Chuan and {Cardiel-Sas}, Laia and {Carlberg}, Raymond G. and {Carton}, Pierre-Henri and {Casas}, Ricard and {Castander}, Francisco J. and {Cervantes-Cota}, Jorge L. and {Claybaugh}, Todd M. and {Close}, Madeline and {Coker}, Carl T. and {Cole}, Shaun and {Comparat}, Johan and {Cooper}, Andrew P. and {Cousinou}, M. -C. and {Crocce}, Martin and {Cuby}, Jean-Gabriel and {Cunningham}, Daniel P. and {Davis}, Tamara M. and {Dawson}, Kyle S. and {de la Macorra}, Axel and {De Vicente}, Juan and {Delubac}, Timoth{\'e}e and {Derwent}, Mark and {Dey}, Arjun and {Dhungana}, Govinda and {Ding}, Zhejie and {Doel}, Peter and {Duan}, Yutong T. and {Ealet}, Anne and {Edelstein}, Jerry and {Eftekharzadeh}, Sarah and {Eisenstein}, Daniel J. and {Elliott}, Ann and {Escoffier}, St{\'e}phanie and {Evatt}, Matthew and {Fagrelius}, Parker and {Fan}, Xiaohui and {Fanning}, Kevin and {Farahi}, Arya and {Farihi}, Jay and {Favole}, Ginevra and {Feng}, Yu and {Fernandez}, Enrique and {Findlay}, Joseph R. and {Finkbeiner}, Douglas P. and {Fitzpatrick}, Michael J. and {Flaugher}, Brenna and {Flender}, Samuel and {Font-Ribera}, Andreu and {Forero-Romero}, Jaime E. and {Fosalba}, Pablo and {Frenk}, Carlos S. and {Fumagalli}, Michele and {Gaensicke}, Boris T. and {Gallo}, Giuseppe and {Garcia-Bellido}, Juan and {Gaztanaga}, Enrique and {Pietro Gentile Fusillo}, Nicola and {Gerard}, Terry and {Gershkovich}, Irena and {Giannantonio}, Tommaso and {Gillet}, Denis and {Gonzalez-de-Rivera}, Guillermo and {Gonzalez-Perez}, Violeta and {Gott}, Shelby and {Graur}, Or and {Gutierrez}, Gaston and {Guy}, Julien and {Habib}, Salman and {Heetderks}, Henry and {Heetderks}, Ian and {Heitmann}, Katrin and {Hellwing}, Wojciech A. and {Herrera}, David A. and {Ho}, Shirley and {Holland}, Stephen and {Honscheid}, Klaus and {Huff}, Eric and {Hutchinson}, Timothy A. and {Huterer}, Dragan and {Hwang}, Ho Seong and {Illa Laguna}, Joseph Maria and {Ishikawa}, Yuzo and {Jacobs}, Dianna and {Jeffrey}, Niall and {Jelinsky}, Patrick and {Jennings}, Elise and {Jiang}, Linhua and {Jimenez}, Jorge and {Johnson}, Jennifer and {Joyce}, Richard and {Jullo}, Eric and {Juneau}, St{\'e}phanie and {Kama}, Sami and {Karcher}, Armin and {Karkar}, Sonia and {Kehoe}, Robert and {Kennamer}, Noble and {Kent}, Stephen and {Kilbinger}, Martin and {Kim}, Alex G. and {Kirkby}, David and {Kisner}, Theodore and {Kitanidis}, Ellie and {Kneib}, Jean-Paul and {Koposov}, Sergey and {Kovacs}, Eve and {Koyama}, Kazuya and {Kremin}, Anthony and {Kron}, Richard and {Kronig}, Luzius and {Kueter-Young}, Andrea and {Lacey}, Cedric G. and {Lafever}, Robin and {Lahav}, Ofer and {Lambert}, Andrew and {Lampton}, Michael and {Landriau}, Martin and {Lang}, Dustin and {Lauer}, Tod R. and {Le Goff}, Jean-Marc and {Le Guillou}, Laurent and {Le Van Suu}, Auguste and {Lee}, Jae Hyeon and {Lee}, Su-Jeong and {Leitner}, Daniela and {Lesser}, Michael and {Levi}, Michael E. and {L'Huillier}, Benjamin and {Li}, Baojiu and {Liang}, Ming and {Lin}, Huan and {Linder}, Eric and {Loebman}, Sarah R. and {Luki{\'c}}, Zarija and {Ma}, Jun and {MacCrann}, Niall and {Magneville}, Christophe and {Makarem}, Laleh and {Manera}, Marc and {Manser}, Christopher J. and {Marshall}, Robert and {Martini}, Paul and {Massey}, Richard and {Matheson}, Thomas and {McCauley}, Jeremy and {McDonald}, Patrick and {McGreer}, Ian D. and {Meisner}, Aaron and {Metcalfe}, Nigel and {Miller}, Timothy N. and {Miquel}, Ramon and {Moustakas}, John and {Myers}, Adam and {Naik}, Milind and {Newman}, Jeffrey A. and {Nichol}, Robert C. and {Nicola}, Andrina and {Nicolati da Costa}, Luiz and {Nie}, Jundan and {Niz}, Gustavo and {Norberg}, Peder and {Nord}, Brian and {Norman}, Dara and {Nugent}, Peter and {O'Brien}, Thomas and {Oh}, Minji and {Olsen}, Knut A.~G.},
        title = "{The DESI Experiment Part II: Instrument Design}",
      journal = {arXiv e-prints},
         year = 2016,
        month = oct,
          eid = {arXiv:1611.00037},
        pages = {arXiv:1611.00037},
          doi = {10.48550/arXiv.1611.00037},
archivePrefix = {arXiv},
       eprint = {1611.00037},
 primaryClass = {astro-ph.IM},
       adsurl = {https://ui.adsabs.harvard.edu/abs/2016arXiv161100037D}
}

@ARTICLE{FocalPlane.Silber.2023,
       author = {{Silber}, Joseph Harry and {Fagrelius}, Parker and {Fanning}, Kevin and {Schubnell}, Michael and {Aguilar}, Jessica Nicole and {Ahlen}, Steven and {Ameel}, Jon and {Ballester}, Otger and {Baltay}, Charles and {Bebek}, Chris and {Benton Beard}, Dominic and {Besuner}, Robert and {Cardiel-Sas}, Laia and {Casas}, Ricard and {Castander}, Francisco Javier and {Claybaugh}, Todd and {Dobson}, Carl and {Duan}, Yutong and {Dunlop}, Patrick and {Edelstein}, Jerry and {Emmet}, William T. and {Elliott}, Ann and {Evatt}, Matthew and {Gershkovich}, Irena and {Guy}, Julien and {Harris}, Stu and {Heetderks}, Henry and {Heetderks}, Ian and {Honscheid}, Klaus and {Illa}, Jose Maria and {Jelinsky}, Patrick and {Jelinsky}, Sharon R. and {Jimenez}, Jorge and {Karcher}, Armin and {Kent}, Stephen and {Kirkby}, David and {Kneib}, Jean-Paul and {Lambert}, Andrew and {Lampton}, Mike and {Leitner}, Daniela and {Levi}, Michael and {McCauley}, Jeremy and {Meisner}, Aaron and {Miller}, Timothy N. and {Miquel}, Ramon and {Mundet}, Juli{\'a} and {Poppett}, Claire and {Rabinowitz}, David and {Reil}, Kevin and {Roman}, David and {Schlegel}, David and {Serrano}, Santiago and {Van Shourt}, William and {Sprayberry}, David and {Tarl{\'e}}, Gregory and {Tie}, Suk Sien and {Weaverdyck}, Curtis and {Zhang}, Kai and {Azzaro}, Marco and {Bailey}, Stephen and {Becerril}, Santiago and {Blackwell}, Tami and {Bouri}, Mohamed and {Brooks}, David and {Buckley-Geer}, Elizabeth and {Castro}, Jose Pe{\~n}ate and {Derwent}, Mark and {Dey}, Arjun and {Dhungana}, Govinda and {Doel}, Peter and {Eisenstein}, Daniel J. and {Fahim}, Nasib and {Garcia-Bellido}, Juan and {Gazta{\~n}aga}, Enrique and {A Gontcho}, Satya Gontcho and {Gutierrez}, Gaston and {H{\"o}rler}, Philipp and {Kehoe}, Robert and {Kisner}, Theodore and {Kremin}, Anthony and {Kronig}, Luzius and {Landriau}, Martin and {Le Guillou}, Laurent and {Martini}, Paul and {Moustakas}, John and {Palanque-Delabrouille}, Nathalie and {Peng}, Xiyan and {Percival}, Will and {Prada}, Francisco and {Allende Prieto}, Carlos and {de Rivera}, Guillermo Gonzalez and {Sanchez}, Eusebio and {Sanchez}, Justo and {Sharples}, Ray and {Soares-Santos}, Marcelle and {Schlafly}, Edward and {Weaver}, Benjamin Alan and {Zhou}, Zhimin and {Zhu}, Yaling and {Zou}, Hu and {DESI Collaboration}},
        title = "{The Robotic Multiobject Focal Plane System of the Dark Energy Spectroscopic Instrument (DESI)}",
      journal = {\aj},
         year = 2023,
        month = jan,
       volume = {165},
       number = {1},
          eid = {9},
        pages = {9},
          doi = {10.3847/1538-3881/ac9ab1},
archivePrefix = {arXiv},
       eprint = {2205.09014},
 primaryClass = {astro-ph.IM},
       adsurl = {https://ui.adsabs.harvard.edu/abs/2023AJ....165....9S}
}

@ARTICLE{Corrector.Miller.2023,
       author = {{Miller}, Timothy N. and {Doel}, Peter and {Gutierrez}, Gaston and {Besuner}, Robert and {Brooks}, David and {Gallo}, Giuseppe and {Heetderks}, Henry and {Jelinsky}, Patrick and {Kent}, Stephen M. and {Lampton}, Michael and {Levi}, Michael E. and {Liang}, Ming and {Meisner}, Aaron and {Sholl}, Michael J. and {Silber}, Joseph Harry and {Sprayberry}, David and {Aguilar}, Jessica Nicole and {de la Macorra}, Axel and {Eisenstein}, Daniel and {Fanning}, Kevin and {Font-Ribera}, Andreu and {Gazta{\~n}aga}, Enrique and {Gontcho A Gontcho}, Satya and {Honscheid}, Klaus and {Jimenez}, Jorge and {Joyce}, Dick and {Kehoe}, Robert and {Kisner}, Theodore and {Kremin}, Anthony and {Landriau}, Martin and {Le Guillou}, Laurent and {Magneville}, Christophe and {Martini}, Paul and {Miquel}, Ramon and {Moustakas}, John and {Nie}, Jundan and {Percival}, Will and {Poppett}, Claire and {Prada}, Francisco and {Rossi}, Graziano and {Schlegel}, David and {Schubnell}, Michael and {Seo}, Hee-Jong and {Sharples}, Ray and {Tarl{\'e}}, Gregory and {Vargas-Maga{\~n}a}, Mariana and {Zhou}, Zhimin and {the DESI Collaboration}},
        title = "{The Optical Corrector for the Dark Energy Spectroscopic Instrument}",
      journal = {\aj},
         year = 2024,
        month = aug,
       volume = {168},
       number = {2},
          eid = {95},
        pages = {95},
          doi = {10.3847/1538-3881/ad45fe},
archivePrefix = {arXiv},
       eprint = {2306.06310},
 primaryClass = {astro-ph.IM},
       adsurl = {https://ui.adsabs.harvard.edu/abs/2024AJ....168...95M}
}

@ARTICLE{FiberSystem.Poppett.2024,
       author = {{Poppett}, Claire and {Tyas}, Luke and {Aguilar}, J. and {Bebek}, Christopher and {Bramall}, D. and {Claybaugh}, T. and {Edelstein}, J. and {Fagrelius}, P. and {Heetderks}, H. and {Jelinsky}, P. and {Jelinsky}, S. and {Lafever}, Robin and {Lambert}, A. and {Lampton}, M. and {Levi}, Michael E. and {Martini}, P. and {Rockosi}, C. and {Schmoll}, J. and {Sharples}, Ray M. and {Sirk}, Martin and {Wishnow}, Edward and {Yu}, Jiaxi and {Ahlen}, S. and {Bault}, A. and {BenZvi}, S. and {Brooks}, D. and {Cole}, S. and {de la Macorra}, A. and {Dey}, Arjun and {Doel}, P. and {Fanning}, K. and {Font-Ribera}, A. and {Forero-Romero}, J.~E. and {Gazta{\~n}aga}, E. and {Gontcho A Gontcho}, S. and {Gonzalez-Morales}, A.~X. and {Hahn}, C. and {Honscheid}, K. and {Jimenez}, J. and {Juneau}, S. and {Kirkby}, D. and {Kremin}, A. and {Landriau}, M. and {Le Guillou}, L. and {Manera}, M. and {Meisner}, A. and {Miquel}, R. and {Moustakas}, J. and {Mueller}, E. and {Mu{\~n}oz-Guti{\'e}rrez}, A. and {Myers}, A.~D. and {Nie}, J. and {Niz}, G. and {Palanque-Delabrouille}, N. and {Percival}, W.~J. and {Prada}, F. and {Rabinowitz}, D. and {Rezaie}, M. and {Rossi}, G. and {Sanchez}, E. and {Schlafly}, Edward F. and {Schlegel}, D. and {Schubnell}, M. and {Seo}, H. and {Sprayberry}, D. and {Tarl{\'e}}, G. and {Vargas-Maga{\~n}a}, M. and {Weaver}, B.~A. and {Zhou}, R.},
        title = "{Overview of the Fiber System for the Dark Energy Spectroscopic Instrument}",
      journal = {\aj},
         year = 2024,
        month = dec,
       volume = {168},
       number = {6},
          eid = {245},
        pages = {245},
          doi = {10.3847/1538-3881/ad76a4},
       adsurl = {https://ui.adsabs.harvard.edu/abs/2024AJ....168..245P}
}

@ARTICLE{TS.Pipeline.Myers.2023,
       author = {{Myers}, Adam D. and {Moustakas}, John and {Bailey}, Stephen and {Weaver}, Benjamin A. and {Cooper}, Andrew P. and {Forero-Romero}, Jaime E. and {Abolfathi}, Bela and {Alexander}, David M. and {Brooks}, David and {Chaussidon}, Edmond and {Chuang}, Chia-Hsun and {Dawson}, Kyle and {Dey}, Arjun and {Dey}, Biprateep and {Dhungana}, Govinda and {Doel}, Peter and {Fanning}, Kevin and {Gazta{\~n}aga}, Enrique and {A Gontcho}, Satya Gontcho and {Gonzalez-Morales}, Alma X. and {Hahn}, ChangHoon and {Herrera-Alcantar}, Hiram K. and {Honscheid}, Klaus and {Ishak}, Mustapha and {Karim}, Tanveer and {Kirkby}, David and {Kisner}, Theodore and {Koposov}, Sergey E. and {Kremin}, Anthony and {Lan}, Ting-Wen and {Landriau}, Martin and {Lang}, Dustin and {Levi}, Michael E. and {Magneville}, Christophe and {Napolitano}, Lucas and {Martini}, Paul and {Meisner}, Aaron and {Newman}, Jeffrey A. and {Palanque-Delabrouille}, Nathalie and {Percival}, Will and {Poppett}, Claire and {Prada}, Francisco and {Raichoor}, Anand and {Ross}, Ashley J. and {Schlafly}, Edward F. and {Schlegel}, David and {Schubnell}, Michael and {Tan}, Ting and {Tarle}, Gregory and {Wilson}, Michael J. and {Y{\`e}che}, Christophe and {Zhou}, Rongpu and {Zhou}, Zhimin and {Zou}, Hu},
        title = "{The Target-selection Pipeline for the Dark Energy Spectroscopic Instrument}",
      journal = {\aj},
         year = 2023,
        month = feb,
       volume = {165},
       number = {2},
          eid = {50},
        pages = {50},
          doi = {10.3847/1538-3881/aca5f9},
archivePrefix = {arXiv},
       eprint = {2208.08518},
 primaryClass = {astro-ph.IM},
       adsurl = {https://ui.adsabs.harvard.edu/abs/2023AJ....165...50M}
}

@ARTICLE{Spectro.Pipeline.Guy.2023,
       author = {{Guy}, J. and {Bailey}, S. and {Kremin}, A. and {Alam}, Shadab and {Alexander}, D.~M. and {Allende Prieto}, C. and {BenZvi}, S. and {Bolton}, A.~S. and {Brooks}, D. and {Chaussidon}, E. and {Cooper}, A.~P. and {Dawson}, K. and {de la Macorra}, A. and {Dey}, A. and {Dey}, Biprateep and {Dhungana}, G. and {Eisenstein}, D.~J. and {Font-Ribera}, A. and {Forero-Romero}, J.~E. and {Gazta{\~n}aga}, E. and {Gontcho A Gontcho}, S. and {Green}, D. and {Honscheid}, K. and {Ishak}, M. and {Kehoe}, R. and {Kirkby}, D. and {Kisner}, T. and {Koposov}, Sergey E. and {Lan}, Ting-Wen and {Landriau}, M. and {Le Guillou}, L. and {Levi}, Michael E. and {Magneville}, C. and {Manser}, Christopher J. and {Martini}, P. and {Meisner}, Aaron M. and {Miquel}, R. and {Moustakas}, J. and {Myers}, Adam D. and {Newman}, Jeffrey A. and {Nie}, Jundan and {Palanque-Delabrouille}, N. and {Percival}, W.~J. and {Poppett}, C. and {Prada}, F. and {Raichoor}, A. and {Ravoux}, C. and {Ross}, A.~J. and {Schlafly}, E.~F. and {Schlegel}, D. and {Schubnell}, M. and {Sharples}, Ray M. and {Tarl{\'e}}, Gregory and {Weaver}, B.~A. and {Y{\'e}che}, Christophe and {Zhou}, Rongpu and {Zhou}, Zhimin and {Zou}, H.},
        title = "{The Spectroscopic Data Processing Pipeline for the Dark Energy Spectroscopic Instrument}",
      journal = {\aj},
         year = 2023,
        month = apr,
       volume = {165},
       number = {4},
          eid = {144},
        pages = {144},
          doi = {10.3847/1538-3881/acb212},
archivePrefix = {arXiv},
       eprint = {2209.14482},
 primaryClass = {astro-ph.IM},
       adsurl = {https://ui.adsabs.harvard.edu/abs/2023AJ....165..144G}
}

@ARTICLE{FBA.Raichoor.2024,
   author = {{Raichoor et al.}},
    year = 2025,
  journal = {, in preparation}
}

@ARTICLE{SurveyOps.Schlafly.2023,
       author = {{Schlafly}, Edward F. and {Kirkby}, David and {Schlegel}, David J. and {Myers}, Adam D. and {Raichoor}, Anand and {Dawson}, Kyle and {Aguilar}, Jessica and {Allende Prieto}, Carlos and {Bailey}, Stephen and {BenZvi}, Segev and {Bermejo-Climent}, Jose and {Brooks}, David and {de la Macorra}, Axel and {Dey}, Arjun and {Doel}, Peter and {Fanning}, Kevin and {Font-Ribera}, Andreu and {Forero-Romero}, Jaime E. and {Garc{\'\i}a-Bellido}, Juan and {Gontcho A Gontcho}, Satya and {Guy}, Julien and {Hahn}, ChangHoon and {Honscheid}, Klaus and {Ishak}, Mustapha and {Juneau}, St{\'e}phanie and {Kehoe}, Robert and {Kisner}, Theodore and {Kremin}, Anthony and {Landriau}, Martin and {Lang}, Dustin A. and {Lasker}, James and {Levi}, Michael E. and {Magneville}, Christophe and {Manser}, Christopher J. and {Martini}, Paul and {Meisner}, Aaron M. and {Miquel}, Ramon and {Moustakas}, John and {Newman}, Jeffrey A. and {Nie}, Jundan and {Palanque-Delabrouille}, Nathalie. and {Percival}, Will J. and {Poppett}, Claire and {Rockosi}, Constance and {Ross}, Ashley J. and {Rossi}, Graziano and {Tarl{\'e}}, Gregory and {Weaver}, Benjamin A. and {Y{\`e}che}, Christophe and {Zhou}, Rongpu and {DESI Collaboration}},
        title = "{Survey Operations for the Dark Energy Spectroscopic Instrument}",
      journal = {\aj},
         year = 2023,
        month = dec,
       volume = {166},
       number = {6},
          eid = {259},
        pages = {259},
          doi = {10.3847/1538-3881/ad0832},
archivePrefix = {arXiv},
       eprint = {2306.06309},
 primaryClass = {astro-ph.CO},
       adsurl = {https://ui.adsabs.harvard.edu/abs/2023AJ....166..259S}
}

@ARTICLE{LRG.TS.Zhou.2023,
       author = {{Zhou}, Rongpu and {Dey}, Biprateep and {Newman}, Jeffrey A. and {Eisenstein}, Daniel J. and {Dawson}, K. and {Bailey}, S. and {Berti}, A. and {Guy}, J. and {Lan}, Ting-Wen and {Zou}, H. and {Aguilar}, J. and {Ahlen}, S. and {Alam}, Shadab and {Brooks}, D. and {de la Macorra}, A. and {Dey}, A. and {Dhungana}, G. and {Fanning}, K. and {Font-Ribera}, A. and {Gontcho}, S. Gontcho A. and {Honscheid}, K. and {Ishak}, Mustapha and {Kisner}, T. and {Kov{\'a}cs}, A. and {Kremin}, A. and {Landriau}, M. and {Levi}, Michael E. and {Magneville}, C. and {Manera}, Marc and {Martini}, P. and {Meisner}, Aaron M. and {Miquel}, R. and {Moustakas}, J. and {Myers}, Adam D. and {Nie}, Jundan and {Palanque-Delabrouille}, N. and {Percival}, W.~J. and {Poppett}, C. and {Prada}, F. and {Raichoor}, A. and {Ross}, A.~J. and {Schlafly}, E. and {Schlegel}, D. and {Schubnell}, M. and {Tarl{\'e}}, Gregory and {Weaver}, B.~A. and {Wechsler}, R.~H. and {Y{\'e}che}, Christophe and {Zhou}, Zhimin},
        title = "{Target Selection and Validation of DESI Luminous Red Galaxies}",
      journal = {\aj},
         year = 2023,
        month = feb,
       volume = {165},
       number = {2},
          eid = {58},
        pages = {58},
          doi = {10.3847/1538-3881/aca5fb},
archivePrefix = {arXiv},
       eprint = {2208.08515},
 primaryClass = {astro-ph.CO},
       adsurl = {https://ui.adsabs.harvard.edu/abs/2023AJ....165...58Z}
}

@ARTICLE{ELG.TS.Raichoor.2023,
       author = {{Raichoor}, A. and {Moustakas}, J. and {Newman}, Jeffrey A. and {Karim}, T. and {Ahlen}, S. and {Alam}, Shadab and {Bailey}, S. and {Brooks}, D. and {Dawson}, K. and {de la Macorra}, A. and {de Mattia}, A. and {Dey}, A. and {Dey}, Biprateep and {Dhungana}, G. and {Eftekharzadeh}, S. and {Eisenstein}, D.~J. and {Fanning}, K. and {Font-Ribera}, A. and {Garc{\'\i}a-Bellido}, J. and {Gazta{\~n}aga}, E. and {A Gontcho}, S. Gontcho and {Guy}, J. and {Honscheid}, K. and {Ishak}, M. and {Kehoe}, R. and {Kisner}, T. and {Kremin}, Anthony and {Lan}, Ting-Wen and {Landriau}, M. and {Le Guillou}, L. and {Levi}, Michael E. and {Magneville}, C. and {Manera}, M. and {Martini}, P. and {Meisner}, Aaron M. and {Myers}, Adam D. and {Nie}, Jundan and {Palanque-Delabrouille}, N. and {Percival}, W.~J. and {Poppett}, C. and {Prada}, F. and {Ross}, A.~J. and {Ruhlmann-Kleider}, V. and {Sabiu}, C.~G. and {Schlafly}, E.~F. and {Schlegel}, D. and {Tarl{\'e}}, Gregory and {Weaver}, B.~A. and {Y{\`e}che}, Christophe and {Zhou}, Rongpu and {Zhou}, Zhimin and {Zou}, H.},
        title = "{Target Selection and Validation of DESI Emission Line Galaxies}",
      journal = {\aj},
         year = 2023,
        month = mar,
       volume = {165},
       number = {3},
          eid = {126},
        pages = {126},
          doi = {10.3847/1538-3881/acb213},
archivePrefix = {arXiv},
       eprint = {2208.08513},
 primaryClass = {astro-ph.CO},
       adsurl = {https://ui.adsabs.harvard.edu/abs/2023AJ....165..126R}
}

@ARTICLE{DESI.DR2.BAO.cosmo,
  title = {DESI DR2 results. II. Measurements of baryon acoustic oscillations and cosmological constraints},
  author = {Abdul Karim, M. and Aguilar, J. and Ahlen, S. and Alam, S. and Allen, L. and Prieto, C. Allende and Alves, O. and Anand, A. and Andrade, U. and Armengaud, E. and Aviles, A. and Bailey, S. and Baltay, C. and Bansal, P. and Bault, A. and Behera, J. and BenZvi, S. and Bianchi, D. and Blake, C. and Brieden, S. and Brodzeller, A. and Brooks, D. and Buckley-Geer, E. and Burtin, E. and Calderon, R. and Canning, R. and Rosell, A. Carnero and Carrilho, P. and Casas, L. and Castander, F. J. and Charles, M. and Chaussidon, E. and Chaves-Montero, J. and Chebat, D. and Chen, X. and Claybaugh, T. and Cole, S. and Cooper, A. P. and Cuceu, A. and Dawson, K. S. and de la Macorra, A. and de Mattia, A. and Deiosso, N. and Della Costa, J. and Demina, R. and Dey, A. and Dey, B. and Ding, Z. and Doel, P. and Edelstein, J. and Eisenstein, D. J. and Elbers, W. and Fagrelius, P. and Fanning, K. and Fern\'andez-Garc\'{\i}a, E. and Ferraro, S. and Font-Ribera, A. and Forero-Romero, J. E. and Frenk, C. S. and Garcia-Quintero, C. and Garrison, L. H. and Gazta\~naga, E. and Gil-Mar\'{\i}n, H. and Gontcho, S. Gontcho A. and Gonzalez, D. and Gonzalez-Morales, A. X. and Gordon, C. and Green, D. and Gutierrez, G. and Guy, J. and Hadzhiyska, B. and Hahn, C. and He, S. and Herbold, M. and Herrera-Alcantar, H. K. and Ho, M.-F. and Honscheid, K. and Howlett, C. and Huterer, D. and Ishak, M. and Juneau, S. and Kamble, N. V. and Kara\ifmmode \mbox{\c{c}}\else \c{c}\fi{}ayl��, N. G. and Kehoe, R. and Kent, S. and Kim, A. G. and Kirkby, D. and Kisner, T. and Koposov, S. E. and Kremin, A. and Krolewski, A. and Lahav, O. and Lamman, C. and Landriau, M. and Lang, D. and Lasker, J. and Le Goff, J. M. and Le Guillou, L. and Leauthaud, A. and Levi, M. E. and Li, Q. and Li, T. S. and Lodha, K. and Lokken, M. and Lozano-Rodr\'{\i}guez, F. and Magneville, C. and Manera, M. and Martini, P. and Matthewson, W. L. and Meisner, A. and Mena-Fern\'andez, J. and Menegas, A. and Mergulh\~ao, T. and Miquel, R. and Moustakas, J. and Mu\~noz-Guti\'errez, A. and Mu\~noz-Santos, D. and Myers, A. D. and Nadathur, S. and Naidoo, K. and Napolitano, L. and Newman, J. A. and Niz, G. and Noriega, H. E. and Paillas, E. and Palanque-Delabrouille, N. and Pan, J. and Peacock, J. A. and Ibanez, M. P. and Percival, W. J. and P\'erez-Fern\'andez, A. and P\'erez-R\`afols, I. and Pieri, M. M. and Poppett, C. and Prada, F. and Rabinowitz, D. and Raichoor, A. and Ram\'{\i}rez-P\'erez, C. and Rashkovetskyi, M. and Ravoux, C. and Rich, J. and Rocher, A. and Rockosi, C. and Rohlf, J. and Rom\'an-Herrera, J. O. and Ross, A. J. and Rossi, G. and Ruggeri, R. and Ruhlmann-Kleider, V. and Samushia, L. and Sanchez, E. and Sanders, N. and Schlegel, D. and Schubnell, M. and Seo, H. and Shafieloo, A. and Sharples, R. and Silber, J. and Sinigaglia, F. and Sprayberry, D. and Tan, T. and Tarl\'e, G. and Taylor, P. and Turner, W. and Ure\~na-L\'opez, L. A. and Vaisakh, R. and Valdes, F. and Valogiannis, G. and Vargas-Maga\~na, M. and Verde, L. and Walther, M. and Weaver, B. A. and Weinberg, D. H. and White, M. and Wolfson, M. and Y\`eche, C. and Yu, J. and Zaborowski, E. A. and Zarrouk, P. and Zhai, Z. and Zhang, H. and Zhao, C. and Zhao, G. B. and Zhou, R. and Zou, H.},
  collaboration = {DESI Collaboration},
  journal = {Phys. Rev. D},
  volume = {112},
  issue = {8},
  pages = {083515},
  numpages = {40},
  year = {2025},
  month = {Oct},
  publisher = {American Physical Society},
  doi = {10.1103/tr6y-kpc6},
  url = {https://link.aps.org/doi/10.1103/tr6y-kpc6},
  archivePrefix = {arXiv},
  eprint = {2503.14738},
  primaryClass = {astro-ph.CO}
}

@article{Chuang:2014vfa,
	Archiveprefix = {arXiv},
	Author = {Chuang, Chia-Hsun and Kitaura, Francisco-Shu and Prada, Francisco and Zhao, Cheng and Yepes, Gustavo},
	Doi = {10.1093/mnras/stu2301},
	Eprint = {1409.1124},
	Journal = {\mnras},
	Month = {Jan},
	Pages = {2621-2628},
	Primaryclass = {astro-ph.CO},
	Slaccitation = {%%CITATION = ARXIV:1409.1124;%%},
	Title = {{EZmocks: extending the Zel'dovich approximation to generate mock galaxy catalogues with accurate clustering statistics}},
	Volume = {446},
	Year = {2015}}

@ARTICLE{2001MNRAS.327.1297P,
       author = {{Percival}, Will J. and {Baugh}, Carlton M. and {Bland-Hawthorn}, Joss and {Bridges}, Terry and {Cannon}, Russell and {Cole}, Shaun and {Colless}, Matthew and {Collins}, Chris and {Couch}, Warrick and {Dalton}, Gavin and {De Propris}, Roberto and {Driver}, Simon P. and {Efstathiou}, George and {Ellis}, Richard S. and {Frenk}, Carlos S. and {Glazebrook}, Karl and {Jackson}, Carole and {Lahav}, Ofer and {Lewis}, Ian and {Lumsden}, Stuart and {Maddox}, Steve and {Moody}, Stephen and {Norberg}, Peder and {Peacock}, John A. and {Peterson}, Bruce A. and {Sutherland}, Will and {Taylor}, Keith},
        title = "{The 2dF Galaxy Redshift Survey: the power spectrum and the matter content of the Universe}",
      journal = {\mnras},
         year = 2001,
        month = nov,
       volume = {327},
       number = {4},
        pages = {1297-1306},
          doi = {10.1046/j.1365-8711.2001.04827.x},
archivePrefix = {arXiv},
       eprint = {astro-ph/0105252},
 primaryClass = {astro-ph},
       adsurl = {https://ui.adsabs.harvard.edu/abs/2001MNRAS.327.1297P}
}

@ARTICLE{1987MNRAS.226..655B,
       author = {{Bond}, J.~R. and {Efstathiou}, G.},
        title = "{The statistics of cosmic background radiation fluctuations}",
      journal = {\mnras},
         year = 1987,
        month = jun,
       volume = {226},
        pages = {655-687},
          doi = {10.1093/mnras/226.3.655},
       adsurl = {https://ui.adsabs.harvard.edu/abs/1987MNRAS.226..655B}
}

@ARTICLE{2001MNRAS.328.1039C,
       author = {{Colless}, Matthew and {Dalton}, Gavin and {Maddox}, Steve and {Sutherland}, Will and {Norberg}, Peder and {Cole}, Shaun and {Bland-Hawthorn}, Joss and {Bridges}, Terry and {Cannon}, Russell and {Collins}, Chris and {Couch}, Warrick and {Cross}, Nicholas and {Deeley}, Kathryn and {De Propris}, Roberto and {Driver}, Simon P. and {Efstathiou}, George and {Ellis}, Richard S. and {Frenk}, Carlos S. and {Glazebrook}, Karl and {Jackson}, Carole and {Lahav}, Ofer and {Lewis}, Ian and {Lumsden}, Stuart and {Madgwick}, Darren and {Peacock}, John A. and {Peterson}, Bruce A. and {Price}, Ian and {Seaborne}, Mark and {Taylor}, Keith},
        title = "{The 2dF Galaxy Redshift Survey: spectra and redshifts}",
      journal = {\mnras},
         year = 2001,
        month = dec,
       volume = {328},
       number = {4},
        pages = {1039-1063},
          doi = {10.1046/j.1365-8711.2001.04902.x},
archivePrefix = {arXiv},
       eprint = {astro-ph/0106498},
 primaryClass = {astro-ph},
       adsurl = {https://ui.adsabs.harvard.edu/abs/2001MNRAS.328.1039C}
}

@ARTICLE{2009ApJS..182..543A,
       author = {{Abazajian}, Kevork N. and {Adelman-McCarthy}, Jennifer K. and {Ag{\"u}eros}, Marcel A. and {Allam}, Sahar S. and {Allende Prieto}, Carlos and {An}, Deokkeun and {Anderson}, Kurt S.~J. and {Anderson}, Scott F. and {Annis}, James and {Bahcall}, Neta A. and {Bailer-Jones}, C.~A.~L. and {Barentine}, J.~C. and {Bassett}, Bruce A. and {Becker}, Andrew C. and {Beers}, Timothy C. and {Bell}, Eric F. and {Belokurov}, Vasily and {Berlind}, Andreas A. and {Berman}, Eileen F. and {Bernardi}, Mariangela and {Bickerton}, Steven J. and {Bizyaev}, Dmitry and {Blakeslee}, John P. and {Blanton}, Michael R. and {Bochanski}, John J. and {Boroski}, William N. and {Brewington}, Howard J. and {Brinchmann}, Jarle and {Brinkmann}, J. and {Brunner}, Robert J. and {Budav{\'a}ri}, Tam{\'a}s and {Carey}, Larry N. and {Carliles}, Samuel and {Carr}, Michael A. and {Castander}, Francisco J. and {Cinabro}, David and {Connolly}, A.~J. and {Csabai}, Istv{\'a}n and {Cunha}, Carlos E. and {Czarapata}, Paul C. and {Davenport}, James R.~A. and {de Haas}, Ernst and {Dilday}, Ben and {Doi}, Mamoru and {Eisenstein}, Daniel J. and {Evans}, Michael L. and {Evans}, N.~W. and {Fan}, Xiaohui and {Friedman}, Scott D. and {Frieman}, Joshua A. and {Fukugita}, Masataka and {G{\"a}nsicke}, Boris T. and {Gates}, Evalyn and {Gillespie}, Bruce and {Gilmore}, G. and {Gonzalez}, Belinda and {Gonzalez}, Carlos F. and {Grebel}, Eva K. and {Gunn}, James E. and {Gy{\"o}ry}, Zsuzsanna and {Hall}, Patrick B. and {Harding}, Paul and {Harris}, Frederick H. and {Harvanek}, Michael and {Hawley}, Suzanne L. and {Hayes}, Jeffrey J.~E. and {Heckman}, Timothy M. and {Hendry}, John S. and {Hennessy}, Gregory S. and {Hindsley}, Robert B. and {Hoblitt}, J. and {Hogan}, Craig J. and {Hogg}, David W. and {Holtzman}, Jon A. and {Hyde}, Joseph B. and {Ichikawa}, Shin-ichi and {Ichikawa}, Takashi and {Im}, Myungshin and {Ivezi{\'c}}, {\v{Z}}eljko and {Jester}, Sebastian and {Jiang}, Linhua and {Johnson}, Jennifer A. and {Jorgensen}, Anders M. and {Juri{\'c}}, Mario and {Kent}, Stephen M. and {Kessler}, R. and {Kleinman}, S.~J. and {Knapp}, G.~R. and {Konishi}, Kohki and {Kron}, Richard G. and {Krzesinski}, Jurek and {Kuropatkin}, Nikolay and {Lampeitl}, Hubert and {Lebedeva}, Svetlana and {Lee}, Myung Gyoon and {Lee}, Young Sun and {French Leger}, R. and {L{\'e}pine}, S{\'e}bastien and {Li}, Nolan and {Lima}, Marcos and {Lin}, Huan and {Long}, Daniel C. and {Loomis}, Craig P. and {Loveday}, Jon and {Lupton}, Robert H. and {Magnier}, Eugene and {Malanushenko}, Olena and {Malanushenko}, Viktor and {Mandelbaum}, Rachel and {Margon}, Bruce and {Marriner}, John P. and {Mart{\'\i}nez-Delgado}, David and {Matsubara}, Takahiko and {McGehee}, Peregrine M. and {McKay}, Timothy A. and {Meiksin}, Avery and {Morrison}, Heather L. and {Mullally}, Fergal and {Munn}, Jeffrey A. and {Murphy}, Tara and {Nash}, Thomas and {Nebot}, Ada and {Neilsen}, Eric H., Jr. and {Newberg}, Heidi Jo and {Newman}, Peter R. and {Nichol}, Robert C. and {Nicinski}, Tom and {Nieto-Santisteban}, Maria and {Nitta}, Atsuko and {Okamura}, Sadanori and {Oravetz}, Daniel J. and {Ostriker}, Jeremiah P. and {Owen}, Russell and {Padmanabhan}, Nikhil and {Pan}, Kaike and {Park}, Changbom and {Pauls}, George and {Peoples}, John, Jr. and {Percival}, Will J. and {Pier}, Jeffrey R. and {Pope}, Adrian C. and {Pourbaix}, Dimitri and {Price}, Paul A. and {Purger}, Norbert and {Quinn}, Thomas and {Raddick}, M. Jordan and {Re Fiorentin}, Paola and {Richards}, Gordon T. and {Richmond}, Michael W. and {Riess}, Adam G. and {Rix}, Hans-Walter and {Rockosi}, Constance M. and {Sako}, Masao and {Schlegel}, David J. and {Schneider}, Donald P. and {Scholz}, Ralf-Dieter and {Schreiber}, Matthias R. and {Schwope}, Axel D. and {Seljak}, Uro{\v{s}} and {Sesar}, Branimir and {Sheldon}, Erin and {Shimasaku}, Kazu and {Sibley}, Valena C. and {Simmons}, A.~E. and {Sivarani}, Thirupathi and {Allyn Smith}, J. and {Smith}, Martin C. and {Smol{\v{c}}i{\'c}}, Vernesa and {Snedden}, Stephanie A. and {Stebbins}, Albert and {Steinmetz}, Matthias and {Stoughton}, Chris and {Strauss}, Michael A. and {SubbaRao}, Mark and {Suto}, Yasushi and {Szalay}, Alexander S. and {Szapudi}, Istv{\'a}n and {Szkody}, Paula and {Tanaka}, Masayuki and {Tegmark}, Max and {Teodoro}, Luis F.~A. and {Thakar}, Aniruddha R. and {Tremonti}, Christy A. and {Tucker}, Douglas L. and {Uomoto}, Alan and {Vanden Berk}, Daniel E. and {Vandenberg}, Jan and {Vidrih}, S. and {Vogeley}, Michael S. and {Voges}, Wolfgang and {Vogt}, Nicole P. and {Wadadekar}, Yogesh and {Watters}, Shannon and {Weinberg}, David H. and {West}, Andrew A. and {White}, Simon D.~M. and {Wilhite}, Brian C. and {Wonders}, Alainna C. and {Yanny}, Brian and {Yocum}, D.~R. and {York}, Donald G. and {Zehavi}, Idit and {Zibetti}, Stefano and {Zucker}, Daniel B.},
        title = "{The Seventh Data Release of the Sloan Digital Sky Survey}",
      journal = {\apjs},
         year = 2009,
        month = jun,
       volume = {182},
       number = {2},
        pages = {543-558},
          doi = {10.1088/0067-0049/182/2/543},
archivePrefix = {arXiv},
       eprint = {0812.0649},
 primaryClass = {astro-ph},
       adsurl = {https://ui.adsabs.harvard.edu/abs/2009ApJS..182..543A}
}

@ARTICLE{2020ApJ...901..153D,
       author = {{du Mas des Bourboux}, H{\'e}lion and {Rich}, James and {Font-Ribera}, Andreu and {de Sainte Agathe}, Victoria and {Farr}, James and {Etourneau}, Thomas and {Le Goff}, Jean-Marc and {Cuceu}, Andrei and {Balland}, Christophe and {Bautista}, Julian E. and {Blomqvist}, Michael and {Brinkmann}, Jonathan and {Brownstein}, Joel R. and {Chabanier}, Sol{\`e}ne and {Chaussidon}, Edmond and {Dawson}, Kyle and {Gonz{\'a}lez-Morales}, Alma X. and {Guy}, Julien and {Lyke}, Brad W. and {de la Macorra}, Axel and {Mueller}, Eva-Maria and {Myers}, Adam D. and {Nitschelm}, Christian and {Mu{\~n}oz Guti{\'e}rrez}, Andrea and {Palanque-Delabrouille}, Nathalie and {Parker}, James and {Percival}, Will J. and {P{\'e}rez-R{\`a}fols}, Ignasi and {Petitjean}, Patrick and {Pieri}, Matthew M. and {Ravoux}, Corentin and {Rossi}, Graziano and {Schneider}, Donald P. and {Seo}, Hee-Jong and {Slosar}, An{\v{z}}e and {Stermer}, Julianna and {Vivek}, M. and {Y{\`e}che}, Christophe and {Youles}, Samantha},
        title = "{The Completed SDSS-IV Extended Baryon Oscillation Spectroscopic Survey: Baryon Acoustic Oscillations with Ly{\ensuremath{\alpha}} Forests}",
      journal = {\apj},
         year = 2020,
        month = oct,
       volume = {901},
       number = {2},
          eid = {153},
        pages = {153},
          doi = {10.3847/1538-4357/abb085},
archivePrefix = {arXiv},
       eprint = {2007.08995},
 primaryClass = {astro-ph.CO},
       adsurl = {https://ui.adsabs.harvard.edu/abs/2020ApJ...901..153D}
}

@ARTICLE{2007MNRAS.375.1329G,
       author = {{Guzik}, Jacek and {Bernstein}, Gary and {Smith}, Robert E.},
        title = "{Systematic effects in the sound horizon scale measurements}",
      journal = {\mnras},
         year = 2007,
        month = mar,
       volume = {375},
       number = {4},
        pages = {1329-1337},
          doi = {10.1111/j.1365-2966.2006.11385.x},
archivePrefix = {arXiv},
       eprint = {astro-ph/0605594},
 primaryClass = {astro-ph},
       adsurl = {https://ui.adsabs.harvard.edu/abs/2007MNRAS.375.1329G}
}

@ARTICLE{2013PhRvD..88j3520Y,
       author = {{Yoo}, Jaiyul and {Seljak}, Uro{\v{s}}},
        title = "{Signatures of first stars in galaxy surveys: Multitracer analysis of the supersonic relative velocity effect and the constraints from the BOSS power spectrum measurements}",
      journal = {\prd},
         year = 2013,
        month = nov,
       volume = {88},
       number = {10},
          eid = {103520},
        pages = {103520},
          doi = {10.1103/PhysRevD.88.103520},
archivePrefix = {arXiv},
       eprint = {1308.1401},
 primaryClass = {astro-ph.CO},
       adsurl = {https://ui.adsabs.harvard.edu/abs/2013PhRvD..88j3520Y}
}

@ARTICLE{2010PhRvD..82h3520T,
       author = {{Tseliakhovich}, Dmitriy and {Hirata}, Christopher},
        title = "{Relative velocity of dark matter and baryonic fluids and the formation of the first structures}",
      journal = {\prd},
         year = 2010,
        month = oct,
       volume = {82},
       number = {8},
          eid = {083520},
        pages = {083520},
          doi = {10.1103/PhysRevD.82.083520},
archivePrefix = {arXiv},
       eprint = {1005.2416},
 primaryClass = {astro-ph.CO},
       adsurl = {https://ui.adsabs.harvard.edu/abs/2010PhRvD..82h3520T}
}

@ARTICLE{2003ApJ...594..665B,
       author = {{Blake}, Chris and {Glazebrook}, Karl},
        title = "{Probing Dark Energy Using Baryonic Oscillations in the Galaxy Power Spectrum as a Cosmological Ruler}",
      journal = {\apj},
         year = 2003,
        month = sep,
       volume = {594},
       number = {2},
        pages = {665-673},
          doi = {10.1086/376983},
archivePrefix = {arXiv},
       eprint = {astro-ph/0301632},
 primaryClass = {astro-ph},
       adsurl = {https://ui.adsabs.harvard.edu/abs/2003ApJ...594..665B}
}

@ARTICLE{Y3BAO-Sanders,
	title = "{Optimizing combined tracers for DESI DR2 BAO}", 
	 author = {{Sanders et al.}},
 	   year = 2025,
	journal = {in preparation}
}

\appendix

% \newpage

\section{Constructing a multi-tracer density field}
\label{construct_field}

Approximating the fields as Gaussian, we have the following log-likelihood function:
\begin{align}
\sL =& - \frachalf \vdelta^t \vS^{-1} \vdelta
\\ \nonumber
& - \frachalf\left[\vn_1 -\vbn_1 \left(\vone + \vB_1 \vdelta\right)\right]^t
\vN_1^{-1} \left[\vn_1 -\vbn_1 \left(\vone + \vB_1 \vdelta\right)\right] 
\\ \nonumber
& -\frachalf\left[\vn_2 -\vbn_2 \left(\vone + \vB_2 \vdelta\right)\right]^t
\vN_2^{-1} \left[\vn_2 -\vbn_2 \left(\vone + \vB_2 \vdelta\right)\right]
\end{align}
where $\vdelta$ is the underlying density field we would like to estimate,
$\vS$ is the covariance (power spectrum) of this, 
$\vn_i$ is the counts of tracer $i$, $\vbn_i$ is the mean of this (left as a matrix for now to allow for, e.g., redshift evolution),
$\vB_i$ is the bias (left as a matrix to allow for, e.g., damping or 
redshift space distortions),
and $\vN_i$ is the noise covariance (e.g., $\bn_i$ in the simplest case (not $\bn_i^{-1}$ because this is for the raw field $\vn_i$)).
Taking a derivative with respect to $\vdelta$ and setting it to zero, we find a maximum likelihood solution
\be
\vdelta_{\rm ML}= \vM \left[
\left(\vbn_1 \vB_1\right)^t \vN_1^{-1} \left(\vn_1
-\vbn_1\right)
+
\left(\vbn_2 \vB_2\right)^t \vN_2^{-1} \left(\vn_2
-\vbn_2\right)
\right]
\ee
Setting aside $\vM$ (which does not contain $\vn_i$s) for the moment, and taking the simplest scenario of constant bias and 
standard shot noise, we have 
\begin{align}
\left(\vbn_1 \vB_1\right)^t \vN_1^{-1} \left(\vn_1
-\vbn_1\right)
&+
\left(\vbn_2 \vB_2\right)^t \vN_2^{-1} \left(\vn_2
-\vbn_2\right) \\ \nonumber
&\rightarrow
b_1 \left(\vn_1
-\vbn_1\right)
+
b_2 \left(\vn_2
-\vbn_2\right)
\end{align}
i.e., we see that the basic field we want to construct is simply the 
bias weighted counts (where note that $\vn_i$ is a set of delta functions at the positions of objects --- e.g., constructing this field on a grid amounts to summing over bias-weighted galaxies --- the $\vbn$ part would be similarly constructed from randoms).

$\vM$ is like a Wiener filter kernel:
\be
\vM \equiv 
\left[\vS^{-1}
+\left(\vbn_1 \vB_1\right)^t \vN_1^{-1}\vbn_1 \vB_1
+\left(\vbn_2 \vB_2\right)^t \vN_2^{-1}\vbn_2 \vB_2
\right]^{-1}
\ee
In the simplest case, it is just a diagonal constant, i.e., related to the normalization of the estimated $\vdelta$ (in any case, we can see it separates from the issue of how to combine the two tracers, i.e., it just tells us what to do with the combination).

For the simplest high signal-to-noise case the full $\vdelta$ is then 
\be
\delta_{\rm ML} = 
\frac{b_1 \Delta n_1 + b_2 \Delta n_2}{
b_1^2 \bn_1 + b_2^2 \bn_2}
\ee
where $\Delta n_i \equiv n_i - \bn_i$, {which is \cref{eq:deltaML}}.
The denominator is related to the fact that you are dividing out the effective bias of your $b_1 n_1+b_2 n_2$ field to get to $\delta$.
The bias in this simplest case is
\be
b_{\rm eff}= \frac{b_1^2 \bn_1 + b_2^2 \bn_2}{b_1 \bn_1 + b_2 \bn_2},
\ee
giving \cref{eq:beff}.

With full symmetry, even the full matrix equations would diagonalize in Fourier space, which would allow one to include RSD and possibly different damping in $\vB_i$ and do a $k$ and $\mu$-dependent combination.
Also, the $\vS$ term, which gives a Wiener filter-like suppression of low S/N modes, turns into a simple $P(k)$ here.
For a simple hack combination in redshift space, $b_i$ could just be replaced by e.g., $b_i+f 0.6^2$ or the redshift space bias at some typical $\mu\sim 0.6$.
We will test these improvements in future data analysis.

\section{Author Affiliations}
\label{sec:affiliations}

\noindent \hangindent=.5cm $^{1}${Department of Physics \& Astronomy, Ohio University, 139 University Terrace, Athens, OH 45701, USA}

\noindent \hangindent=.5cm $^{2}${University of California, Berkeley, 110 Sproul Hall \#5800 Berkeley, CA 94720, USA}

\noindent \hangindent=.5cm $^{3}${Center for Astrophysics $|$ Harvard \& Smithsonian, 60 Garden Street, Cambridge, MA 02138, USA}

\noindent \hangindent=.5cm $^{4}${Center for Cosmology and AstroParticle Physics, The Ohio State University, 191 West Woodruff Avenue, Columbus, OH 43210, USA}

\noindent \hangindent=.5cm $^{5}${Department of Physics, The Ohio State University, 191 West Woodruff Avenue, Columbus, OH 43210, USA}

\noindent \hangindent=.5cm $^{6}${Institute for Astronomy, University of Edinburgh, Royal Observatory, Blackford Hill, Edinburgh EH9 3HJ, UK}

\noindent \hangindent=.5cm $^{7}${Lawrence Berkeley National Laboratory, 1 Cyclotron Road, Berkeley, CA 94720, USA}

\noindent \hangindent=.5cm $^{8}${IRFU, CEA, Universit\'{e} Paris-Saclay, F-91191 Gif-sur-Yvette, France}

\noindent \hangindent=.5cm $^{9}${Department of Astronomy, The Ohio State University, 4055 McPherson Laboratory, 140 W 18th Avenue, Columbus, OH 43210, USA}

\noindent \hangindent=.5cm $^{10}${The Ohio State University, Columbus, 43210 OH, USA}

\noindent \hangindent=.5cm $^{11}${Physics Department, Yale University, P.O. Box 208120, New Haven, CT 06511, USA}

\noindent \hangindent=.5cm $^{12}${Department of Physics, Boston University, 590 Commonwealth Avenue, Boston, MA 02215 USA}

\noindent \hangindent=.5cm $^{13}${Leinweber Center for Theoretical Physics, University of Michigan, 450 Church Street, Ann Arbor, Michigan 48109-1040, USA}

\noindent \hangindent=.5cm $^{14}${University of Michigan, 500 S. State Street, Ann Arbor, MI 48109, USA}

\noindent \hangindent=.5cm $^{15}${Dipartimento di Fisica ``Aldo Pontremoli'', Universit\`a degli Studi di Milano, Via Celoria 16, I-20133 Milano, Italy}

\noindent \hangindent=.5cm $^{16}${INAF-Osservatorio Astronomico di Brera, Via Brera 28, 20122 Milano, Italy}

\noindent \hangindent=.5cm $^{17}${Department of Physics \& Astronomy, University College London, Gower Street, London, WC1E 6BT, UK}

\noindent \hangindent=.5cm $^{18}${Institute for Advanced Study, 1 Einstein Drive, Princeton, NJ 08540, USA}

\noindent \hangindent=.5cm $^{19}${Department of Physics and Astronomy, The University of Utah, 115 South 1400 East, Salt Lake City, UT 84112, USA}

\noindent \hangindent=.5cm $^{20}${Instituto de F\'{\i}sica, Universidad Nacional Aut\'{o}noma de M\'{e}xico,  Circuito de la Investigaci\'{o}n Cient\'{\i}fica, Ciudad Universitaria, Cd. de M\'{e}xico  C.~P.~04510,  M\'{e}xico}

\noindent \hangindent=.5cm $^{21}${Department of Astronomy \& Astrophysics, University of Toronto, Toronto, ON M5S 3H4, Canada}

\noindent \hangindent=.5cm $^{22}${Department of Physics \& Astronomy and Pittsburgh Particle Physics, Astrophysics, and Cosmology Center (PITT PACC), University of Pittsburgh, 3941 O'Hara Street, Pittsburgh, PA 15260, USA}

\noindent \hangindent=.5cm $^{23}${University of Chinese Academy of Sciences, Nanjing 211135, People's Republic of China.}

\noindent \hangindent=.5cm $^{24}${Institut de F\'{i}sica d’Altes Energies (IFAE), The Barcelona Institute of Science and Technology, Edifici Cn, Campus UAB, 08193, Bellaterra (Barcelona), Spain}

\noindent \hangindent=.5cm $^{25}${Departamento de F\'isica, Universidad de los Andes, Cra. 1 No. 18A-10, Edificio Ip, CP 111711, Bogot\'a, Colombia}

\noindent \hangindent=.5cm $^{26}${Observatorio Astron\'omico, Universidad de los Andes, Cra. 1 No. 18A-10, Edificio H, CP 111711 Bogot\'a, Colombia}

\noindent \hangindent=.5cm $^{27}${Institut d'Estudis Espacials de Catalunya (IEEC), c/ Esteve Terradas 1, Edifici RDIT, Campus PMT-UPC, 08860 Castelldefels, Spain}

\noindent \hangindent=.5cm $^{28}${Institute of Cosmology and Gravitation, University of Portsmouth, Dennis Sciama Building, Portsmouth, PO1 3FX, UK}

\noindent \hangindent=.5cm $^{29}${Institute of Space Sciences, ICE-CSIC, Campus UAB, Carrer de Can Magrans s/n, 08913 Bellaterra, Barcelona, Spain}

\noindent \hangindent=.5cm $^{30}${University of Virginia, Department of Astronomy, Charlottesville, VA 22904, USA}

\noindent \hangindent=.5cm $^{31}${Fermi National Accelerator Laboratory, PO Box 500, Batavia, IL 60510, USA}

\noindent \hangindent=.5cm $^{32}${Steward Observatory, University of Arizona, 933 N. Cherry Avenue, Tucson, AZ 85721, USA}

\noindent \hangindent=.5cm $^{33}${School of Mathematics and Physics, University of Queensland, Brisbane, QLD 4072, Australia}

\noindent \hangindent=.5cm $^{34}${Department of Physics, The University of Texas at Dallas, 800 W. Campbell Rd., Richardson, TX 75080, USA}

\noindent \hangindent=.5cm $^{35}${Department of Physics, Southern Methodist University, 3215 Daniel Avenue, Dallas, TX 75275, USA}

\noindent \hangindent=.5cm $^{36}${Department of Physics and Astronomy, University of California, Irvine, 92697, USA}

\noindent \hangindent=.5cm $^{37}${Departament de F\'{i}sica, Serra H\'{u}nter, Universitat Aut\`{o}noma de Barcelona, 08193 Bellaterra (Barcelona), Spain}

\noindent \hangindent=.5cm $^{38}${NSF NOIRLab, 950 N. Cherry Ave., Tucson, AZ 85719, USA}

\noindent \hangindent=.5cm $^{39}${Laboratoire de Physique Subatomique et de Cosmologie, 53 Avenue des Martyrs, 38000 Grenoble, France}

\noindent \hangindent=.5cm $^{40}${Instituci\'{o} Catalana de Recerca i Estudis Avan\c{c}ats, Passeig de Llu\'{\i}s Companys, 23, 08010 Barcelona, Spain}

\noindent \hangindent=.5cm $^{41}${Department of Physics and Astronomy, Siena College, 515 Loudon Road, Loudonville, NY 12211, USA}

\noindent \hangindent=.5cm $^{42}${Instituto de Estudios Astrof\'isicos, Facultad de Ingenier\'ia y Ciencias, Universidad Diego Portales, Av. Ej\'ercito Libertador 441, Santiago, Chile}

\noindent \hangindent=.5cm $^{43}${Department of Physics and Astronomy, University of Waterloo, 200 University Ave W, Waterloo, ON N2L 3G1, Canada}

\noindent \hangindent=.5cm $^{44}${Perimeter Institute for Theoretical Physics, 31 Caroline St. North, Waterloo, ON N2L 2Y5, Canada}

\noindent \hangindent=.5cm $^{45}${Waterloo Centre for Astrophysics, University of Waterloo, 200 University Ave W, Waterloo, ON N2L 3G1, Canada}

\noindent \hangindent=.5cm $^{46}${Instituto de Astrof\'{i}sica de Andaluc\'{i}a (CSIC), Glorieta de la Astronom\'{i}a, s/n, E-18008 Granada, Spain}

\noindent \hangindent=.5cm $^{47}${Departament de F\'isica, EEBE, Universitat Polit\`ecnica de Catalunya, c/Eduard Maristany 10, 08930 Barcelona, Spain}

\noindent \hangindent=.5cm $^{48}${Department of Physics and Astronomy, Sejong University, 209 Neungdong-ro, Gwangjin-gu, Seoul 05006, Republic of Korea}

\noindent \hangindent=.5cm $^{49}${Queensland University of Technology,  School of Chemistry \& Physics, George St, Brisbane 4001, Australia}

\noindent \hangindent=.5cm $^{50}${Abastumani Astrophysical Observatory, Tbilisi, GE-0179, Georgia}

\noindent \hangindent=.5cm $^{51}${Department of Physics, Kansas State University, 116 Cardwell Hall, Manhattan, KS 66506, USA}

\noindent \hangindent=.5cm $^{52}${Faculty of Natural Sciences and Medicine, Ilia State University, 0194 Tbilisi, Georgia}

\noindent \hangindent=.5cm $^{53}${CIEMAT, Avenida Complutense 40, E-28040 Madrid, Spain}

\noindent \hangindent=.5cm $^{54}${Max Planck Institute for Extraterrestrial Physics, Gie\ss enbachstra\ss e 1, 85748 Garching, Germany}

\noindent \hangindent=.5cm $^{55}${Department of Physics, University of Michigan, 450 Church Street, Ann Arbor, MI 48109, USA}

\noindent \hangindent=.5cm $^{56}${Kavli IPMU (WPI), UTIAS, The University of Tokyo, Kashiwa, Chiba 277-8583, Japan}

\noindent \hangindent=.5cm $^{57}${National Astronomical Observatories, Chinese Academy of Sciences, A20 Datun Road, Chaoyang District, Beijing, 100101, P.~R.~China}

\end{document}